\documentclass{aa}  
\bibpunct{(}{)}{;}{a}{}{,} 
\usepackage{graphicx}
\usepackage{txfonts}
\usepackage{hyperref}
\usepackage{multirow}
\usepackage{rotating}
\usepackage[flushleft]{threeparttable}
\usepackage{subfloat}
\usepackage{xcolor}
\usepackage{tablefootnote}
\usepackage{rotating}

\def\co{$^{12}$CO}
\def\tco{$^{13}$CO}
\def\ceo{C$^{18}$O}

\newcommand{\hi}{H~\textsc{i}}
\newcommand{\hii}{H~\textsc{ii}}
\newcommand{\msun}{$\rm M_\odot$}
\newcommand{\lsun}{$\rm L_\odot$}
\newcommand{\kms}{km~s$^{-1}$}

\newcommand{\hmole}{H$_2$}

\newcommand{\nthp}{N$_2$H$^+$}

\newcommand{\cth}{C$_2$H}

\newcommand{\micron}{$\mu$m}

\newcommand{\tcoone}{$^{13}$CO $J=1-0$}
\newcommand{\tcotwo}{$^{13}$CO $J=2-1$}
\newcommand{\ceotwo}{C$^{18}$O $J=2-1$}

\newcommand{\degree}{$^{\circ}$}

\newcommand{\nhtcd}{$N_{\rm H_{2}}$}
\newcommand{\dustt}{$T_{\rm dust}$}

\newcommand{\pnhtcd}{$p_{N_{\rm H_{2}}}$}
\newcommand{\pdustt}{$p_{T_{\rm dust}}$}
\newcommand{\vlsr}{$\rm v_{lsr}$}
\definecolor{blue}{rgb}{0.0,0.0,1}
\definecolor{green}{rgb}{0.0,1,0.0}
\hypersetup{
    colorlinks,
    linkcolor=blue,
    citecolor=blue,
    urlcolor=blue
    }
\usepackage{etoolbox}
\makeatletter
\patchcmd\@combinedblfloats{\box\@outputbox}{\unvbox\@outputbox}{}{\errmessage{\noexpand patch failed}}
\makeatother
\begin{document} 
   \title{\hii~regions and high-mass starless clump candidates}

   \subtitle{I: Catalogs and properties \thanks{Table 4 that describes the basic properties of high-mass starless clump candidates is only available in electronic form at the CDS via anonymous ftp to \url{cdsarc.u-strasbg.fr} (\url{130.79.128.5}) or via \url{http://cdsweb.u-strasbg.fr/cgi-bin/qcat?J/A+A/.}}\fnmsep\thanks{This publication is based on the SEDIGISM and PPMAP data.} }

   \author{S. Zhang \inst{1}
          \and
           A. Zavagno \inst{1,2}
           \and
           J. Yuan \inst{3}
           \and
           H. Liu  \inst{4, 5} 
           \and
           M. Figueira  \inst{6}    
           \and
           D. Russeil \inst{1} 
           \and 
           F. Schuller \inst{7}                
           \and
           K. A. Marsh \inst{8}
           \and
           Y. Wu \inst{9}         
          }

   \institute{Aix Marseille Univ, CNRS, CNES, LAM, Marseille, France\\
              \email{siju.zhang@lam.fr, sijuzhangastro@gmail.com}
         \and
             {Institut Universitaire de France (IUF)}
         \and
             {National Astronomical Observatories, Chinese Academy of Sciences, 20A Datun Road, Chaoyang District, Beijing 100012, China}  
          \and
             {CASSACA, China-Chile Joint Center for Astronomy, Camino El Observatorio \#1515, Las Condes, Santiago, Chile}
          \and
             {Departamento de Astronom\'ia, Universidad de Concepci\'on, Av. Esteban Iturra s/n, Distrito Universitario, 160-C, Chile} 
          \and
             {National Centre for Nuclear Research, ul. Pasteura 7, 02-093, Warszawa} 
  \       \and
             {Max-Planck-Institut f\"ur Radioastronomie, Auf dem H\"ugel 69, D-53121 Bonn} 
          \and
             {Infrared Processing and Analysis Center, California Institute of Technology, Pasadena, California 91125}        
          \and
             {Department of Astronomy, Peking University, 100871 Beijing, China} \\
             }
 
   \date{Received September 26, 2019; accepted }

 
  \abstract
   {The role of ionization feedback on high-mass (> 8 \msun) star formation is still highly debated. Questions remain concerning the presence of nearby \hii~regions changes the properties of early high-mass star formation and whether \hii~regions promote or inhibit the formation of high-mass stars.}
   {To characterize the role of \hii~regions on the formation of high-mass stars, we study the properties of a sample of candidates high-mass starless clumps (HMSCs), of which  about 90\% have masses larger than 100 \msun. These high-mass objects probably represent the earliest stages of high-mass star formation; we search if (and how) their properties are modified by the presence of an \hii~region.}
   {We took advantage of the recently published catalog of HMSC candidates. By cross matching the HMSCs and \hii~regions, we classified HMSCs into three categories: 1) The HMSCs associated with \hii~regions both in the position in the projected plane of the sky and in velocity; 2) HMSCs associated in the plane of the sky, but not in velocity; and 3) HMSCs far away from any \hii~regions in the projected sky plane. We carried out comparisons between associated and nonassociated HMSCs based on statistical analyses of multiwavelength data from infrared to radio.}
   {We show that there are systematic differences of the properties of HMSCs in different environments. Statistical analyses suggest that HMSCs associated with \hii~regions are warmer, more luminous, more centrally-peaked and turbulent. We also clearly show, for the first time, that the ratio of bolometric luminosity to envelope mass of HMSCs ($L/M$) could not be a reliable evolutionary probe for early massive star formation due to the external heating effects of the \hii~regions.}
   {We show HMSCs associated with \hii~regions present statistically significant differences from HMSCs far away from \hii~regions, especially for dust temperature and $L/M$. More centrally peaked and turbulent properties of HMSCs associated with \hii~regions may promote the formation of high-mass stars by limiting fragmentation. High-resolution interferometric surveys toward HMSCs are crucial to reveal how \hii~regions impact the star formation process inside HMSCs.}


   \maketitle
%

\section{Introduction} \label{Intro}

The formation of high-mass (>~8~\msun) stars is still a mystery as a consequence of their rapid evolution, larger distance, and violent feedback, which are all deeply embedded in molecular clouds. With the benefits of higher resolution and sensitivity from millimeter interferometers such as Atacama Large Millimeter/Submillimeter Array (ALMA), the details happening at the very early evolutionary stages of high-mass star formation (HMSF) have been investigated in recent years. Two popular models for HMSF have been discussed widely: (1) monolithic collapse \citep{mck03, kru09, cse18, san18, mot19} and (2) competitive accretion \citep{bon07, cyg17, fon18}. It is possible that these models are not strongly and mutually incompatible as described by many current studies; these two models of HMSF are expected to be merged into a unified and consistent star formation model with the progress of observations and theories \citep{sch15}. The new evolutionary scenario proposed by \citet{mot18} suggests that the large-scale (0.1 to 1~pc) gas reservoirs, known as starless massive dense cores (MDCs) or starless clumps, could replace the high-mass analogs of prestellar cores (about 0.02~pc). These mass reservoirs could concentrate their mass into high-mass cores at the same time when stellar embryos are accreting.

The complexities of HMSF result from not only the controversially basic pictures of formation mechanism but also from the environmental impacts on the formation process. The powerful UV radiation of massive stars ionizes hydrogen to form \hii~regions. Because high-mass stars form in the densest parts of molecular clouds, \hii~regions are expected to become a typical environment of HMSF. Previous studies found that at least 30\% of HMSF in the Galaxy are observed at the edges of \hii~regions \citep{deh10, ken16, pal17}. 

The possible impacts of ionization radiation on star formation have been investigated in many publications. \hii~regions could trigger star formation with several mechanisms, mainly depending on the density and configuration of the surrounding medium (see Sect.~5 in \citealt{deh10}), such as  the collect and collapse (C\&C) mechanism in IR dust bubbles \citep{zav10, sam14, dur17} or the radiation-driven implosion (RDI) mechanism in pillars or bright rimmed clouds \citep{ber89, ima17, marshall19}. 

The interferometric studies of HMSF generally focus on physical and/or chemical phenomena inside the massive clumps or in a single field of view. This could produce a blind zone for the possible interactions between massive clumps, HMSF, and their various environments. For example, \citet{cse17} used ALMA\ to survey fragmentation at the early evolutionary stage of IR-quiet massive clumps and ignored possible environmental factors, even though environmental factors such as the presence of nearby \hii~region could impact the fragmentation process \citep{bus16}.

In a series of forthcoming papers, we try to systematically explore how the environment of the \hii~region changes HMSF, particularly for the HMSF at the very early evolutionary stages. High-mass starless clumps (HMSCs), which are very cold (dust temperature of around 10 to 20~K) and massive (from a few hundreds to several thousands of solar mass at a typical scale of 0.1 pc to 1~pc), are thought to be the possible precursors of high-mass stars. These sources are likely at the earliest stage of HMSF (quiescent prestellar stage), and only less than 10\% of the clumps with the ability to probably form massive stars are at this stage, indicating their short lifetime \citep{urq18}. Observationally, HMSCs are dark against their background environment from optical to mid-infrared (mid-IR) owing to the absence of the embedded protostellar objects.

Systematic studies of HMSCs under the effect of \hii~regions are very crucial to characterize the influence of the \hii~regions and how the influence impacts very early HMSF and the associated gas dynamics. We investigate the impacts of \hii~regions on a large sample of 463 HMSC candidates recently selected by \citet{yua17} (written briefly as Y17 in the following) in the inner Galactic plane. As in the first paper, we cross match HMSCs and \hii~regions based on their mutual correlation in the plane of sky to ascertain their association and then investigate the differences of the physical properties between the HMSCs associated with and those not associated with \hii~regions.  

This paper is organized as follows: The sample selection and archival data acquisition are described in Sect.~\ref{DA}. Sect.~\ref{RIPP} shows how we determine the basic parameters of HMSCs with updated data, followed by cross matching of \hii~regions and HMSCs in Sect.~\ref{HRCM}. In Sect.~\ref{DIG} to \ref{TVS}, we explore and compare the properties of different HMSCs. In Sect.~\ref{discussion}, we discuss the results to shed light on the role of ionization feedback on modifying the physical properties of HMSCs. The conclusions are finally presented in Sect.~\ref{conclusions}.\\

\section{Sample and data} \label{DA}
\subsection{Sample of HMSC candidates} \label{HS}
There are several studies dedicated to the search of starless clumps in the Milky Way. \citet{eli17} explore the physical properties of Herschel infrared Galactic Plane Survey (Hi-GAL, see \citealt{mol10}) compact sources in the inner Galaxy and termed the starless cores or clumps when they are not associated with 70 \micron~counterpart. A large number of 76338 starless clump/core candidates are found. About 65\%, 52\%, and 3\% of these clumps are compatible with massive star formation according to the different thresholds, which are $M(r)$~>~870~\msun$(r/{\rm pc})^{1.33}$, or ${\Sigma}_{\rm crit}$~=~0.2 and 1~g~cm$^{-2}$ suggested by \citet{kau10}, \citet{but12}, and \citet{kru08}, respectively. With Bolocam Galactic Plane Survey data (BGPS; \citealt{agu11}), \citet{svo16} (S16) obtained a sample of 2223 starless clump candidates in 10\degree~<~l~<~65\degree~based on the absence of a series of observational signatures of star formation (see Table \ref{criteria_table}).

The recent search for starless clumps by Y17, which targeted at \textbf{only massive starless clumps ($\gtrsim$~100~\msun~for 90\%~of these clumps at a typical scale of 0.2 to 0.8~pc)} in the inner Galactic plane ($|l|$~<~60\degree, $|b|$~<~1\degree), resulted in a sample of 463 HMSC candidates by extracting APEX Telescope Large Area Survey of the Galaxy 870~\micron~clumps (ATLASGAL; \citealt{cse14}) based on a series of selection criteria. Firstly, Y17 selected clumps with a 870~\micron~peak intensity higher than 0.5~Jy~beam$^{-1}$ to ensure that the column density is sufficient to form a massive star. Then, these authors removed clumps with star formation indicators such as young stellar objects (YSOs) identified by Galactic Legacy Infrared Midplane Survey Extraordinaire data (GLIMPSE; \citealt{chu09}), MIPS Galactic Plane Survey 24~\micron~data (MIPSGAL; \citealt{car09}), and Herschel 70~\micron~data \citep{mol10}. Besides, various kinds of star formation indicators in the SIMBAD database\footnote{\url{http://simbad.u-strasbg.fr/simbad/sim-display?data=otypes}} were also used. Finally, all candidates were inspected in Spitzer images by eye. 

A comparison between the samples of Y17 and S16 shows that only 20\% of ATLASGAL counterparts of BGPS starless clumps are in the HMSC sample of Y17. This could be due to denser or more reliably starless properties of HMSCs in the sample of Y17. Additional star formation indicators such as outflows (candidates) are considered. Furthermore, the inclusion of extended 24~\micron~could mitigate false identifications due to a strong background. Another factor could be the better resolution of ATLASGAL (about 20\arcsec) compared to BGPS (about 33\arcsec), which may resolve one BGPS clump to more individual structures in ATLASGAL considering that BGPS could detect the similar mass range as ATLASGAL \citep{ell15}. 

We chose the sample by Y17 as our \textbf{HMSC candidates} because it is probably more robust and focused on massive clumps. Even though a larger number of star formation indicators are cross matched, the selected HMSCs are still starless candidates rather than absolutely starless \citep{pil19}. One of the examples is the massive 70~\micron~quiet clump, which was considered a more reliable starless candidate in previous, while evidences of embedded low- to intermediate-mass star formation have been found in such clump recently \citep{tra17, svo19, li19, san19}. Basically, HMSCs hereafter represent HMSC candidates unless there are special explanations.\\

      \begin{table}[ht]
      \begin{threeparttable}
      \tiny
      \caption{Star formation indicators.} 
      \label{criteria_table} 
       \begin{tabular}{c c c c} 
       \hline\hline 
             Search                   & IR point sources                    & Maser               & Others        \\ 
       \hline 
       \multirow{2}*{S16}   &  Hi-GAL 70~\micron      &   H$_{2}$O masers    &   UC\hii~regions          \\
                                      &  2-24~\micron~YSOs              &   CH$_{3}$OH masers  &                       \\
       \hline                                                        
       \multirow{8}*{Y17}    & Hi-GAL 70~\micron       &   \textbf{masers}\tnote{a}      &  \textbf{\hii~regions}      \\
                                      &  MIPSGAL sources              &                     &  extended 24~\micron     \\    
                                      &  GLIMPSE YSOs                 &                     &  \textbf{cm radio sources}    \\
                                      &  \textbf{IR sources}            &                     &  \textbf{HH objects}          \\
                                      &  \textbf{YSOs (C.)\tnote{b}}       &                      &  \textbf{outflows (C.)} \\
                                      &  \textbf{PMS (C.)}   &                       &                              \\
                                      &  \textbf{T Tau (C.)} &                     &                             \\
                                      &  \textbf{HAEBE (C.)}&              &                            \\             
        \hline                                         
      \end{tabular} 
      \begin{tablenotes}
      \item [a] The sources in bold are the star formation indicators in SIMBAD database.
      \item [b] The abbreviation ``C.'' means candidate.
      \end{tablenotes}
      \end{threeparttable}
      \end{table}

\subsection{Ionized region sample} \label{IRS}
The ionized regions created during HMSF produce a shell or layer of photodissociation region (PDR) that can be observed by the PAH\footnote{Polycyclic aromatic hydrocarbon (PAH) commonly exits in the PDR \citep{fle10}.} emission in the mid-IR, for example, via GLIMPSE 8~\micron~or Wide-field Infrared Survey Explorer 12~\micron~emission (WISE; \citealt{wri10}). This shell or layer surrounds the hot thermal dust emission of ionized regions, which can be observed in MIPSGAL 24~\micron~or WISE 22~\micron~emission. The photodissociation shell and the inner hot thermal dust emission in mid-IR images make up the objects that are generally named as IR dust bubbles \citep{deh10}. 

The ionized regions are originally traced by the emission from ionized gas such as the free-free centimeter continuum or radio recombination lines (RRL). The close correlations between the mid-IR hot dust emission and the centimeter radio free-free emission of \hii~regions have been revealed by several studies \citep{ing14, mak17}, which indicate a tight spatial association between \hii~regions and IR dust bubbles. \citet{chu06} cataloged 322 IR bubbles by visual examination of GLIMPSE mid-IR images. These authors argued that most of the bubbles are formed by hot young stars in HMSF regions. Only three of their bubbles are identified as known supernova remnants (SNRs). The overlapping fraction between \hii~regions and bubbles could approach $\simeq$~85\% or even more in other similar studies \citep{deh10, ban10, and11}, which shed light on the association between \hii~regions, bubbles, and HMSF.

We used the IR bubbles identified in mid-IR images as our candidates of \hii~regions. Based on GLIMPSE images, a large catalog of more than 5000 IR bubbles is attained through the visual identification by tens of thousands citizens within Milky Way Project (MWP) \citep{sim12, ken12}. This catalog consists of the following two types of bubbles based on angular size: large (>~30\arcsec) and small (<~30\arcsec) bubbles. In this work, we take advantage of the sample of 3744 large bubbles as our \hii~region candidates. The small bubble candidates are ignored because of their high level of confusion with other astronomical objects unrelated to star formation such as AGB. Another reason is that the resolutions of millimeter and far-infrared (far-IR) images in our studies are close to the scale of small bubble (20\arcsec~to 30\arcsec). The unresolved \hii~regions could mix with nearby HMSCs, leading to the unreal properties of HMSCs. 

In addition, \citet{and14} constructed a catalog of over 8000 Galactic \hii~regions (candidates), around 2000 of which are known \hii~regions. They developed an online tool for viewing WISE emission of these \hii~regions\footnote{\url{http://astro.phys.wvu.edu/wise/}}, which is updated when there are new radio continuum and RRL observations such as \hii~Region Discovery Survey (HRDS) \citep{and15,and18, wen19}. The distant Galactic \hii~regions are also included in HRDS \citep{and15}. \citet{and14} propose that WISE has a sensitivity to detect all Galactic \hii~regions (see Figure 2 in their paper). The size and position of \hii~region (bubble) candidates in this online tool are outlined by characteristic morphology of WISE mid-IR emission. We use this catalog and associated online tool as supplements in searching for our \hii~region candidates. \\

\subsection{Survey data} \label{SD}
The main observing parameters of the surveys we used for this work are listed in Table~\ref{table:1}. Several types of data are included as follows: \textbf{(1) Infrared data:} GLIMPSE mapped the Galactic plane in mid-IR and we used this survey to show the spatial distribution of PAH, hot dust, and YSOs. \textbf{(2) (Sub)millimeter and far-IR continuum data:} ATLASGAL 870 \micron~survey ($|l|$~<~60\degree, $|b|$~<~1.5\degree) and Point Process Mapping for Hi-GAL (Hi-GAL PPMAP; \citealt{mar15}) column density and dust temperature data are used. A more detailed description about Hi-GAL PPMAP is given in Sect.~\ref{CDT}. \textbf{(3) (Sub)millimeter molecular line data:} Structure, Excitation, and Dynamics of the Inner Galactic InterStellar Medium survey (SEDIGISM; \citealt{sch17}) mapped the inner Galactic plane ($-$60\degree~$\leq$~$l$~$\leq$~18\degree, $|b|$~$\leq$~0.5\degree) in the $J = 2 - 1$ rotational transition of \tco~and \ceo~using the SHFI instrument on APEX with a spatial resolution of about 30\arcsec~and a velocity resolution of 0.25~\kms~\citep{sch17}. We used this survey to explore kinematics of the HMSCs. The survey is described in more detail in Sect.~\ref{TVS}. \textbf{(4) Centimeter radio images}: Multi-Array Galactic Plane Imaging Survey (MAGPIS, first quadrant the Galactic plane; \citealt{hel06}), Sydney University Molonglo Sky Survey (SUMSS, southern sky; \citealt{boc99}), and VLA Galactic Plane Survey (VGPS, 18\degree~<~$l$~<~67\degree~and $b$~<~1.3\degree~to 3.3\degree; \citealt{sti06}) are used to indicate the \hii~regions. \\

      \begin{table*}[ht]
      \centering 
      \tiny
      \caption{Information about the surveys used in this work.} 
      \label{table:1} 
       \begin{tabular}{c c c c c c} 
       \hline\hline 
             Survey                   &               Band & Resolutions               & Pixel Sizes                    & Facilities                  & References\\ 
       \hline 
        \multirow{4}*{GLIMPSE}        & 3.6 \micron        & \multirow{4}*{$\simeq$2\arcsec} & \multirow{4}*{1.2\arcsec}  &  \multirow{4}*{\textit{Spitzer} IRAC} &  \multirow{4}*{\citet{chu09}}  \\   
                                      & 4.5 \micron        &                            &                                 &                             &            \\
                                      & 5.8 \micron        &                            &                                 &                             &            \\
                                      & 8.0 \micron        &                            &                                 &                             &         \\                         
       \hline
        ATLASGAL                      & 870 \micron        &   19.2\arcsec              & 6\arcsec                        & APEX LABOCA                 &  \citet{sch09}       \\
       \hline
      SEDIGISM                        & 220 GHz            &    30\arcsec               &  9.5\arcsec                     &  APEX  SHFI                  &  \citet{sch17}              \\
        \hline  
    PPMAP                             &  -                 &    12\arcsec               &  6\arcsec                       &  \textit{Herschel}             &  \citet{mar15}          \\
       \hline       
        \multirow{3}*{MAGPIS}         &  6 cm              &   ~1.8\arcsec              & 0.6\arcsec                        & \multirow{3}*{VLA}          &\multirow{3}*{\citet{hel06}}     \\   
                                      & 21 cm              &   ~6.0\arcsec              & 2\arcsec                        &                             &             \\                 
                                      & 90 cm              &   ~25\arcsec               & 6\arcsec                        &                             &                          \\
       \hline
           SUMSS                      & 35 cm & ~43\arcsec $\times$ 43\arcsec cosec|$\delta$| & 11\arcsec                   &  MOST                       & \citet{boc99}          \\ 
       \hline  
          VGPS                        & 20 cm              &    45\arcsec                &    18\arcsec                   &     VLA                     &  \citet{sti06}         \\
        \hline                                         
      \end{tabular}
      \end{table*}

\section{Physical parameters} \label{RIPP}
\subsection{Velocity and distance}  \label{VD} 
\begin{figure}
      \centering
    \includegraphics[width=0.45\textwidth]{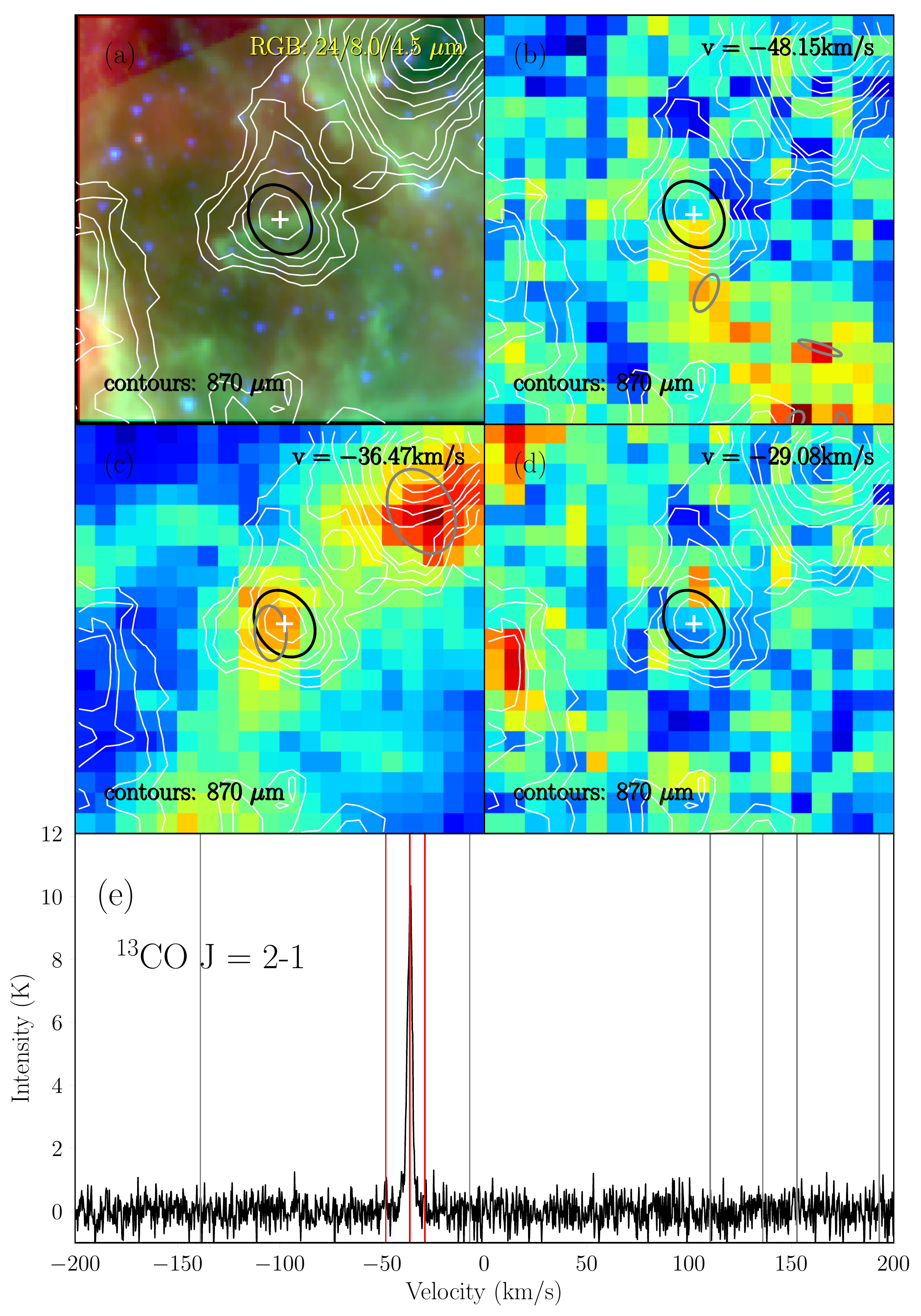}
       \caption{Example of the \vlsr~determination with the results of Hi-GAL distance project for the HMSC G305.1543+0.0477. Panel (a) shows the ATLASGAL contours overlaid on RGB image constructed from GLIMPSE 24~\micron~(in red), 8~\micron~(in green), and 4.5~\micron~(in blue) images. Panels (b), (c), and (d) show the SEDIGISM \tcotwo~intensity at the velocity of $-$48.15, $-$36.47, and $-$29.08~\kms, respectively. Black ellipses indicate the size and position of the HMSCs taken from Y17. Gray ellipses indicate the clumps identified in the corresponding intensity maps of the Hi-GAL distance project. Gray and red solid lines in (e) indicate the \vlsr~components identified with \tcotwo~spectra of the HMSC by the Hi-GAL distance project. The red lines highlight the \vlsr~for the corresponding intensity maps (b), (c), and (d). According to the mapping results, the most proper \vlsr~for the clump is $-$36.47~\kms.}
        \label{velocity_determination}
    \end{figure}

A reliable velocity determination is crucial to estimate a reliable set of other physical properties of the clumps. We take advantage of the results of velocity component identification from the Hi-GAL distance project (M\`{e}ge et al., in prep.) to revise the systematic velocity (\vlsr) estimated by Y17. This project aims to determine the \vlsr~and distance of about 150000 Hi-GAL-selected compact sources as accurately as possible by using all the molecular tracers available for the sources. In practice, we determine the most reliable \vlsr~of a clump by carefully inspecting the spatial association and morphology between dust and molecular line emission. The velocity component with a morphology most similar to ATLASGAL emission of the HMSCs is selected. An example of the velocity determination of the HMSCs with the results of Hi-GAL distance project is shown in Fig. \ref{velocity_determination}.

After revising the velocity, we used the same methods as that adopted in Y17 to derive the distance from renewed velocity, which is a parallax-based distance estimator developed by \citet{rei16} based on the Bayesian maximum-likelihood approach.  We carefully considered kinematic distance ambiguities (KDA) when determining the distance. We applied the KDA solution results of \citet{urq18} to ascertain the distance. If the HMSCs are not included in \citet{urq18}, we followed their work flows to solve the ambiguities, which use a series of evidence such as the scale height of the Galactic plane, \hi~self-absorption (\hi~SA), and extinction (IR dark clouds). More detailed information about KDA solutions is given in Appendix \ref{appendix-KDA}.

\subsection{Kinematic uncertainty} \label{KAZ}
\begin{figure}
      \centering
    \includegraphics[width=0.45\textwidth]{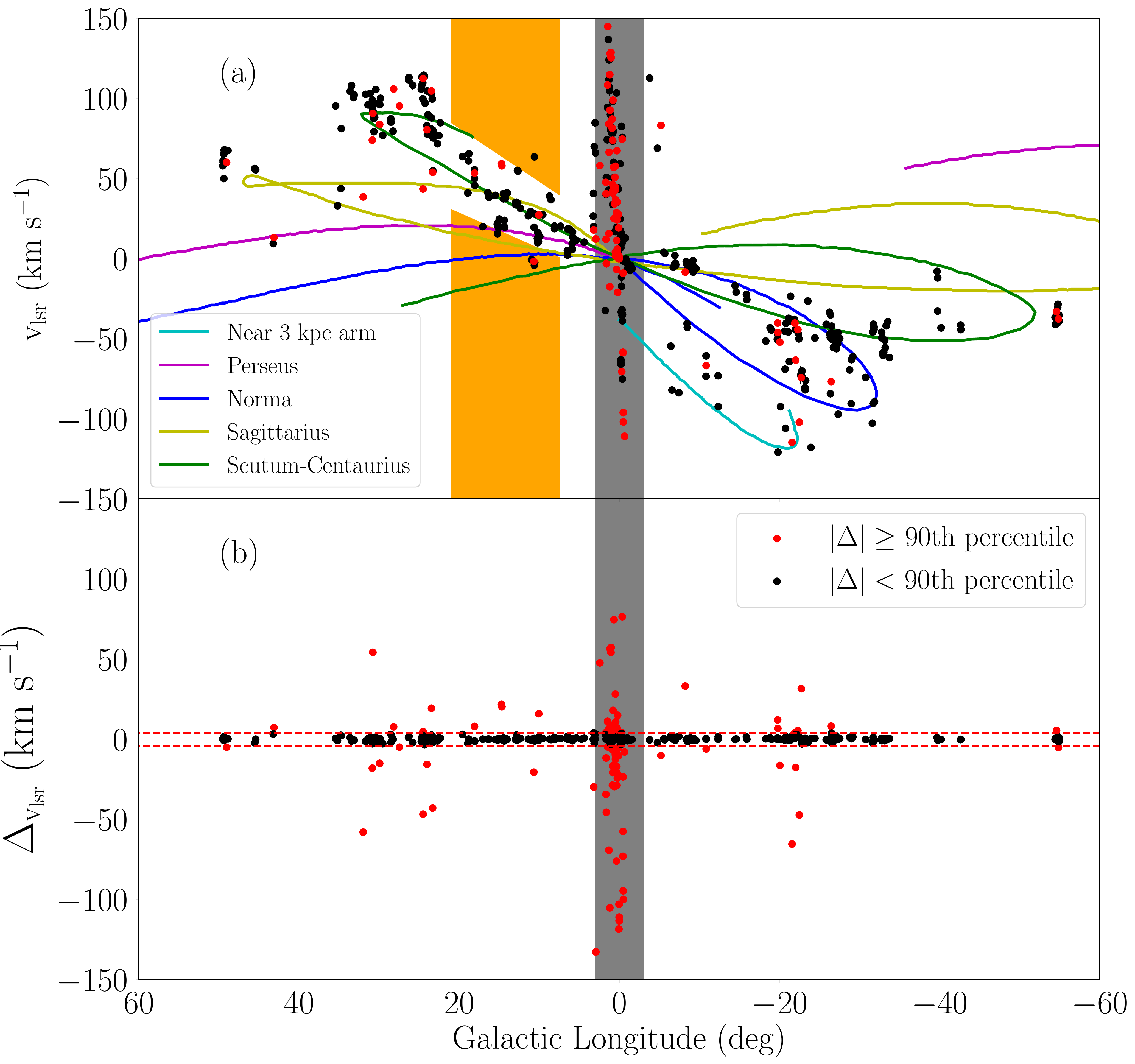}
       \caption{New HMSC \vlsr~(this work) and those in Y17. Panels (a) and (b) show the renewed \vlsr~and the difference between renewed \vlsr~and those in Y17, ${\Delta}_{\rm v_{lsr}} = {\rm v_{lsr}}_{\rm new} - {\rm v_{lsr}}_{\rm Y17}$, respectively. The 90th percentile of $|{\Delta}_{\rm v_{lsr}}|$ for HMSCs located at higher longitude ($|l| > 3$\degree) is 3.9 \kms, shown as red dashed lines in (b).  The red and black dots represent HMSCs with  $|{\Delta}_{\rm v_{lsr}}|$ values larger and smaller than 3.9 \kms, respectively. The gray region indicates the lower-longitude region, $|l| < 3$\degree, where the HMSCs are excluded from following analyses. Orange regions are considered by \citet{ell15}. The four spiral and local arms are shown by different colors and their positions are taken from \citet{urq18} and its references \citet{tay93} and \citet{bro00}. Most of the errors of \vlsr~are less than 2~\kms~and the associated error bars are smaller than the point size in (a).}
          \label{KAZ_image}
    \end{figure}

The renewed \vlsr~and its comparisons with that in Y17 are shown in Fig.~\ref{KAZ_image}. Most of the HMSCs except for the low-longitude sources are located at or close to spiral arms as shown in Fig.~\ref{KAZ_image} (a). The HMSCs toward the Galactic center have a radial component of the velocity that is only a small fraction of the entire velocity of the clump, making the rotation curve and resulted kinematic distance are just not reliable. Using any kinematic information in extremely low longitude is challenging, as shown by the significant $l$ - \vlsr~structure blending in $|l|$~<~3\degree~region in Fig.~\ref{KAZ_image} (a). Furthermore, the commonly existing broad line width (several dozens of \kms) of molecular lines in this region increases the difficulties to identify a proper \vlsr~\citep{kru15, gin16}. The 90th percentiles of the identified \vlsr~difference between Y17 and this paper are 3.9~\kms~and 64.6~\kms~for $|l|$~>~3\degree~and $|l|$~$\leq$ 3\degree, respectively. This shows the large uncertainty in \vlsr~identification for low-longitude clumps. 

The strong radial streaming motions caused by the Galactic bar, which could also impact the $l$ - \vlsr~structure, are carefully considered by \citet{ell15}. Fig. \ref{KAZ_image} (a) shows that only few HMSCs are located in the exclusion regions proposed by \citet{ell15} and that these HMSCs are close to the boundaries of exclusion regions. Thus, this kind of uncertainty is not dominating for our HMSCs. Considering the unreliable velocity and kinematic distance in lower longitude, we exclude a total of 125 HMSCs located in $|l|$~$\leq$~3\degree~from the following analyses and visualization.

The uncertain kinematics is not the only reason to exclude the lower-longitude HMSCs.  Possible different star formation mechanism and history compared to other molecular cloud--star formation complexes are also possible reasons. The central molecular zone (CMZ) is about 400 pc in size (angular scale at the distance of the Earth is about 0.4/8.34~rad~$\simeq$3\degree). The star formation rate (SFR) in the CMZ is an order of magnitude lower than expected \citep{lu19}. One of the explanations for the low SFR is that the clouds in CMZ are without active star formation at very early evolutionary stages because of the co-impacts of high turbulence, magnetic field, and a number of other processes \citep{kru14}. A higher density is needed for the collapse of the clumps in the CMZ. Therefore, the inhibited HMSF could result in the overproduction of HMSCs (26\% HMSCs are in the region $|l|$~<~3\degree). Our study focuses on the impacts of \hii~regions on HMSCs, thus it is more reliable to only use the higher-longitude HMSCs, which could avoid the complex co-impacts in CMZ.

\begin{figure}
     \centering
   \includegraphics[width=0.45\textwidth]{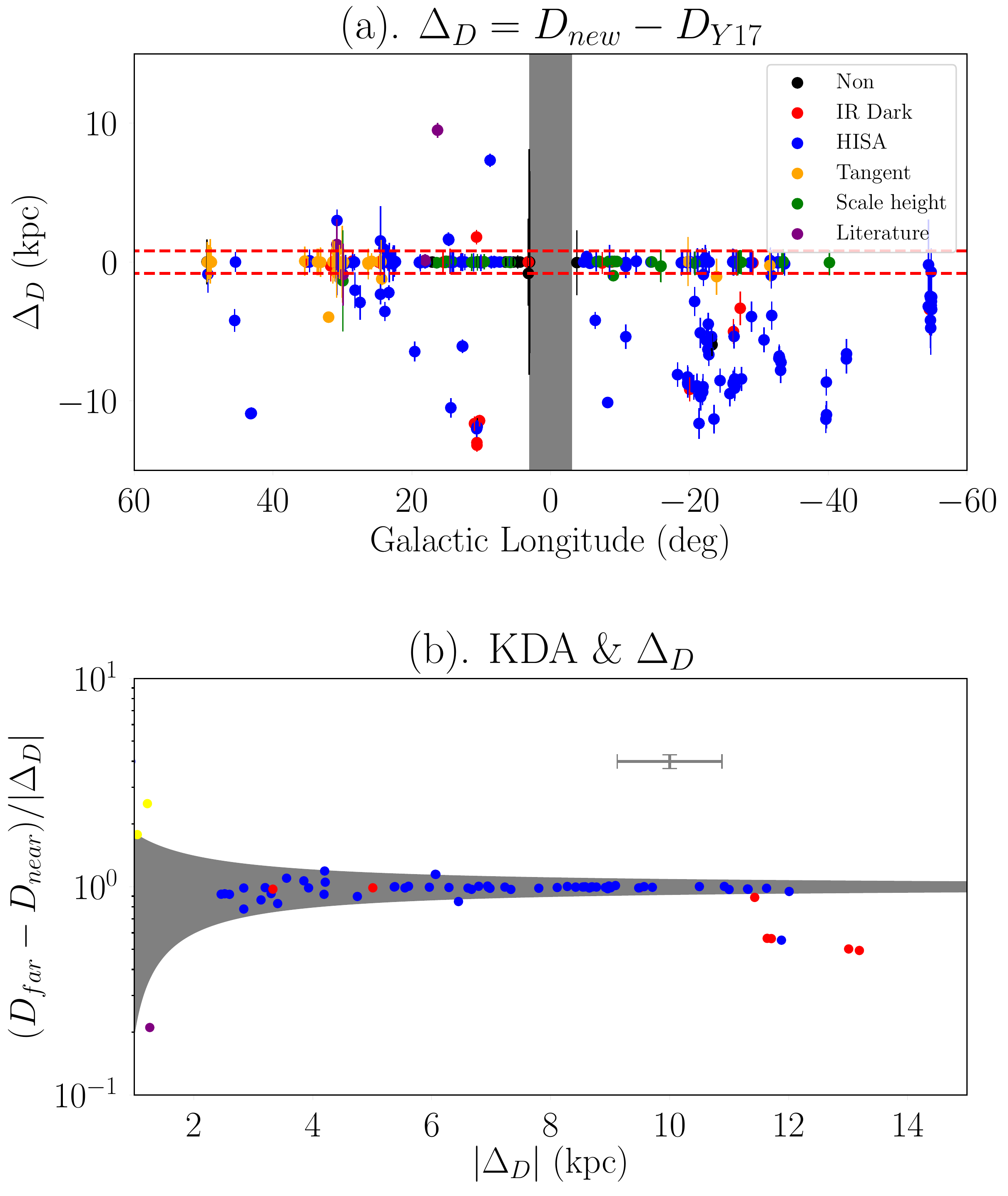}
       \caption{New distance of HMSC (this work) and those in Y17. Panel (a) shows the distance difference ${\Delta}_{D} = D_{new} - D_{Y17}$. The red dashed lines show $|{\Delta}_{D}| = 0.8$~kpc, representing the 70th percentile of $|{\Delta}_{D}|$. The red, blue, orange, green, and purple dots mean that their KDA are solved by IR dark, \hi~SA, tangent, scale height, and literature, respectively (see Table~\ref{phy_hmsc} and Appendix \ref{appendix-KDA}). The black dots are the sources without KDA solution. Panel (b) shows $|{\Delta}_{D}|$ and its relation with the difference between resulted far distance and near distance $D_{far} - D_{near}$. Only HMSCs with $|{\Delta}_{D}| > 0.8$~kpc and $|{\Delta}_{\rm v_{lsr}}| < 3.9$~\kms~are shown. The gray region shows $(D_{far} - D_{near})/|{\Delta}_{D}| = 1 \pm 0.8/|{\Delta}_{D}|$, indicating the error of relation. The gray cross shows the typical error.}
        \label{KDA}
    \end{figure}

The 70th percentile of the distance difference $|{\Delta}_{D}|$ between this paper and Y17 is about 0.8~kpc, which is shown in Fig.~\ref{KDA} (a) in red dashed lines. Owing to the small $|{\Delta}_{\rm v_{lsr}}|$ for most HMSCs, the improvement in solving KDA problem could be the most possible reason for the large difference between our new distance determination and those in Y17. For HMSCs with a similar \vlsr~but large $|{\Delta}_{D}|$,  their $|{\Delta}_{D}|$ are approximately equal to the differences between far and near distance $D_{far} - D_{near}$ in KDA solutions, as shown in Fig. \ref{KDA} (b); this suggests the large $|{\Delta}_{D}|$ are mainly due to the difference of KDA solutions. Nearly 70 HMSCs in total have a different KDA solution compared to those in Y17. \\

\subsection{Column density and dust temperature}  \label{CDT} 
One of the classical ways to investigate the dust properties of HMSC  is to combine far-IR to millimeter multiwavelength data and then fit the spectral energy distribution (SED) pixel by pixel. All images need to be smoothed to the largest beam size in the data set and then fitted with the modified blackbody function at each pixel. Principally, we made use of the \hmole~column density (\nhtcd) and dust temperature (\dustt) resulting from the Y17 SED fitting to update the physical properties according to the renewed distance. Besides, we also utilized the higher resolution PPMAP Hi-GAL data set in investigations of density and temperature profile.

\subsubsection{Results from Y17}
In Y17, the Hi-GAL 160, 250, 350, 500 \micron,~and ATLASGAL 870 \micron~images are smoothed to the largest beam of the data set, which is 36.4\arcsec. After carefully removing the foreground/background flux, gray-body functions are fitted pixel-by-pixel with a dust opacity of ${\kappa}_{\nu} = 3.33({\nu}/{\rm 600 GHz})^{\beta}~\rm cm^2g^{-1}$ and an emissivity index of $\beta = 2$. In this paper, physical properties (e.g., clump mass $M_{\rm clump}$, number density $n_{\rm H_{2}}$, and size $r_{pc}$ ) are basically taken from Y17 but revised according to the new distance determination.

\subsubsection{Higher resolution results from PPMAP} \label{CDT_PPMAP}
To explore the density/temperature structures of HMSCs, data with higher resolution is helpful because HMSF regions are generally located at a distance farther than low mass star formation regions.  To avoid degrading resolution when performing fitting process, \citet{mar15} fit the SED based on the nonhierarchical Bayesian procedure point process mapping technique to utilize full instrumental point source functions (PSF). \citet{mar17} applied the PPMAP technique on the Hi-GAL 70, 160, 250, 350, and 500~\micron~images to get a series of \dustt~- differential \nhtcd~planes, which means the column densities are equally spaced in log space between 8~K and 50~K and the dust temperature is the third dimension of \nhtcd~maps (data cube). The spatial resolution of the resulting PPMAP Hi-GAL \nhtcd~and \dustt~images is 12\arcsec; the value is better than that from the conservative SED fitting for Hi-GAL images ($\simeq$ 36\arcsec). The typical errors in the resulting \nhtcd~and \dustt~are around 10\% to 50\% and around 1~K, respectively.

\begin{figure}
      \centering
   \includegraphics[width=0.4\textwidth]{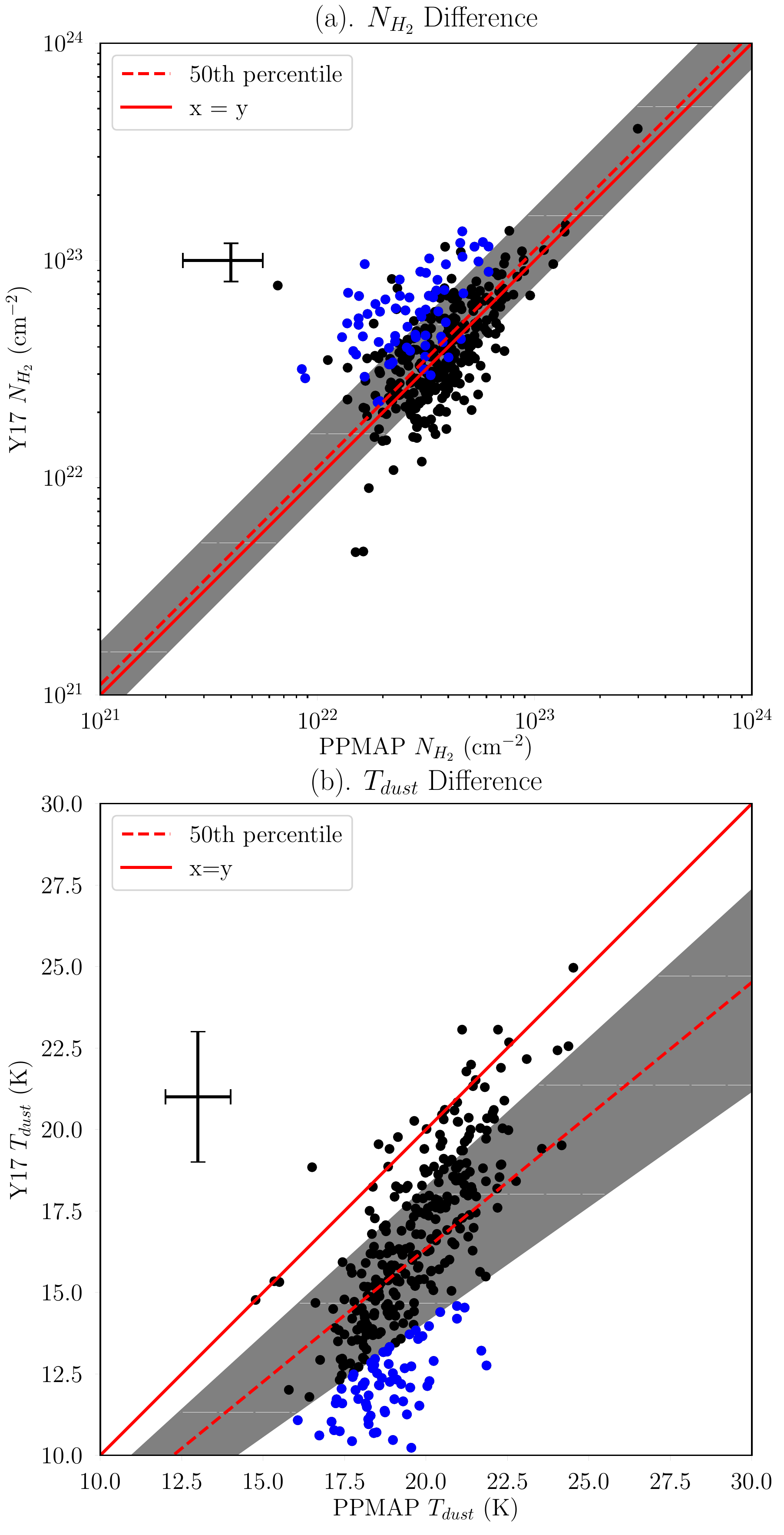}
       \caption{HMSCs \nhtcd~and \dustt~derived from PPMAP Hi-GAL data and those in Y17. The red solid lines show 1:1 relations. The red dashed lines show the 50th percentile of the ratios of Y17 derived parameters to PPMAP Hi-GAL derived parameters. The gray regions show the 20th percentile to 80th percentile of the ratios of Y17's parameters to PPMAP parameters. The blue points are sources with the lowest \dustt~and the largest difference of \dustt~between Y17 and PPMAP.}
        \label{PPMAPparameters}
    \end{figure}

Comparisons between PPMAP Hi-GAL \dustt~and \nhtcd~with those in Y17 are shown in Fig.~\ref{PPMAPparameters}. Generally, the PPMAP Hi-GAL \dustt~value is a higher of 2 to 3~K than the value obtained in Y17. At the low - \dustt~end (<~14~K for \dustt~of Y17), the difference can reach 5 to 7~K. The most likely reason for that is the difference in the wavelength of the data used because similar dust opacity and graybody functions are used by PPMAP and Y17. Y17 excluded Hi-GAL 70 \micron~emission and used additional 870~\micron~data to fit the SED, which could better constrain the shape of the SED at the long-wavelength end and lead to trace more the coldest parts of the clumps. Besides, Y17 carefully excluded the foreground/background flux when fitting the SED while this is not included in PPMAP Hi-GAL SED fitting. 

When the data used in fitting are the same, the PPMAP technique could provide a more accurate estimation of \nhtcd~and \dustt, especially for the maximum of \nhtcd, minimum of \dustt~of the clump \citep{marsh19}. Even PPMAP Hi-GAL data should detect more compact structures than Y17 because of  the higher resolution of the images; Fig.~\ref{PPMAPparameters} shows that clumps \nhtcd~are normally underestimated while \dustt~are overestimated at the low - \dustt~end compared to Y17. This suggests that the errors created during the SED-fitting process are dominated by the different wavelength data used rather than different techniques or resolutions in our cases.

Considering these factors, we only used PPMAP data to estimate the column density/temperature structures (see details in Sect. \ref{Com}) rather than using these data to derive other global properties of HMSCs such as \nhtcd, \dustt,~and the ratio of bolometric luminosity to envelope mass $L/M$.\\

\section{\hii~region cross matching} \label{HRCM}
\subsection{Criteria} \label{Criteria}
To explore differences between HMSCs associated with and nonassociated with \hii~regions, we set two main criteria to ascertain the association relationship.

The first criterion is the position in the plane of the sky. \citet{tho12} studied Red MSX Sources (RMSs; see \citealt{lum13}) YSOs, and Ultra Compact (UC) \hii~regions, which are regarded as the signatures of HMSF, around 322 \textit{Spitzer} mid-IR bubbles. With an angular cross-correlation function of RMSs and bubbles, these authors reveal that the location with the highest probability to find RMSs YSOs is around the effective radius $R_{\rm eff}$ of the bubble. Furthermore, the RMSs are essentially uncorrelated with bubbles and appear as field RMSs when the separations are larger than $2R_{\rm eff}$. A similar relationship for cold clumps has also been found in the MWP bubbles \citep{ken16}. If our HMSCs are precursors of HMSF that could form massive stars in the future, we can simply assume that the HMSCs associated with bubbles have a distance correlation similar to the RMSs or UC \hii~regions. A study that looked into 70,000 star-forming objects around bubbles reveals that either the radial source density profiles of starless clumps or those of protostars become quite flat and near the field values when beyond $2R_{\rm eff}$ \citep{pal17}. This suggests that the nonassociated properties of the clumps outside $2R_{\rm eff}$ of \hii~region are independent of the evolutionary stages of the clumps. Combining these facts, we propose that $2R_{\rm eff}$ is a reasonable value for declaring association in sky position between HMSCs and \hii~regions. We make use of the $R_{\rm eff}$ in \citet{ken12} and \citet{and14}, which is from the measurement of the mid-IR PAH~emission of the bubble (see Sect. \ref{IRS}). 

The second criterion is the \vlsr~difference $|{\delta}_{\rm v_{lsr}}| = |\rm v_{HMSC} - \rm v_{\hii~region}|$ between HMSCs and the \hii~regions. A proper limitation on $|{\delta}_{\rm v_{lsr}}|$ could mitigate mismatching between \hii~region and HMSCs due to the projection in the plane of sky. If a HMSC is indeed associated with an \hii~region,  ${\delta}_{\rm v_{lsr}}$  mainly comes from the molecular cloud turbulence and the expansion of \hii~region. By Larson's law \citep{lar81}, the velocity dispersion $\sigma_{\rm v}$ of a molecular cloud is related with its size. If we assume a cloud size of 5~pc, which is similar to the typical size of IR bubbles, the velocity dispersion $\sigma_{\rm v}$ is about $0.9 \times {5}^{0.56 \pm 0.02}$~\kms~$\approx$ 2~\kms~\citep{hey04}. This value is a bit smaller than the \tcoone~observations of \citet{yan16} toward 13 Galactic bubbles, which resulted in a value of about 2.5 to 5~\kms. The larger $\sigma_{\rm v}$ may be due to the additional energy injection from massive stellar feedback such as stellar winds and UV radiation. Another factor that enlarges $|{\delta}_{\rm v_{lsr}}|$ is the expansion of the \hii~region. The simulation of \citet{tre14} for the expanding process of ionized gas in turbulent molecular clouds indicates an averaged expanding velocity $\simeq$ 2~\kms. Molecular spectral observations toward expanding shells surrounding \hii~regions indicate an expanding velocity on the order of one to several \kms~\citep{elm11}. 

\begin{figure}
   \includegraphics[width=0.45\textwidth]{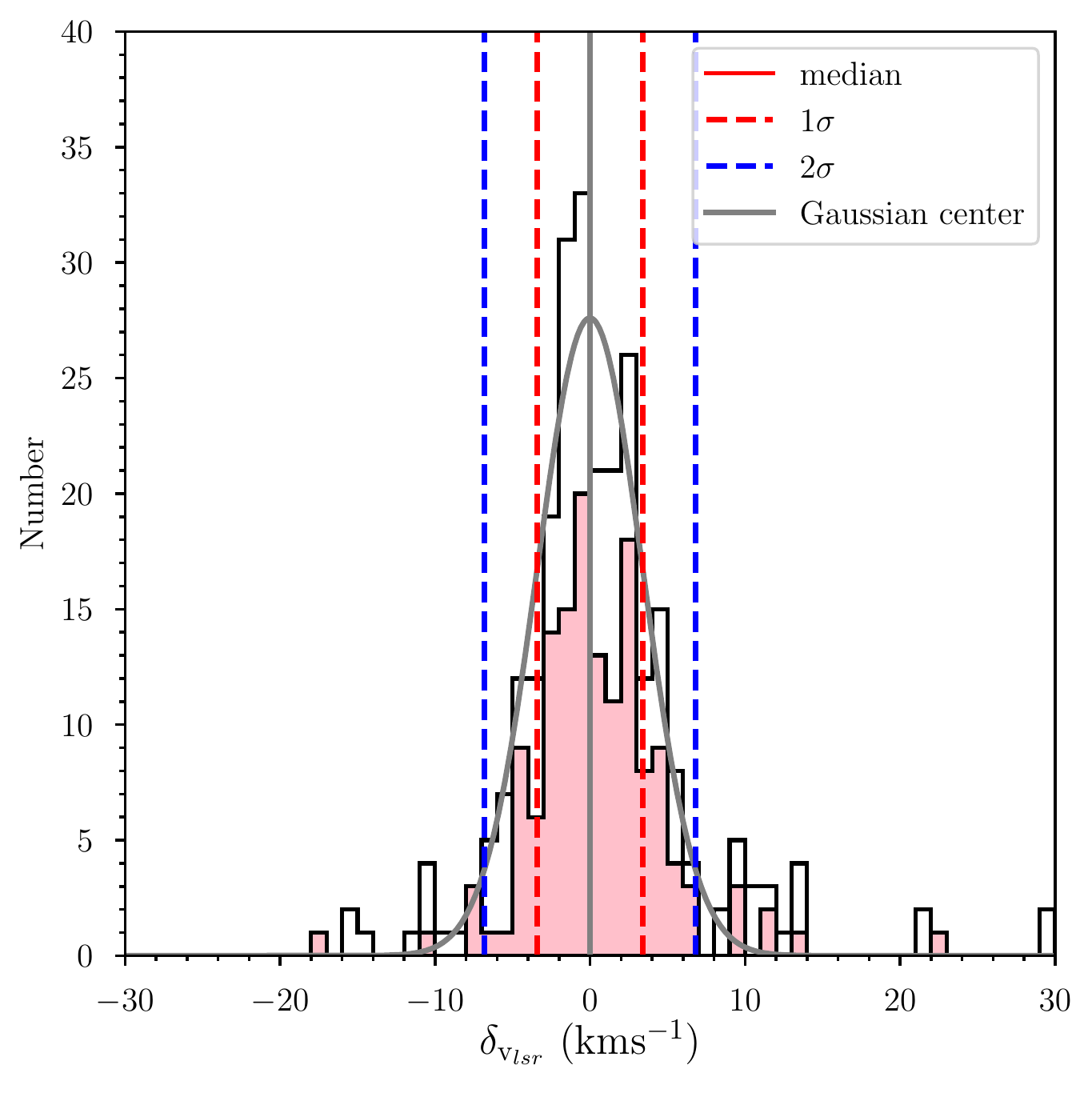}
       \caption{Number distribution of \vlsr~difference between \hii~region and HMSC ${\delta}_{\rm v_{lsr}} = \rm v_{HMSC} - v_{\hii~region}$ in the $-$30 to 30~\kms~range. The gray lines are the fitted Gaussian and its center. The pink histogram highlights the distribution for S-type HMSCs (see Sect. \ref{DMDM}). The red and blue dashed lines represent 1$\sigma$ and 2$\sigma$ of Gaussian, respectively. The median value of ${\delta}_{\rm v_{lsr}}$ overlaps with Gaussian center in the figure.}
        \label{Velocity}
    \end{figure}

To ascertain a proper limitation, ${\delta}_{\rm v_{lsr}}$ for all HMSCs in $2R_{\rm eff}$ of \hii~regions with known \vlsr~are shown in Fig.~\ref{Velocity}. There are only few sources whose $|{\delta}_{\rm v_{lsr}}| > 30$~\kms, thus they are not shown in Fig.~\ref{Velocity}. Gaussian fitting to ${\delta}_{\rm v_{lsr}}$ distribution gives a Gaussian standard deviation $\sigma$ of $\simeq 3.5$~\kms. Considering all the facts mentioned above, we assume that a \vlsr~difference limit of 7~\kms, about $2\sigma$ of fitted Gaussian distribution, is a reasonable value to reduce the misclassification due to the projection on the line of sight. We made use of \hii~region \vlsr~information from the work of \citet{hou14} and \citet{and14}, which are usually derived from H$\alpha$, RRL observations or the average molecular lines of \hii~regions.

\subsection{Workflow} \label{workflow}
According to the two criteria mentioned, we classified HMSCs into three types that reflect the association relations between HMSCs and \hii~regions. Generally, when a HMSC is located outside $2R_{\rm eff}$ of \hii~regions or bubbles, we classify it as \textbf{nonassociated} HMSC (written as \textbf{NA HMSC} or \textbf{NA}), otherwise it is classified as \textbf{possibly associated} HMSC (written as \textbf{PA HMSC} or \textbf{PA}) or \textbf{associated} HMSC (written as \textbf{AS HMSC} or \textbf{AS}).
The criterion for discriminating between PA and AS is whether the absolute value of $\rm v_{lsr}$ difference $|{\delta}_{\rm v_{lsr}}|$ between HMSC and \hii~region is less than 7~\kms. The HMSC is classified as AS if this is the case, otherwise it is classified as PA. When a HMSC is located in $2R_{\rm eff}$ but there is no \vlsr~information for the \hii~region, we also put it in the category of PA HMSCs.

\begin{figure}
   \includegraphics[width=0.4\textwidth]{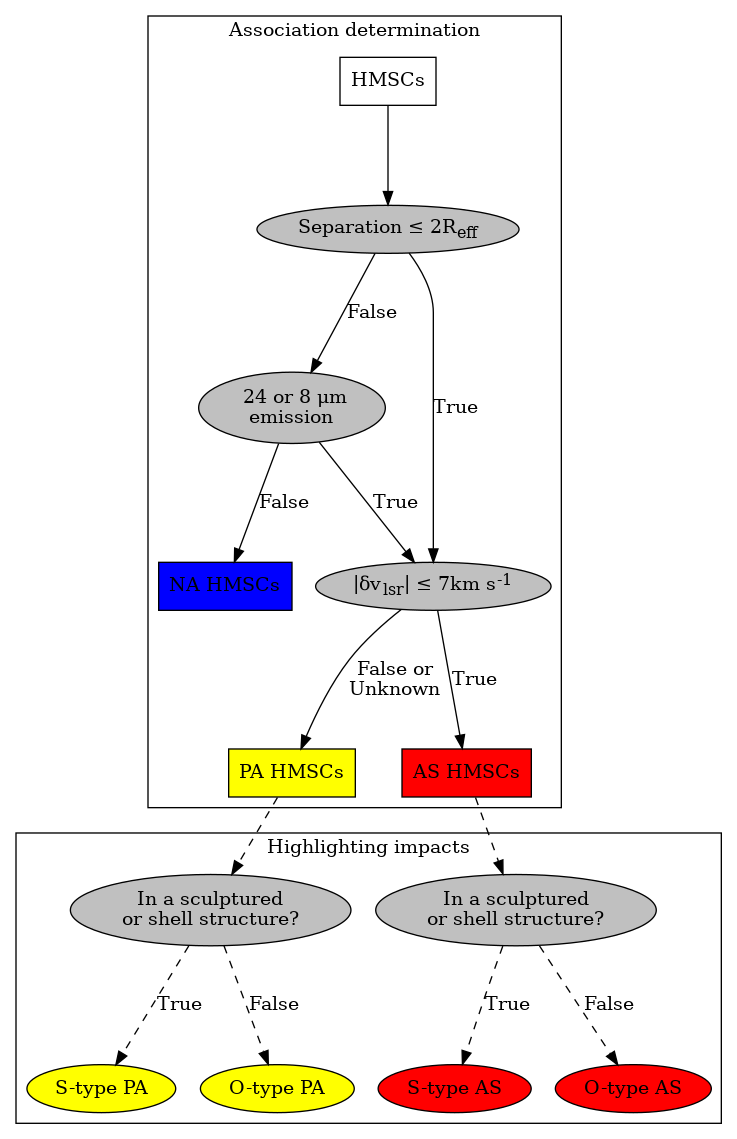}   
       \caption{Workflow chart. The first part is the determination of association as described in Sect. \ref{workflow} and the second part is for highlighting impact of \hii~regions, as described in Sect. \ref{DMDM}.}
        \label{workflow_chart}
    \end{figure}

\begin{figure}
   \includegraphics[width=0.46\textwidth]{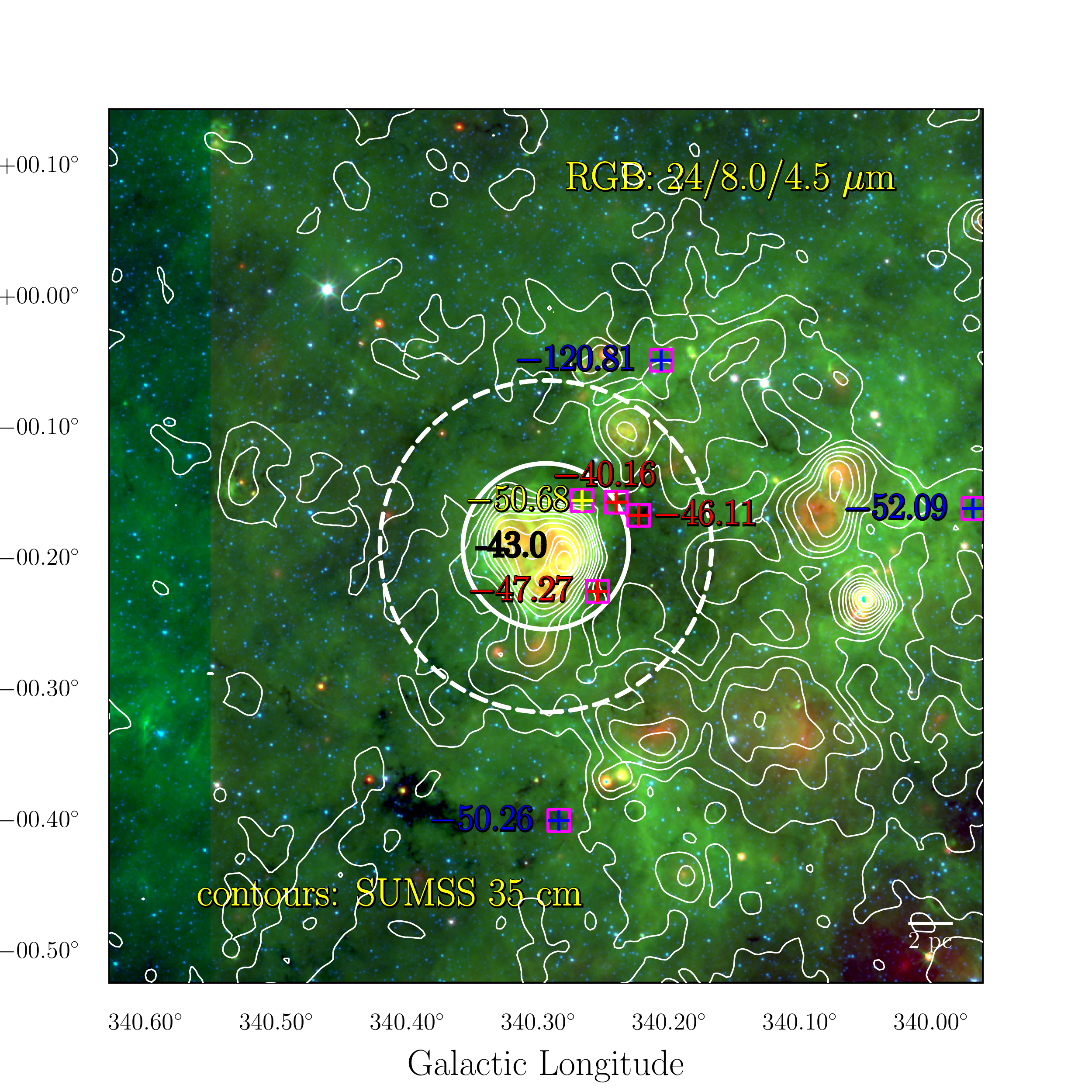}
       \caption{\hii~region G340.294$-$00.193. The $R_{\rm eff}$ and $2R_{\rm eff}$ of the \hii~region are indicated as solid and dashed white circles, respectively. The purple squares highlight the position of HMSCs. The red, yellow, and blue crosses represent AS, PA, and NA HMSCs, respectively. The corresponding \vlsr~in \kms~is shown close to the mark. Number $-$43.0 is the \vlsr~of \hii~region in \kms. The RGB image is constructed in the same method as Fig.~\ref{velocity_determination}. The white contours represent SUMSS 35 cm emission. The \vlsr~of the \hii~region is taken from \citet{hou14} and \citet{and14}, see Sect. \ref{Criteria}.}
        \label{example}
    \end{figure}

Considering that some of the \hii~regions or bubbles have an irregular or elliptical shape, a single circular description with $R_{\rm eff}$ is not accurate enough. We check GLIMPSE 24~\micron~hot dust emission and 8~\micron~PAH emission images to confirm that all NA HMSCs classified by the criterion of $2R_{\rm eff}$ are really far away from any 8~\micron~emission shell structure or diffuse 24~\micron~emission. If this is the case, we keep it in the NA category, otherwise we reclassify it as PA or AS. A flow chart describing how classification process works is shown in Fig. \ref{workflow_chart}. An example for the identifications of AS, PA, and NA HMSCs surrounding the \hii~region G340.294$-$00.193 are shown in Fig.~\ref{example}.

\subsection{Bias and results}
Large (>~30\arcmin) and very diffuse (weak at 24~\micron) \hii~regions are not considered when cross matching \hii~regions and HMSCs. Some AS and PA could be misclassified as NA owing to the omission of the very diffuse \hii~regions. We suggest that the omission of very diffuse \hii~regions does not largely change our comparison results between different HMSCs because the impacts of ionized gas of this kind of \hii~regions are too weak to significantly modify the properties of HMSCs.

Another bias is that some PA with $|{\delta}_{\rm v_{lsr}}|$~>~7~\kms~could be due to the very powerful feedback of \hii~regions, which strongly changes the overall kinematics of molecular clouds, making $|{\delta}_{\rm v_{lsr}}|$ larger than 7~\kms, rather than the properties of nonassociation. It is possible that the chemistry properties of HMSCs could help in our classification. For example, \cth~is a typical molecular tracer for photodissociation, indicating the impacts of \hii~region. In a forthcoming paper, we find that the \cth~detection rates between AS and NA are significantly different (Zhang et al., in prep.). Detailed case-studies will reveal whether these PA are AS or NA. In this paper, we only use the simplest methods with the velocity and sky position to classify the association of HMSCs and leave the more complex and detailed classifications to future studies.

To summarize, we classified HMSCs into three categories according to their association with \hii~region: (1) \textbf{AS}: the HMSC is in $2R_{\rm eff}$ of the \hii~region and its $\rm v_{lsr}$ difference with the \hii~region is $\leq$~7~\kms; (2) \textbf{PA}: the HMSC is in $2R_{\rm eff}$ of the \hii~region, but we do not know $\rm v_{lsr}$ of the \hii~region or the $\rm v_{lsr}$ difference is >~7~\kms; (3) \textbf{NA}: the separation between the HMSC and the \hii~region is >~$2R_{\rm eff}$. 

A summary of the classification results is shown in Table~\ref{association_table}. About 60\% to 80\% of HMSCs are associated with \hii~regions, which reveals the close relation between the \hii~regions and HMSCs. We only use the simplest method with position and velocity to determine the association. The real associated number could be larger if we consider more features, like \cth.  Classification results and physical parameters of HMSCs are given in Table~\ref{phy_hmsc}.\\

      \begin{table}[ht]
      \tiny
      \caption{Association of HMSCs and \hii~regions.} 
      \label{association_table} 
       \begin{tabular}{c c c c} 
       \hline\hline 
            Sample       & AS HMSCs                    & PA HMSCs             & NA HMSCs       \\ 
       \hline 
        338 HMSCs in $|l| > 3$ \degree         &  193 (57\%)    &   64 (19\%)   &   81 (24\%)      \\            
        \hline                                         
      \end{tabular} 
      \end{table}

\begin{sidewaystable*}
  \tiny
   \centering 
    \caption{Classification and physical parameters of HMSCs.} 
    \label{phy_hmsc} 
    \begin{tabular}{c c c c c c c c c c c c c} 
    \hline\hline 
    Name\footnotemark[1]         &\dustt\footnotemark[2]& $n_{\rm H_{2}}$     &   \nhtcd\footnotemark[3]               & $M_{\rm clump}$     & $L_{\rm clump}$ & $L/M$       & $r_{\rm pc}$ & $\rm v_{\rm lsr}$  &  Distance\footnotemark[4] & KDA\footnotemark[5] & Association & Morphology\footnotemark[6] \\ 

                 &  (K) &(10$^{3}$cm$^{-3}$)  &   (10$^{21}$cm$^{-2}$) &    (\msun)                          & (\lsun)      &  (\lsun/\msun) &  (pc)  &(\kms)       & (kpc)     &         &    &  \\  
   \hline 
 
 G003.0371$-$0.0582 & 13.8 & 1.17(0.35)          & 53.6(8.0)              & 1.21(0.64)$\times 10^{4}$  & 8.28(5.07)$\times10^{3}$ & 0.68(0.23)     & 3.3    &  85         & 13(3.3)   &         & NA & C \\     
 G003.0928+0.1680 & 14.2 & 1.96(0.58)          & 80.2(12.0)             & 2.42(1.28)$\times 10^{4}$  & 1.75(1.06)$\times10^{4}$ & 0.72(0.24)     & 3.5    &  150(0.12)  & 16(4.1)   &         & NA & C \\     
 G003.1447+0.4733 & 14.3 & 1.90(0.29)          & 15.4(2.3)              & 9.17(1.39)$\times 10^{2}$  & 7.16(2.70)$\times10^{2}$ & 0.78(0.29)     & 1.2    &  70(1.9 )   & 11(0.12)  &         & NA & F \\     
 G003.2110+0.6465 & 14.5 & 5.06(0.80)          & 22.9(3.4)              & 1.89(0.34)$\times 10^{2}$  & 1.59(0.62)$\times10^{2}$ & 0.84(0.32)     & 0.51   &  18(0.25)   & 2.9(0.15) &         & NA & F \\     
 G003.2278+0.4924 & 14.4 & 6.14(0.97)          & 31.4(4.7)              & 2.34(0.43)$\times 10^{2}$  & 1.88(0.72)$\times10^{2}$ & 0.8(0.29)      & 0.51   &  20(0.88)   & 2.9(0.15) & IR DARK & NA & F \\     
 G003.2413+0.6334 & 13   & 19.2(8.1)           & 59.7(9.0)              & 4.99(4.03)$\times 10^{2}$  & 2.15(1.83)$\times10^{2}$ & 0.43(0.14)     & 0.45   &  41         & 4(1.6)    &         & NA & F \\     
 G003.2702+0.4446 & 13.8 & 6.02(0.95)          & 36.9(5.5)              & 3.87(0.70)$\times 10^{2}$  & 2.34(0.85)$\times10^{2}$ & 0.6(0.21)      & 0.61   &  27         & 2.9(0.15) &         & NA & C \\     
 G004.4076+0.0993 & 18.1 & 123(19)             & 20.8(3.1)              & 8.99(1.63)$\times 10^{1}$  & 3.60(1.71)$\times10^{2}$ & 4(1.9)         & 0.14   &  10(0.35)   & 2.9(0.15) &         & AS & S \\     
 G004.7473$-$0.7623 & 12.7 & 6.53(0.99)          & 71.0(10.7)             & 4.89(0.78)$\times 10^{3}$  & 1.81(0.56)$\times10^{3}$ & 0.37(0.11)     & 1.4    &  210        & 9.4(0.24) &         & NA & I \\     
 G005.3876+0.1874 & 12.9 & 52.7(8.3)           & 38.3(5.7)              & 1.76(0.32)$\times 10^{2}$  & 7.44(2.52)$\times10^{1}$ & 0.42(0.14)     & 0.23   &  11(0.74)   & 2.9(0.15) &         & AS & F \\     
 G005.8357$-$0.9958 & 15.4 & 6.59(1.05)          & 23.9(3.6)              & 1.25(0.23)$\times 10^{2}$  & 1.50(0.62)$\times10^{2}$ & 1.2(0.48)      & 0.4    &  13         & 2.9(0.15) & Z H     & PA & I \\     
 G005.8523$-$0.2397 & 12.1 & 12.2(1.9)           & 45.0(6.7)              & 2.30(0.42)$\times 10^{2}$  & 7.28(2.29)$\times10^{1}$ & 0.32(0.09)     & 0.4    &  17(0.13)   & 3(0.15)   & HISA    & NA & I \\     
 G005.8799$-$0.3591 & 16.5 & 30.0(4.8)           & 51.2(7.7)              & 2.43(0.44)$\times 10^{2}$  & 4.38(1.80)$\times10^{2}$ & 1.8(0.72)      & 0.3    &  5.8(0.1 )  & 2.9(0.15) & HISA    & AS & S \\     
 G005.8893$-$0.4565 & 19.4 & 97.0(15.4)          & 46.6(7.0)              & 2.16(0.39)$\times 10^{2}$  & 1.16(0.53)$\times10^{3}$ & 5.4(2.4)       & 0.2    &  9.7(0.23)  & 2.9(0.15) & IR DARK & AS & S \\     
 G005.8917$-$0.3567 & 17   & 20.6(3.3)           & 35.4(5.3)              & 2.46(0.45)$\times 10^{2}$  & 5.44(2.36)$\times10^{2}$ & 2.2(0.93)      & 0.35   &  5.8(0.13)  & 2.9(0.15) & IR DARK & AS & S \\     
 G005.9394$-$0.3754 & 17.2 & 14.6(2.3)           & 21.0(3.2)              & 1.32(0.24)$\times 10^{2}$  & 3.27(1.49)$\times10^{2}$ & 2.5(1.1)       & 0.31   &  6.1(0.1 )  & 2.9(0.15) & HISA    & AS & S \\     
 G006.2130$-$0.5937 & 15.8 & 23.6(3.7)           & 34.9(5.2)              & 1.67(0.30)$\times 10^{2}$  & 2.44(1.00)$\times10^{2}$ & 1.5(0.58)      & 0.29   &  18(0.11)   & 3(0.15)   & Z H     & AS & F \\     
 G006.4609$-$0.3902 & 17.9 & 6.96(1.10)          & 4.58(0.69)             & 2.02(0.37)$\times 10^{1}$  & 7.12(3.54)$\times10^{1}$ & 3.5(1.7)       & 0.22   &  19         & 3(0.15)   &         & NA & I \\     
 G006.4916$-$0.3322 & 17.7 & 9.88(1.57)          & 4.54(0.68)             & 2.17(0.40)$\times 10^{1}$  & 6.81(3.35)$\times10^{1}$ & 3.1(1.5)       & 0.2    &  2.2        & 2.9(0.15) &         & NA & I \\     
 G006.4982$-$0.3222 & 17.6 & 1.94(0.31)          & 8.96(1.34)             & 5.66(1.03)$\times 10^{1}$  & 1.65(0.79)$\times10^{2}$ & 2.9(1.4)       & 0.46   &  5.1(0.25)  & 2.9(0.15) &         & NA & I \\

     \hline                                              
      \end{tabular}
      \footnotetext{1. Full table can be found in the electronic version.}
      \footnotetext{2. The typical error of \dustt~estimated by Y17 is about 2 to 3 K.}
      \footnotetext{3. The typical error of \nhtcd~estimated is about 15\% from Y17.}
      \footnotetext{4. The distance error is taken from the results of distance calculator and its typical value is less than 1 kpc. The distance errors are propagated to the calculation of other physical parameters.}
      \footnotetext{5. We solve the KDA with several ways: (1) IR DARK. IR extinction. (2) \hi~SA. \hi~self-absorption. (3) Z H. The scale height in the Galactic plane. (4) TANGENT. The tangent line of sight. (e). LITERAT. The literature in \citet{urq18}. See Appendix \ref{appendix-KDA} for more information.} 
      \footnotetext{6. The detailed explanation about the morphology is in Appendix \ref{appendix-morphology}.}

\end{sidewaystable*}

\subsection{Morphology features of impact of \hii~regions on HMSCs} \label{DMDM}
The main goal of this section is to extract the HMSCs with clear impacts from the \hii~regions using the available data. According to the morphology of cold and hot dust emission, PAH, and ionized gas emission, we check whether HMSC is in a compressed dust shell or in a structure that is being photoionized or photodissociated by UV radiation. Steeper profiles of radio emission, ATLASGAL 870~\micron~emission, or column density toward the direction of the \hii~region and/or a denser dust shell surrounding the \hii~region both indicate the compression by the \hii~region \citep{tre14c, tre15, li18, marsh19}. The bright PAH 8~\micron~emission layer toward the direction of 24~\micron~hot dust emission or radio emission of the \hii~regions suggests that photodissociation is working at the surface of clumps and is ``sculpturing'' the clumps. We classify AS or PA HMSCs with these features as \textbf{S (sculptured or shell) type} AS or S-type PA. For AS and PA without morphology features of significant interaction with the \hii~regions, we just classify these AS or PA as \textbf{O- (other) type} AS or O-type PA. These O-type AS or PA could be in a filament, a clumpy structure, or even isolated without neighbor of clump (see Appendix \ref{appendix-morphology}). The related classification process is shown in the work flow presented in Fig. \ref{workflow_chart}. Table \ref{morpho_table} shows that about half of AS are found to be probably significantly impacted by \hii~regions on the mapping data. \\

   \begin{table}[ht]
  \tiny
   \centering 
    \caption{Morphology statistics.} 
    \label{morpho_table} 
    \begin{tabular}{ccc} 
    \hline\hline 
 Types & S-type & O-type  \\
   \hline 

AS HMSCs (193) & 104 (54\%) & 89 (46\%)\\
PA HMSCs (64)  &  23 (36\%)&  41 (64\%)\\
     \hline                                         
      \end{tabular}
      \end{table}

\section{Distributions in the Galaxy} \label{DIG}

   \begin{figure*}
       \centering
   \includegraphics[width=0.9\textwidth]{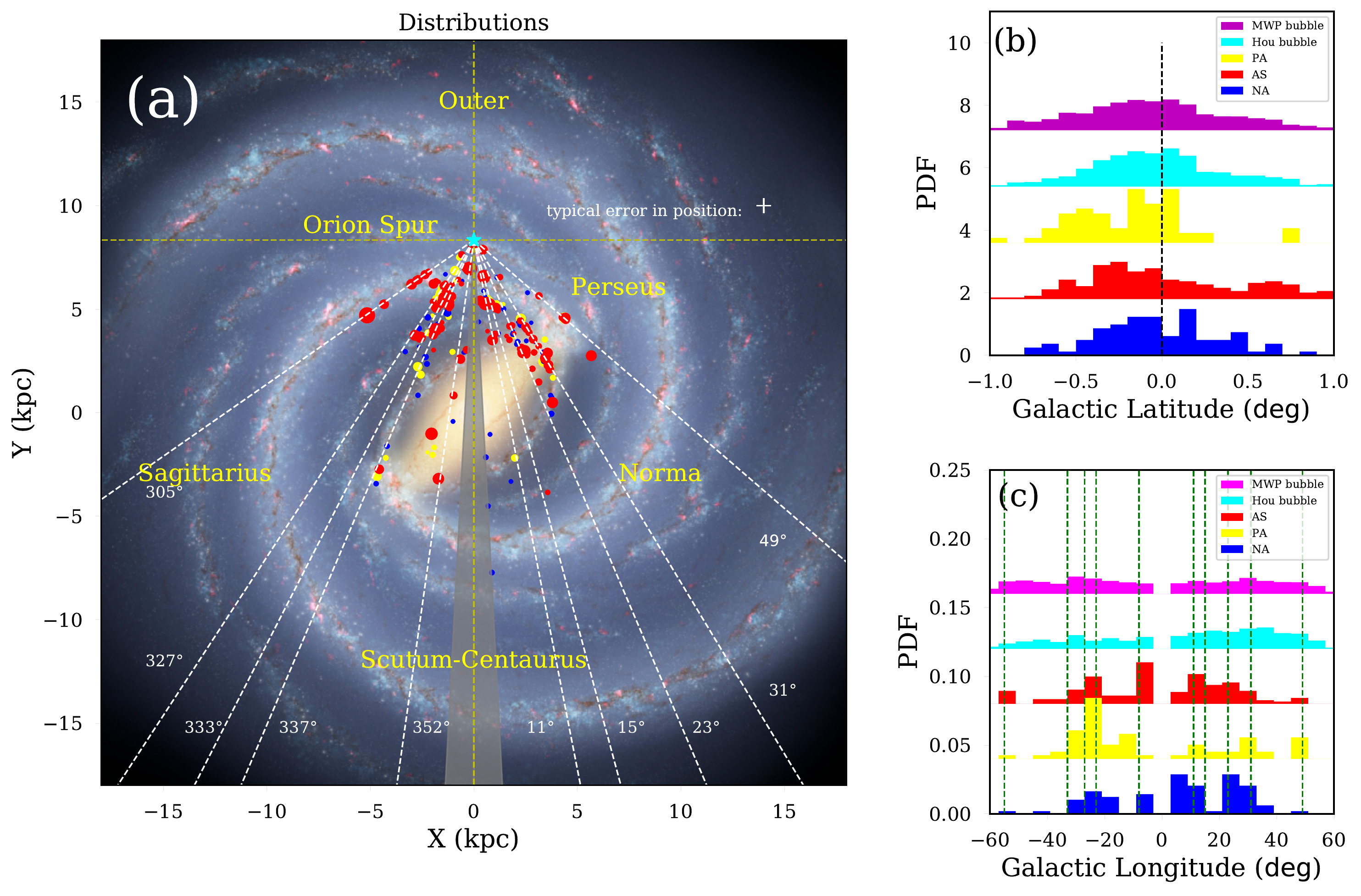}
       \caption{Galactic distributions for different types of HMSCs. (a). Top view of HMSCs in the Galaxy. The red, yellow, and blue dots represent AS, PA, and NA HMSCs, respectively. The size of the dot increases with \dustt~of HMSCs. The gray region represents $|l| < 3$\degree. The typical error in position is shown as a  white cross. Panels (b) and (c) show the latitude and longitude distributions for bubble (see \citealt{ken12, hou14}) and different types of HMSCs. The green dashed lines in panel (c) indicate the tangent directions of the spiral arms, which are shown as white dashed lines in panel (a).}
        \label{Galactic}
    \end{figure*}

The distribution of HMSCs in the Galaxy is shown in Fig.~\ref{Galactic}. The trend is that either AS or NA all highlight the spiral arms and that both of these have a similar distribution in the Milky Way. This trend could be partly due to the coherence between HMSCs and spiral arms in the ${\rm v_{lsr}}$~-~longitude structures shown in Fig.~\ref{KAZ_image}. Another factor could be the nature of \citet{rei16} distance calculator, which links the points in the position-position-space to the specified spiral arms. The peak of latitude distribution of AS shifts to $b = -0.25$\degree. By checking our images of the \hii~regions - HMSCs complex, this shift could be partly due to several star-forming complexes with a number of AS at a relatively higher latitude, such as G010.19$-$0.35 (eight HMSCs) and G333.48$-$0.22 (eight HMSCs). Another factor could be due to the offset between solar system's location and the Galactic midplane. The location of the Sun is thought to be about 20~pc to 25~pc above the Galactic midplane \citep{hum95}.   
    
\begin{figure}
       \centering
  \includegraphics[width=0.4\textwidth]{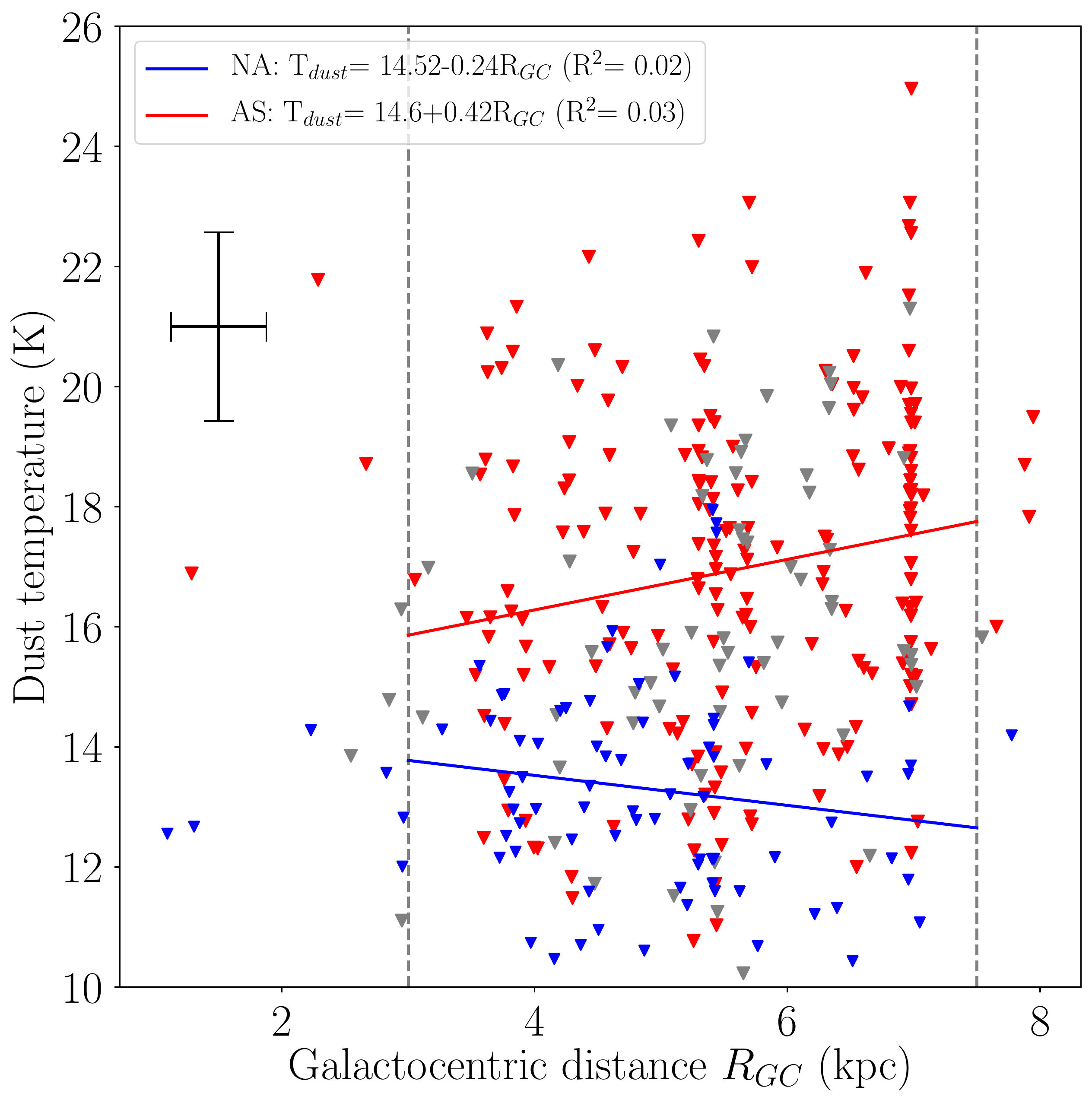}
       \caption{Galactocentric distance $R_{\rm GC}$ and \dustt~of HMSCs. Red, gray, and blue dots represent AS, PA, and NA HMSCs, respectively. The typical error is shown as a black cross. The HSMCs used for performing regression and Spearman's correlations are limited to 3 to 7.5 kpc. The red and blue solid lines represent the results of linear regressions of AS and NA.}
        \label{Galactic-temperature}
    \end{figure}

  The \dustt~of AS and PA seems to be dominated by the local environment rather than the Galactic-scale environment. In the same region of spiral arm, AS and PA usually have a higher \dustt~than that of NA, which probably indicates that \dustt~of AS and PA are deeply impacted by the associated \hii~regions. The relations between \dustt~and the Galactocentric distance $R_{\rm GC}$ for AS and NA HMSCs are shown in Fig.~\ref{Galactic-temperature}. The NA occupy the lower~-~\dustt~end in most distance ranges. Linear regressions to \dustt~for AS and NA indicate the Galactocentric gradient of \dustt~is weakly different between AS and NA HMSCs. The NA show a weak trend that \dustt~decreases with $R_{\rm GC}$, which conforms with the previous results about the Galactic cold dust temperature gradient owing to the decreasing star formation activities from the Galactic inner to outer regions \citep{mis06, par12}, whereas AS HMSCs show a weak positive trend. The Spearman's correlations\footnote{Spearman's correlation is used for assessing how well the relations between two variables could be described by a monotonic function. The value range of the correlation coefficient is from $-1$ to 1. A coefficient of 1 or $-1$ means that the variable is a perfect monotonic function of the other variable.} result in a coefficient of 0.15 and $-0.16$ of $R_{\rm GC}$ - \dustt~relation for AS and NA, respectively, showing the results of different $R_{\rm GC}$~-~\dustt~relations are weak. The probably different \dustt~gradient may indicate the importance of impacts of \hii~regions on the local environment. In the following sections, we confirm the \dustt~difference of HMSCs in different environments. \\

\section{Basic physical properties} \label{BPP}
To explore physical differences of HMSCs in different environments with a more reliable sample, we remove HMSCs with a mass smaller than 100~\msun~to confirm that the selected clumps are at massive end of the sample. The PA are not included in the analyses because of their uncertain association with \hii~regions. AS HMSCs are split into S-type and O-type AS HMSCs to highlight the impacts of \hii~regions as mentioned in Sect. \ref{DMDM}.

\subsection{Distance bias}    \label{DF}   
\begin{figure}
   \includegraphics[width=0.45\textwidth]{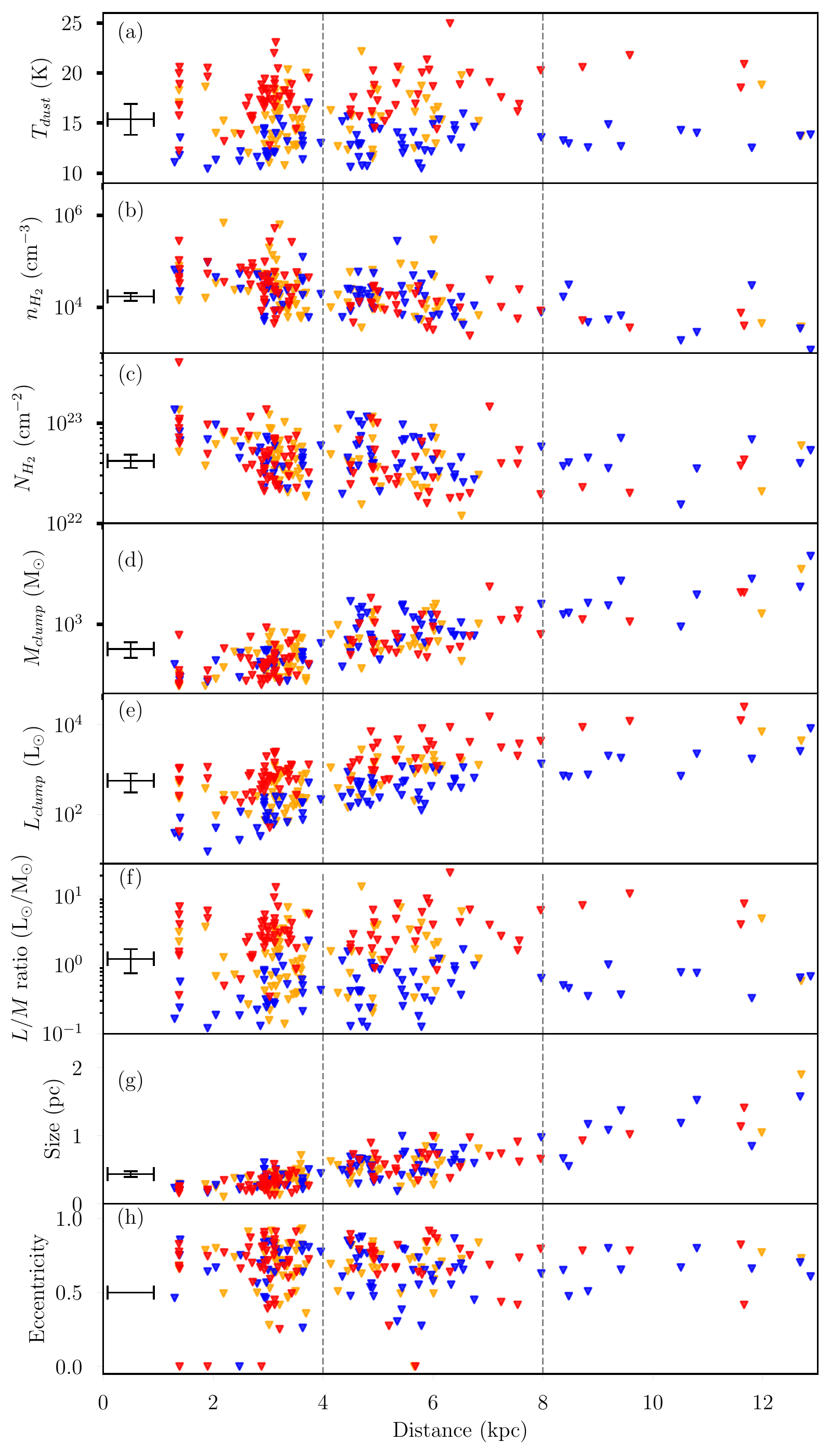}
       \caption{Basic properties and distance. (a), (b), (c), (d), (e), (f), (g), and (h) show the dust temperature (\dustt), \hmole~number density ($n_{\rm H_{2}}$), \hmole~column density (\nhtcd), mass ($M_{\rm clump}$), luminosity ($L_{\rm clump}$), luminosity-to-mass ratio ($L/M$), size, and eccentricity ($e_{\rm clump}$) of HMSCs and their distances. The red, orange, and blue triangles indicate S-type AS, O-type AS, and NA HMSCs. The gray dashed lines indicate the distances of 4~kpc and 8~kpc.}
        \label{distance-scatter}
    \end{figure}

Our samples of HMSCs cover a broad range of distances, from 1~kpc to 13~kpc, which could be a bias when claiming the differences in physical properties of different HMSCs. Eight physical properties and their relations with distance are shown in Fig.~\ref{distance-scatter}. The quantities \dustt, $L/M$ ratio, and eccentricity have a flat relation with distance. Number density $n_{\rm H_{2}}$ and probably also \nhtcd~show a decreasing trend, whereas $M_{\rm clump}$, $L_{\rm clump}$, and size show an increasing trend. The possible reason is that our HMSCs have a similar angular size (20\arcsec~to 40\arcsec) and column density (10$^{22}$ to 10$^{23}$~cm$^{-2}$), while their distances cover a broad range. 

\begin{figure}
   \includegraphics[width=0.45\textwidth]{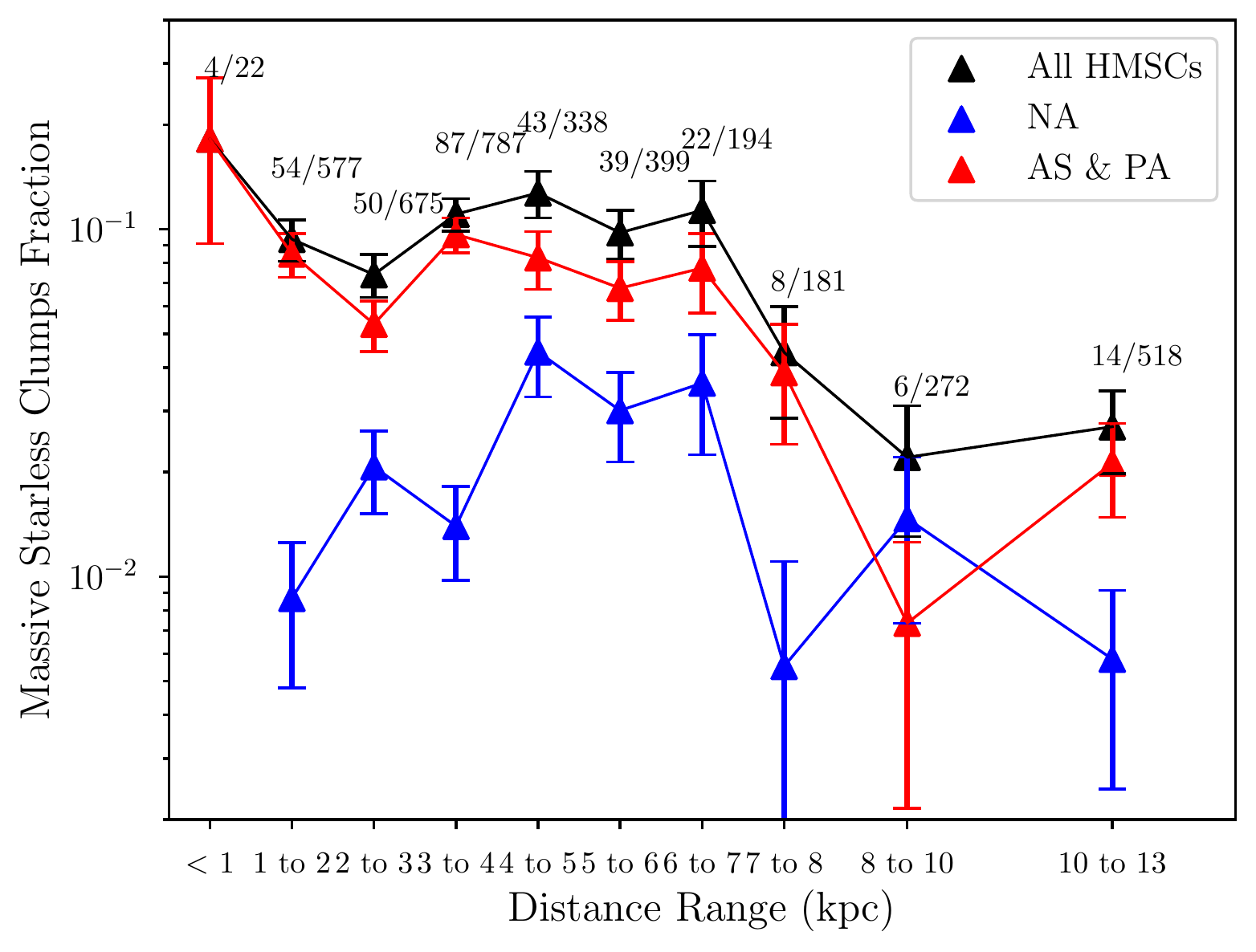}   
       \caption{Fraction of HMSCs. The fraction is calculated by the number ratio of HMSCs to massive ATLASGAL clumps in each distance range, as shown in the numbers in figure. The sample of ATLASGAL clumps is taken from the catalog in \citet{cse14} and the clump distance is taken from \citet{urq18}. ATLASGAL clumps with a 870~\micron~peak intensity higher than 0.5 Jy~beam$^{-1}$ are extracted to be in line with the column density criterion for massive star formation in Sect. \ref{HS}. The black, red, and blue lines indicate the changes of fractions with distance.}
        \label{starless-fraction}
    \end{figure}

\citet{bal17} studied the distance bias when deriving physical properties by fitting SED to Hi-GAL data. They find that the physical properties of the clump, such as temperature, the contribution of inter-core emission in the clumps, and the fractions of starless clumps to protostellar or all clumps, could change with distance. The fractions of HMSCs to massive ATLASGAL clumps in different distance ranges are shown in Fig.~\ref{starless-fraction}. The decreasing trend of starless fraction with distance coincides with \citet{bal17}, who explained it as the combined effects of the unresolved clumps (similar angular size) and their embedded multiple cores (protostellar but also starless). The HMSCs located at large distance are at the large size end (about 1~pc) of all HMSCs and the emission of embedded compact core(s) is more easily confused with the emission of diffuse gas, leading to the error in the determination of physical parameters. We roughly separate HMSCs into three distance ranges, which are <~4~kpc, 4~kpc to 8~kpc, and >~8~kpc according to their distributions. The HMSCs in these distance ranges should be individually considered when comparing the properties of the clump, which show a consistent increasing or decreasing change among distance.

Firstly, we preform Kolmogorov-Smirnov (K-S) tests to the physical properties with a flat relation with distance, which are \dustt, $L/M$ ratio, and possibly also \nhtcd. If the properties have a significant increasing or decreasing trend with distance, it could become a bias for K-S tests. The K-S test conclusions that two considered groups (AS and NA) are different could be just due to the different distance distributions. K-S tests are preformed to three combinations, which are (1) S-type AS and NA, (2) S-type AS and O-type AS, and (3) O-type AS, and NA. When K-S test $p$ value is smaller than 0.05, we reject the null hypothesis that the two samples are from the same distributions. The results of K-S tests shown in Table~\ref{ks_test} suggest that \dustt~and $L/M$ ratio are significantly different for S-type AS, O-type AS, and NA, while \nhtcd~is similar for the different types of HMSCs.

For those properties with a consistent increasing or decreasing trend with distance, we need to carry out the K-S test within the same distance ranges. However, the small size of the sample in a certain distance range decreases the K-S test reliability, for example, only about 20 HMSCs are in the distance range of >~8~kpc. Therefore, we define a dimensionless factor, ${\delta}_{\rm 1,~2} = \sqrt{2}\times(M_{\rm 1}-M_{\rm 2})/\sqrt{{{\sigma}_{\rm 1}}^2+{{\sigma}_{\rm 2}}^2}$, where $M_{\rm 1}$ and $M_{\rm 2}$ are the median values while ${\sigma}_{\rm 1}$ and ${\sigma}_{\rm 2}$ are the standard deviations of the corresponding number distributions of properties. The numbers 1 and 2 are the labels for two types of HMSCs. This factor shows the ratio of median value difference to the combined standard deviations of two distributions, which allows for a simplified comparison between two types of HMSCs. For example, if two Gaussian distributions have the same profile but with a shift of standard deviation $\sigma$ in Gaussian center, the resulted $\delta$ is equal to 1, indicating the difference at 1$\sigma$ level.

\begin{figure}
       \centering
   \includegraphics[width=0.45\textwidth]{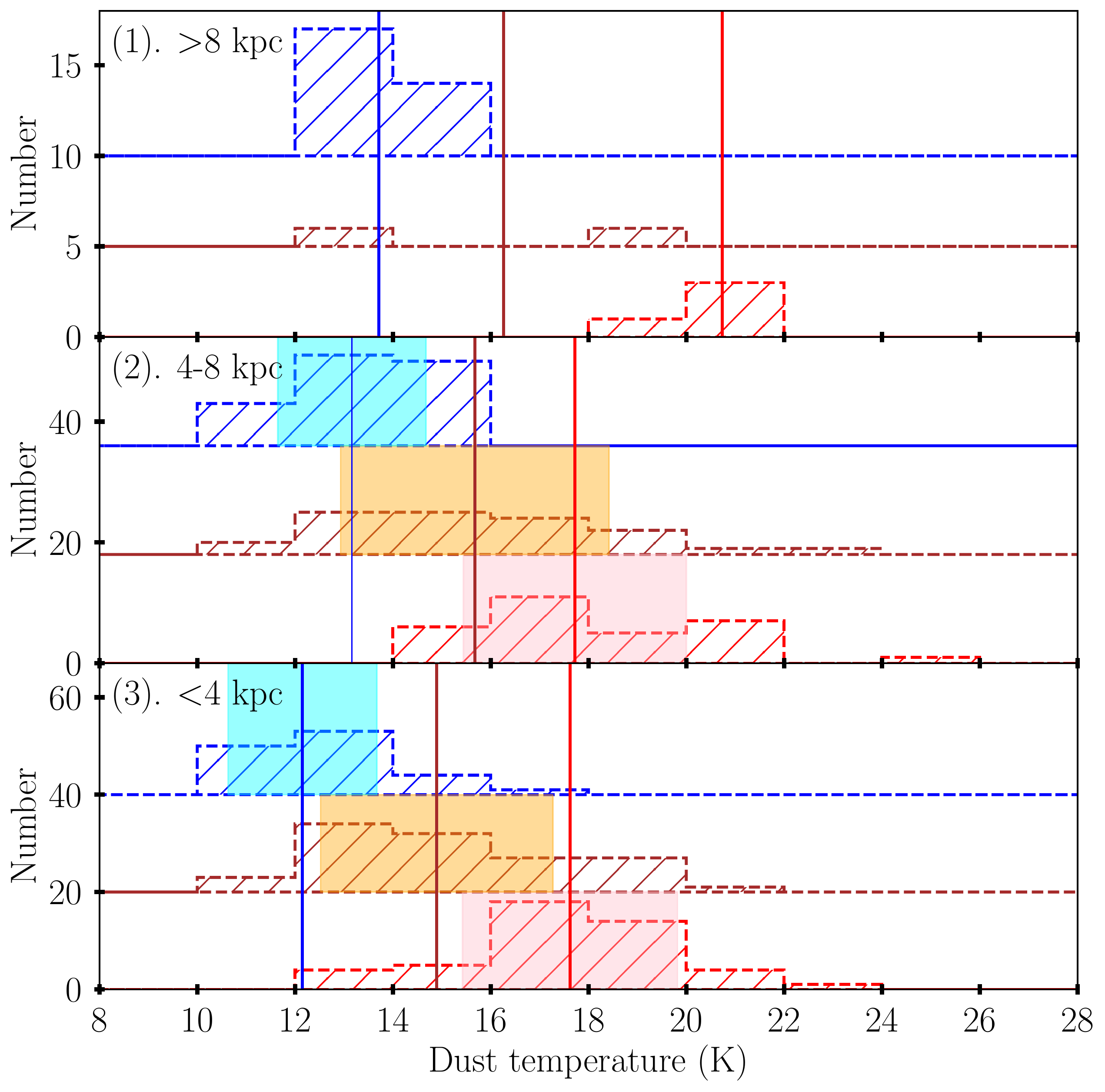}
       \caption{Number distributions of \dustt~of HMSCs.  Three panels from bottom to top represent the number distributions in the distance ranges of <~4~kpc, 4 to 8~kpc, and >~8~kpc, respectively. The red, dark brown, and blue histograms represent the number distributions for S-type AS, O-type AS, and NA, respectively. The solid lines indicate their median values. The pink, brown, and cyan filled color regions have a full width of double standard deviation 2$\sigma$ and center at median value of corresponding types of HMSCs.}
    \label{properties:temperature}
  \end{figure}

\begin{figure}
       \centering
  \includegraphics[width=0.45\textwidth]{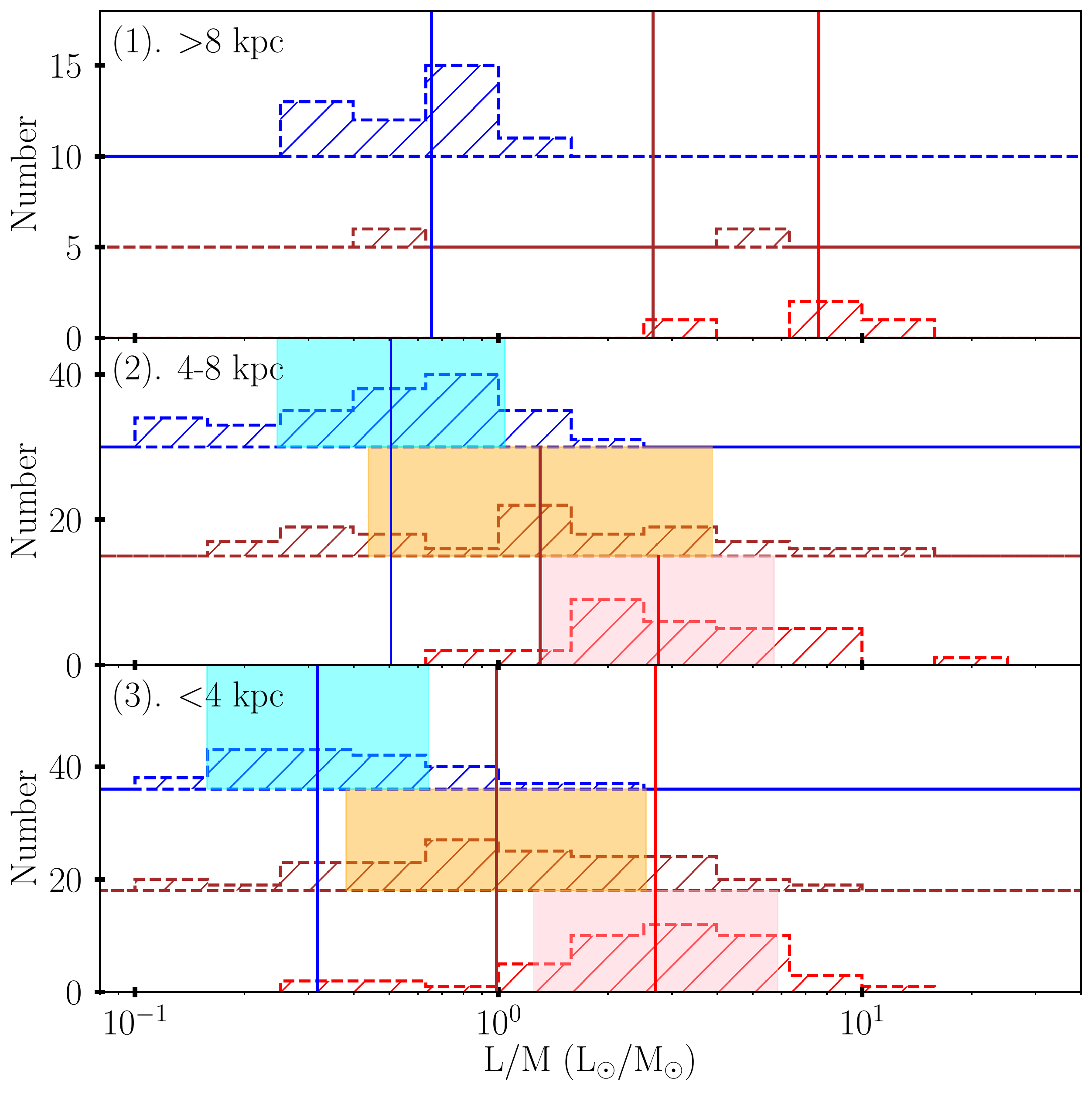} 
       \caption{Number distributions of $L/M$ ratio of HMSCs. The meanings of histograms, lines, and color-filled regions are similar to Fig. \ref{properties:temperature} but for $L/M$ ratio. }
    \label{properties:LM_ratio}
  \end{figure}

The number distributions of eight properties are presented in Fig.~\ref{properties:temperature}, \ref{properties:LM_ratio}, \ref{properties:Lclump} and Fig.~\ref{properties:Nh2} to \ref{properties:Ecc_ratio} (see Appendix \ref{appendix-properties}). Generally, only \dustt, $L_{\rm clump}$, and $L/M$ differences keep on the order of 1$\sigma$ to 2$\sigma$ in each distance range, which could be seen in Fig. \ref{figure:3} by the $\delta$ values of S-type AS and NA (${\delta}_{\rm S-AS,~NA}$) and that of O-type AS and NA (${\delta}_{\rm O-AS,~NA}$) in three ranges of distance. If a physical property is  different under the effect of the \hii~regions, it should show a consistent increasing or decreasing trend from S-type AS to O-type AS and then NA owing to the gradually weaker impacts of the \hii~regions. If we set the criteria of significant difference as $\delta$ larger than 1 or K-S tests suggest that the distributions are different, only \dustt, $L_{\rm clump}$, and $L/M$ show significant differences between different types of HMSCs.\ The AS types have a higher \dustt, $L_{\rm clump}$, and $L/M$ ratio than NA. When we set the criteria of similarity as $\delta~<~0.5$ and there is an inconsistent trend of parameter changing from S-type AS to O-type AS and then to NA, $n_{\rm H_{2}}$, \nhtcd, $M_{\rm clump}$, size, and eccentricity are similar between different HMSCs.\\

   \begin{figure}
       \centering
  \includegraphics[width=0.45\textwidth]{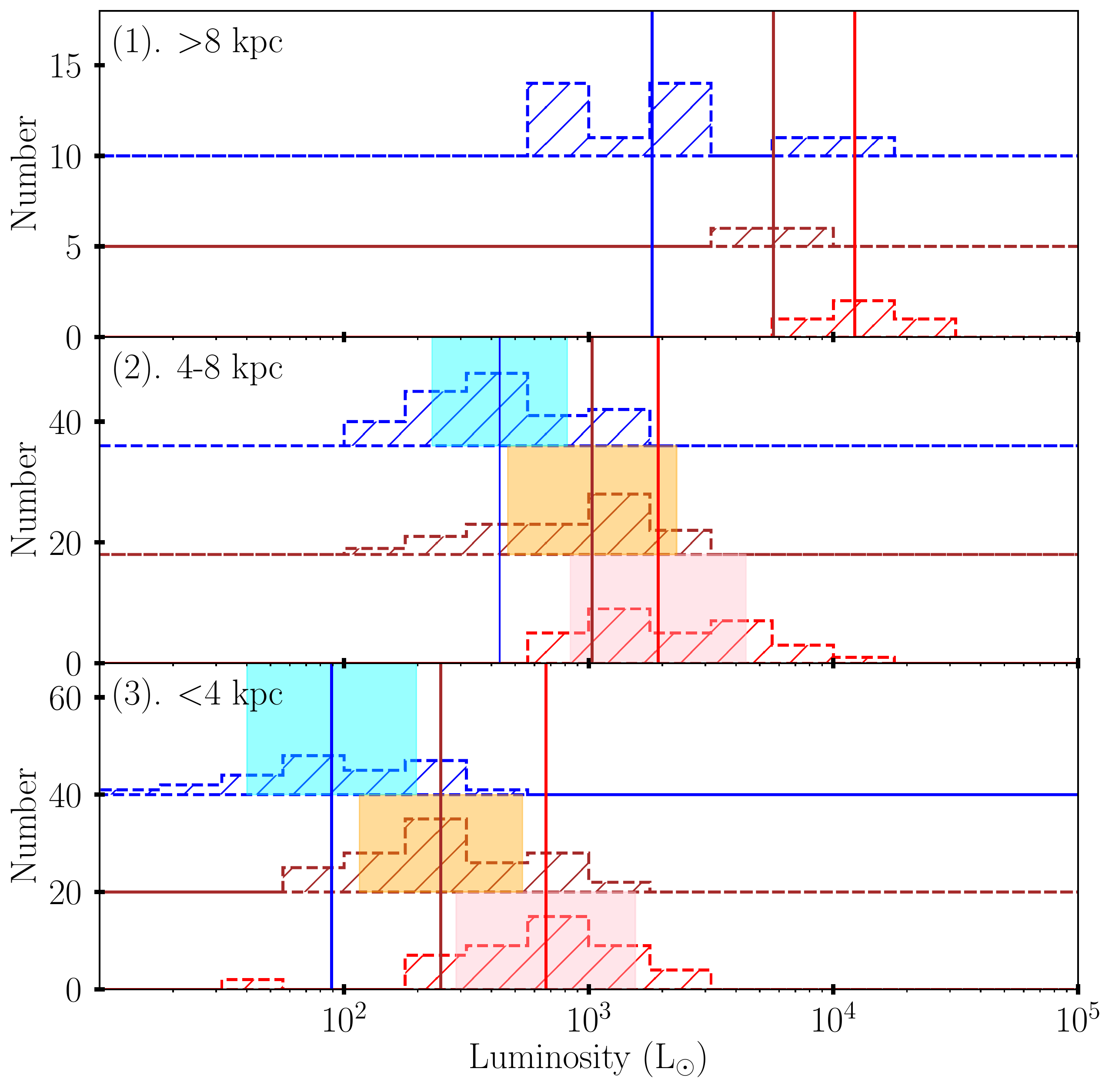}
       \caption{Number distributions of $L_{\rm clump}$ of HMSCs. The meanings of histograms, lines, and color-filled regions are similar to Fig. \ref{properties:temperature} but for $L_{\rm clump}$.}
    \label{properties:Lclump}
  \end{figure}

   \begin{figure}
   \includegraphics[width=0.45\textwidth]{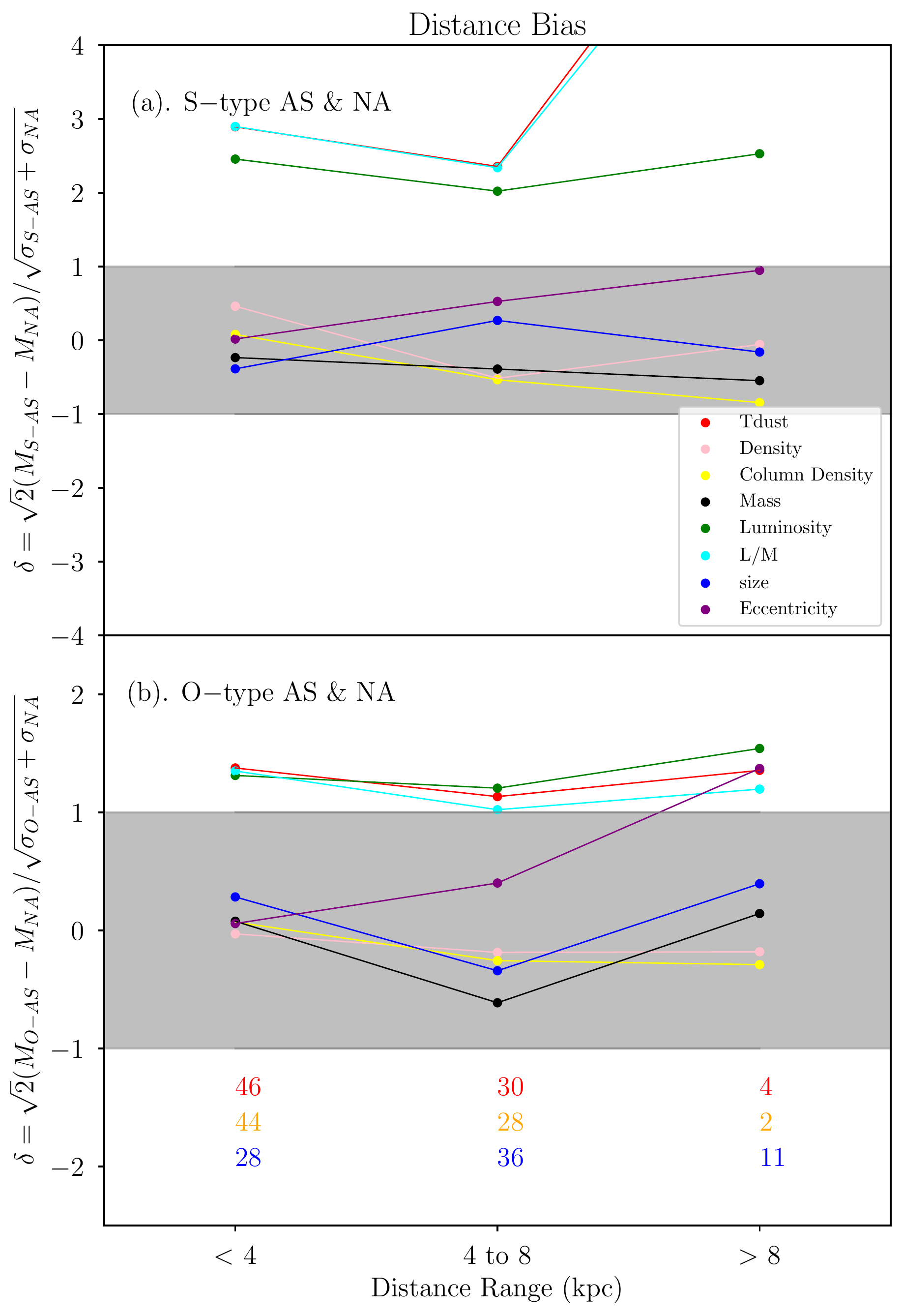}
       \caption{Distance bias. Panels (a) and (b) show $\delta$ values for the combination of S-type AS and NA, and the combination of O-type AS and NA HMSCs, respectively. All sources are grouped into three distance bins, which are <~4~kpc, 4 to 8~kpc, and > 8~kpc. Dots with different colors mean different properties of HMSCs. The red, orange, and blue numbers in (b) are the numbers of S-type AS, O-type AS, and NA HMSCs in the corresponding distance ranges. The gray regions are $|\delta| \leq 1$, where we suggest that the differences are insignificant.}
        \label{figure:3}
    \end{figure}

   \begin{table*}[ht]
    \begin{threeparttable}
  \tiny
   \centering 
    \caption{K-S test for S-type AS, O-type AS, and NA HMSCs. Three combinations are used for the test.} 
    \label{ks_test} 
    \begin{tabular}{c c c c c c c c c c c c c} 
    \hline\hline 

 Properties      & \multicolumn{4}{c}{S-type AS vs. NA}      &  \multicolumn{4}{c}{S-type AS vs. O-type AS}     &  \multicolumn{4}{c}{O-type AS vs. NA}  \\ 
                 & K-S result& K-S $D$\tnote{1} & K-S $p$\tnote{1}  &$\delta$        &   K-S result  &K-S $D$ & K-S $p$  & $\delta$           & K-S result  & K-S $D$ & K-S $p$   & $\delta$  \\%
   \hline 
\dustt           & Different & 0.848      & 4.44$\times10^{-16}$      & 2.5               & Different  & 0.48        & 1.45$\times10^{-8}$  & 1                  & Different  & 0.487       & 1.78$\times10^{-8}$ & 1.14 \\                                                                                                                                                                              
\nhtcd           & Same      & 0.192      & 0.0989                    & $-$0.285            & Same       & 0.0956      & 0.829                & $-$0.197             & Same       & 0.125       & 0.556               & $-$0.0862 \\                                                                                                                                                                           
$L/M$            & Different & 0.847      & 4.44$\times10^{-16}$      & 2.51              & Different  & 0.476       & 1.98$\times10^{-8}$  & 0.97              & Different  & 0.461       & 1.39$\times10^{-7}$ & 1.11 \\

     \hline                                         
      \end{tabular}
      \begin{tablenotes}
      \item [1] $D$ and $p$ are the K-S statistic and the two-tailed $p$ value of K-S test, respectively. 
      \end{tablenotes}
      \end{threeparttable}
      \end{table*}

\subsection{Difference in properties} \label{DPP}
The properties with significant difference between different HMSCs, which are \dustt, $L/M$ ratio, and $L_{\rm clump}$, are described in detail in this section. Other properties that are similar for AS and NA are briefly described and more detailed descriptions for these are given in Appendix \ref{appendix-properties}. 

\paragraph{\textbf{(a). Dust temperature \dustt.}} The changes of \dustt~with distance are very flat according to Fig.~\ref{distance-scatter} and therefore we could compare \dustt~without consideration of distance factor. The K-S tests show that \dustt~distributions are different between S-type AS, O-type AS, and NA. From S-type AS to NA, a decreasing trend of \dustt, from about 18 to 12~K, is shown in Fig. \ref{properties:temperature}. The \dustt~difference is larger than the typical \dustt~error in the SED-fitting process, which is about 2 K to 3 K. Furthermore, $\delta$ values indicate that the differences are statistically significant, from 1$\sigma$ level between O-type AS and NA to 2$\sigma$ level between S-type AS and NA. The results show the probably stronger external heating effect of nearby \hii~regions on S-type AS. We discuss this in more detail in  Sect.~\ref{temperature and L/M}. 

\paragraph{\textbf{(b). Clump luminosity-to-mass ratio $L/M$.}} The $L/M$ ratio has a flat relation with distance in Fig. \ref{distance-scatter}, thus we could directly compare it. The K-S tests suggest that their distributions between AS and NA HMSCs are different. The $L/M$ ratios decrease from about 3~\lsun/\msun~to 1~\lsun/\msun~and then 0.4~\lsun/\msun~for S-type AS, O-type AS, and NA, respectively. The typical error of $L/M$ ratio is less than 40\%, thus the difference is larger than error in the calculation. Meanwhile, $\delta$ values indicate that the difference is on the order of 1$\sigma$ between O-type AS and NA, while it increases to 2$\sigma$ between S-type AS and NA, showing the likely stronger impacts of \hii~regions.

The $L/M$ ratio is not an independent parameter if its calculation is related with far-IR/(sub)millimeter SED fitting because of the similar gray-body functions used in the fitting process \citep{19pit}. The $L/M$ ratios are in proportion to the ratio of integrated flux of gray body function to \nhtcd~(see equation 4 and 7 in Y17), while \nhtcd~is derived by fitting the gray-body function. A higher \dustt~ results in a higher $L/M$ ratio. In the regime of 12 K to 28~K, an increment of 3 to 4~K in \dustt~could increase $L/M$ about 2 to 3~\lsun/\msun~for a clump with a mass of 400~\msun~that is the median mass of the sample. Thus, the $L/M$ ratio difference could be well explained by the \dustt~difference.
	 
With the single-dish observed thermometer molecule CH$_{3}$C$_{2}$H, \citet{mol16} suggest that $L/M$ ratio could be an indicator of star formation evolutionary stage for massive clumps. These authors define three $L/M$ intervals, which are $L/M \lesssim 1$~\lsun/\msun, 1~\lsun/\msun~$\lesssim L/M \lesssim 10$~\lsun/\msun~and $L/M \gtrsim 10$~\lsun/\msun, respectively. These three intervals correspond to three evolutionary stages from (1) very early stages when only low-mass YSOs are forming; to (2) the stage in which relatively low-mass YSOs build up the luminosity of the clump; and then to (3) the stage in which massive stars are forming and warming up the inner clump gas. Most of S-type AS have a $L/M$ ratio larger than the first interval. Some of these could even arrive 10~\lsun/\msun. The $L/M$ ratios of O-type AS are around first interval while most of NA are below the first interval. Recently, \citet{mol19} connected the mass fraction locked in the dense cores of massive clumps to the $L/M$ ratios of clumps. The expression $L/M \lesssim 1$~\lsun/\msun~classifies the protoclusters with a dense core mass fraction $\lesssim 0.1,$ whereas $L/M \gtrsim 10$~\lsun/\msun~excludes dense core fraction $\lesssim 0.1$. The significant $L/M$ change between AS and NA indicates that it is uncertain to classify the early evolutionary stages of massive clumps with the only observed $L/M$ ratio. We discuss this point in Sect.~\ref{temperature and L/M}.
	 
\paragraph{\textbf{(c). Clump luminosity $L_{\rm clump}$.}}The $L_{\rm clump}$ significantly increases with distance and therefore we could not directly compare it. In the certain distance bins, $\delta$ values show that S-type AS have a larger $L_{\rm clump}$ than NA at 2$\sigma$ level while the difference between O-type AS and NA is at 1$\sigma$ level, suggesting the probably different strength of external heating  of \hii~regions on HMSCs between S-type AS and O-type AS.

\paragraph{\textbf{(d). Other properties.}} The \nhtcd, $n_{\rm H_{2}}$, $M_{\rm clump}$, size, and eccentricity are similar for AS and NA. Their $\delta$ values generally indicate that the difference between AS and NA is less than 0.5$\sigma$. The gradual changes of properties from S-type AS to O-type AS and then NA are not clear for these properties. Owing to the selection criteria of S-type AS, which are the features of being photodissociated/photoionized or being in a compressed structure, we probably could expect that S-type AS could be denser than NA. The results show that the difference is not robust. It could be due to the beam dilution effects that narrow the density differences (see Appendix \ref{appendix-properties}).\\

\section{Column density and temperature structure}  \label{Com}

\subsection{Calculation} \label{Com_cal}
In this section, we compare the differences between the structures of different HMSCs. An isothermal, spherically symmetric prestellar clump with gravitational contraction is suggested have a clump radial density profile consisting of two regions: an inner core $n_{\rm H_{2}} = constant$ and envelope $n_{\rm H_{2}} \propto r^{-2}$ \citep{vaz19}. The corresponding radial \nhtcd~profiles are $N_{\rm H_{2}} \propto r$ and $N_{\rm H_{2}} \propto r^{-1}$, respectively. Owing to the limited resolution, a flat central region and truncated radius are usually set when fitting the \nhtcd~profile to the observational data \citep{juv18, tan18}. The Galactic cold cores \nhtcd~structures investigated with \textit{Herschel} data indicate that radial \nhtcd~distribution follows a power law of $N_{\rm H_{2}} \propto~r^{-1}$ in spite of the broad variety of clump morphology, which suggests the universality of the $r^{-1}$ profile for cold cores with gravitational contraction \citep{juv18, ligx18}.
 
We compare \nhtcd~or \dustt~radial profile of different HMSCs with PPMAP data. Beam size is a crucial factor that could affect the clump structures we observed. A larger beam smooths the steep profile to make it flatter. Furthermore, beam size determines the lower size limit of the flat central region. The resolution of PPMAP Hi-GAL data, which is 12\arcsec, is more likely to detect the envelope rather than the inner core of clumps at a distance of several kpc. We simply fit a single power law radial \nhtcd~profile $N_{\rm H_{2}} = C\times r^{p_{N_{H_{2}}}}$ to the PPMAP \nhtcd~data rather than with a more complex function. 

\begin{figure}
   \includegraphics[width=0.45\textwidth]{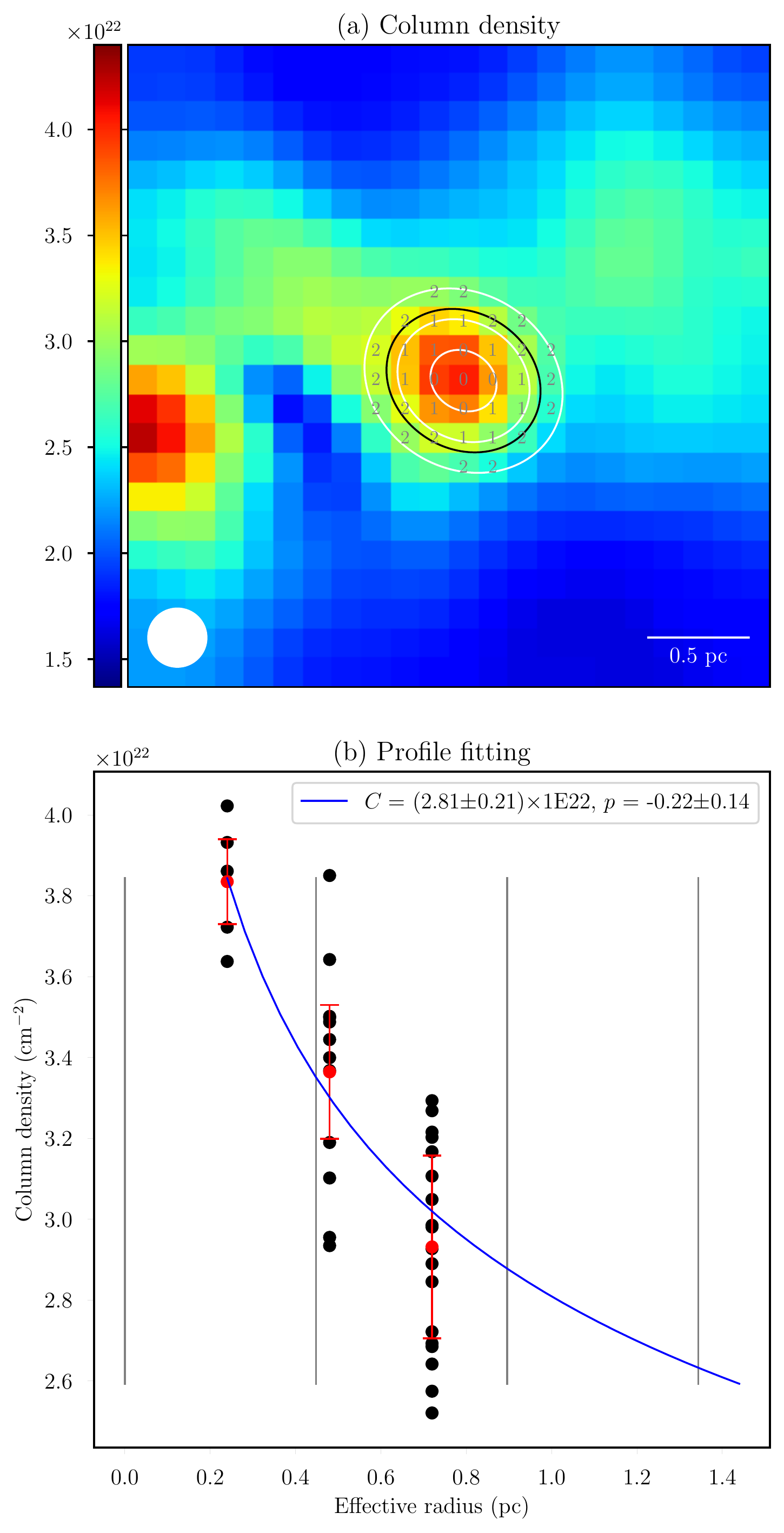}
       \caption{Column density profile fitting for HMSC G010.9823$-$0.3677. Panel (a) shows the PPMAP \nhtcd. The black and white ellipses represent the clump and its shells used for extracting pixels. The gray numbers denote the corresponding shells of the pixels. The beam of PPMAP is shown in bottom left. Panel (b) shows the profile fitting. The black and red dots indicate the PPMAP \nhtcd~and their median values in the corresponding shells. The error of \nhtcd~value of each shell is estimated by the 25th and 75th percentiles of \nhtcd~of each shell, which is shown by the red error bar. The blue line is the fitting result for median values. The gray lines show the triple beams of PPMAP.}
        \label{profile_example}
    \end{figure}

Firstly, only AS and NA HMSCs with a mass larger than 100~\msun~are considered to ensure the selection of massive sources. The HMSCs with an angular size <~24\arcsec~and distance > 8.34~kpc are removed to mitigate the resolution issue. Secondly, we separate the PPMAP pixels into a series of concentric elliptical shells with a shell width of 6\arcsec~in minor axis, which is the pixel size and the half beam size of PPMAP data. The aspect ratios and position angles of each elliptical shell of HMSCs are the same as the ellipses outlining the shapes of HMSCs in Y17. The radius of the largest shell is slightly larger than the clump size. For the pixels crossing the edges of the shells, we simply assign the associated shells according to the centers of these pixels. Thirdly, the median \nhtcd~in one shell is used as the \nhtcd~in this shell whereas the 25th and 75th percentiles of the pixel \nhtcd~of the shell are used as the error ranges of \nhtcd~for this shell. Finally, the median \nhtcd~in different shells are fitted with a power-law profile $N_{\rm H_{2}} = C\times r^{p_{N_{H_{2}}}}$. An example of \nhtcd~profile fitting is shown in Fig.~\ref{profile_example}. Similar operations are carried out on PPMAP \dustt~images to look at the temperature profile and get a power-law index \pdustt.\\ 

\subsection{Properties and results} \label{CPP}
\begin{figure*}
    \centering 
  \includegraphics[width=0.95\textwidth]{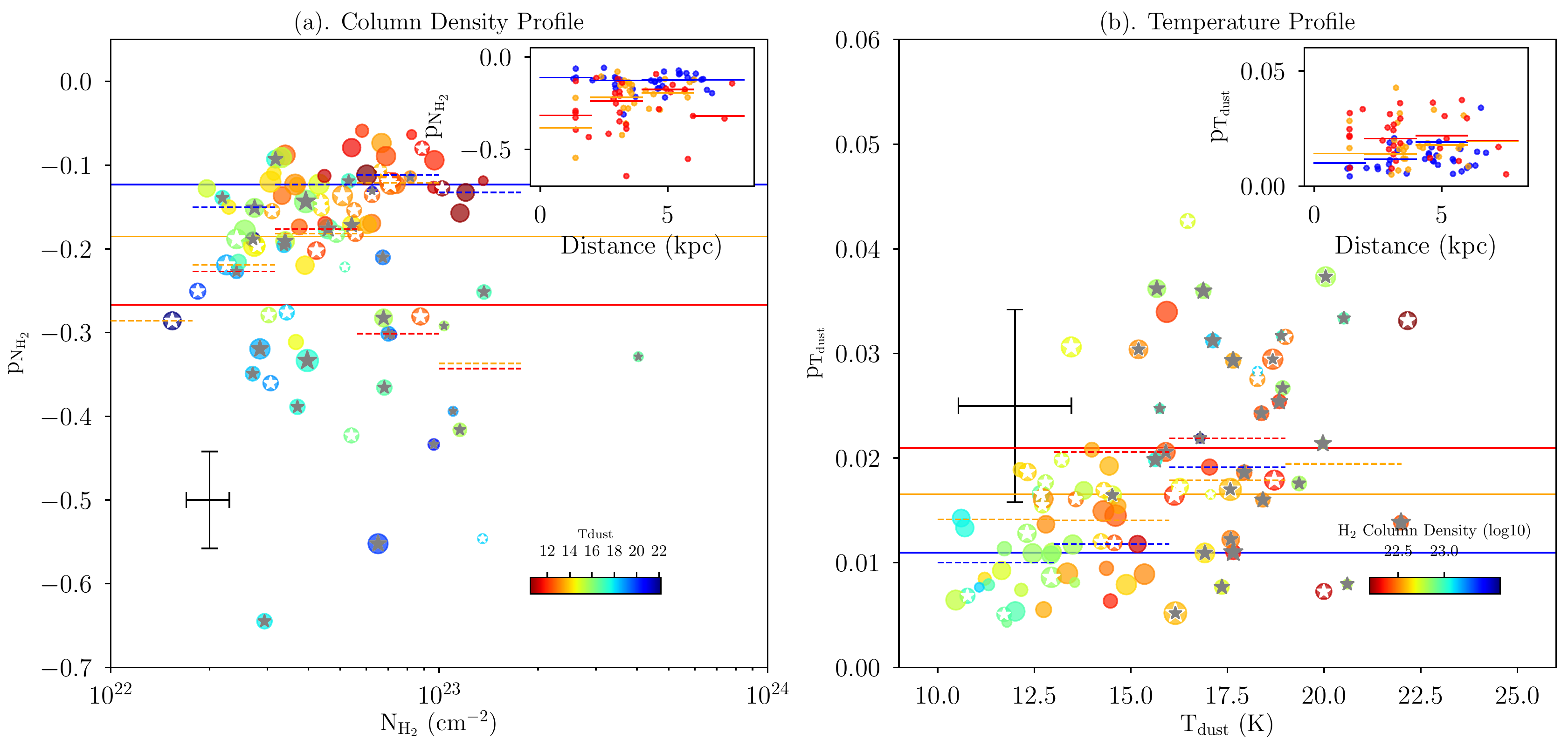}
       \caption{Power-law index of \nhtcd~and \dustt~profile. The whole \dustt~and \nhtcd~of the clump are taken from updated Y17 properties in this paper. The $p$ values are derived from Hi-GAL PPMAP data. Panels (a) and (b) show the column density and dust temperature power-law index $p$, respectively. The crosses represent the typical errors. The dots with gray stars, white stars, and without stars represent the S-type AS, O-type AS, and NA HMSCs, respectively. The red, orange, and blue solid lines represent the median values for S-type AS, O-type AS, and NA, respectively. The red, orange, and blue dashed lines represent the median values for S-type AS, O-type AS, and NA in the corresponding column density or temperature bins. Sub-figures in the top right show the power-law index and its relation with distance. The red, orange, and blue dots represent S-type AS, O-type AS, and NA, respectively. The red, orange, and blue short lines show the median values in different distance bins for the corresponding types of HMSCs.}
        \label{figure:4}
    \end{figure*}

The \nhtcd~and \dustt~power-law indexes (\pnhtcd~and \pdustt)~are shown in Fig.~\ref{figure:4} (a) and (b), respectively. The median value of \pnhtcd~is $-$0.17, which is much larger than the value of $-$0.85 in \citet{juv18}. There are several reasons for this difference. The clumps/cores they studied have a more compact morphology with a typical size of 0.075~pc compared to $\simeq 0.5$~pc for our HMSCs. Their beam size is equal to 0.03~pc with a typical distance of 250~pc, but our beam size is 0.4~pc with a typical distance of 4~kpc. Our relatively poor physical resolution could create a higher \pnhtcd~value. Furthermore, their fitting field is larger than the clump scale, which includes the weaker background emission that results in a smaller \pnhtcd~value. Our \pnhtcd~values are very similar to the results of \citet{gia13}. These authors simply fitted a power-law profile to 1.2~mm continuum emission of high-mass clumps with a beam size of 20\arcsec. They find a typical \pnhtcd~value of $-$0.2 to $-$0.4 for high-mass quiescent clumps, whereas values smaller than $-$1 for high-mass star-forming clumps. Their results indicate that our fitting process is reliable at a similar resolution of 12\arcsec. We note that our main goal is not to derive a specific and exact value of \pnhtcd~but to search for a possible difference between AS and NA HMSCs.

The median values of \pnhtcd~are $-$0.27, $-$0.19, and $-$0.12 for S-type AS (30), O-type AS (26), and NA (34), respectively. Their standard deviations of distributions are 0.13, 0.1, and 0.05, respectively. The $\delta$ values suggest that S-type AS have a smaller value of \pnhtcd~compared with that of O-type AS and NA at 0.7$\sigma$ and 1.5$\sigma$ level, respectively, showing the significant differences between S-type AS and NA. All HMSCs with a \pnhtcd~smaller than $-$0.3 are AS HMSCs except for one NA. A $\chi$$^2$ test to AS and NA with \pnhtcd~larger and smaller than $-$0.2 results in 99\% to reject the hypothesis of independence between the association of \hii~region and \pnhtcd~value. Therefore, a smaller \pnhtcd~value, which suggests a steeper \nhtcd~profile, is closely related with the association of \hii~region. 

The results of \dustt~profile shown in Fig.~\ref{figure:4} (b) are very similar to the results of \nhtcd~profile. The S-type AS have \dustt~profiles steeper than those of NA HMSCs; the median values and standard deviations are 0.021 and 0.01 for S-type AS, 0.016 and 0.01 for O-type AS, and 0.01 and 0.05 for NA HMSCs. The $\delta$ values indicate the differences between AS and NA are at about 1$\sigma$ level. 

In brief, AS HMSCs have steeper \nhtcd~and \dustt~profiles than those of NA HMSCs and the difference between AS and NA types enlarges with a more drastic feedback of the \hii~regions. \\

\subsection{Bias}
The distance bias could impact our comparison of \pnhtcd~values between HMSCs with different distances. The small panel in the top right of Fig.~\ref{figure:4} (a) shows the trend that \pnhtcd~increase with distance. A larger distance indicates a larger physical size of HMSCs when the angular size is similar (about 30\arcsec), thus the profile of the most distant clump could be less easily resolved and more easily mixed with the surrounding density distribution. To check how distance bias affects our results, we separate HMSCs into a series of distance bins with a width of 2~kpc. The results indicate a consistent trend in each distance bin that AS occupy the lowest \pnhtcd~value end compared with NA. The panel in the top right of Fig.~\ref{figure:4} (b) suggests the trend that AS occupy the uppermost \pdustt~value end is also consistent in each distance bin.

The number of shells could affect the \pnhtcd~value because more shells in a clump means that the structure of the clump is better resolved and could result in a steeper observed profile. Of the 90 HMSCs for which we estimated the profile, only six clumps have four shells while the other HMSCs have three shells. Furthermore, these six clumps with four shells occupy the high \pnhtcd~value end ($-$0.2 to 0), which means that their profiles are flatter. Thus, we suggest that the effects of shell number do not significantly change our overall results.

Another factor changing the observed profile is the nature of Hi-GAL PPMAP data. It has been shown in Sect. \ref{CDT_PPMAP} that Hi-GAL PPMAP overestimates \dustt~and underestimate \nhtcd~for the clumps at the coldest end (<~14~K). For starless clump, its center is normally expected to have a lower \dustt~compared to the envelope. The Hi-GAL PPMAP data could overestimate \dustt~and underestimate \nhtcd~for the center of starless clump, making the profile flatter. \\

\subsection{Links between $L/M$ and steeper profile}
A steeper density profile indicates that the clump is more centrally peaked. Observational studies have shown that the density structure of the clump becomes more and more centrally peaked with the evolution from the starless stage to protostellar stage \citep{but12, tan18}. During the formation of protostars, the heating effect of inner protostars and the collapse/accretion make the clump warmer and more compact, thus a positive correlation between $L/M$ and centrally peaked properties of the clump is expected.

The HMSCs impacted more deeply by nearby \hii~region~are compressed and heated more deeply by the ionized gas, thus the clumps are warmer and more centrally peaked. A weak positive relation between \dustt~and steepness of profile can be found in Fig.~\ref{figure:4} (a) by the color of dots and the \pnhtcd~value. There is also a weak trend that \pnhtcd~values become smaller with increasing \nhtcd~as shown in Fig.~\ref{figure:4} (a). The median \nhtcd~and \dustt~are larger for the most centrally peaked AS (\pnhtcd~<~$-$0.3, 6.7 $\times$ 10$^{22}$~cm$^{-2}$ and 18.3~K) than those of other AS (4.7 $\times$ 10$^{22}$~cm$^{-2}$ and 15.7~K) and NA (4.5 $\times$ 10$^{22}$~cm$^{-2}$ and 12.9~K).

\begin{figure}
    \centering 
  \includegraphics[width=0.4\textwidth]{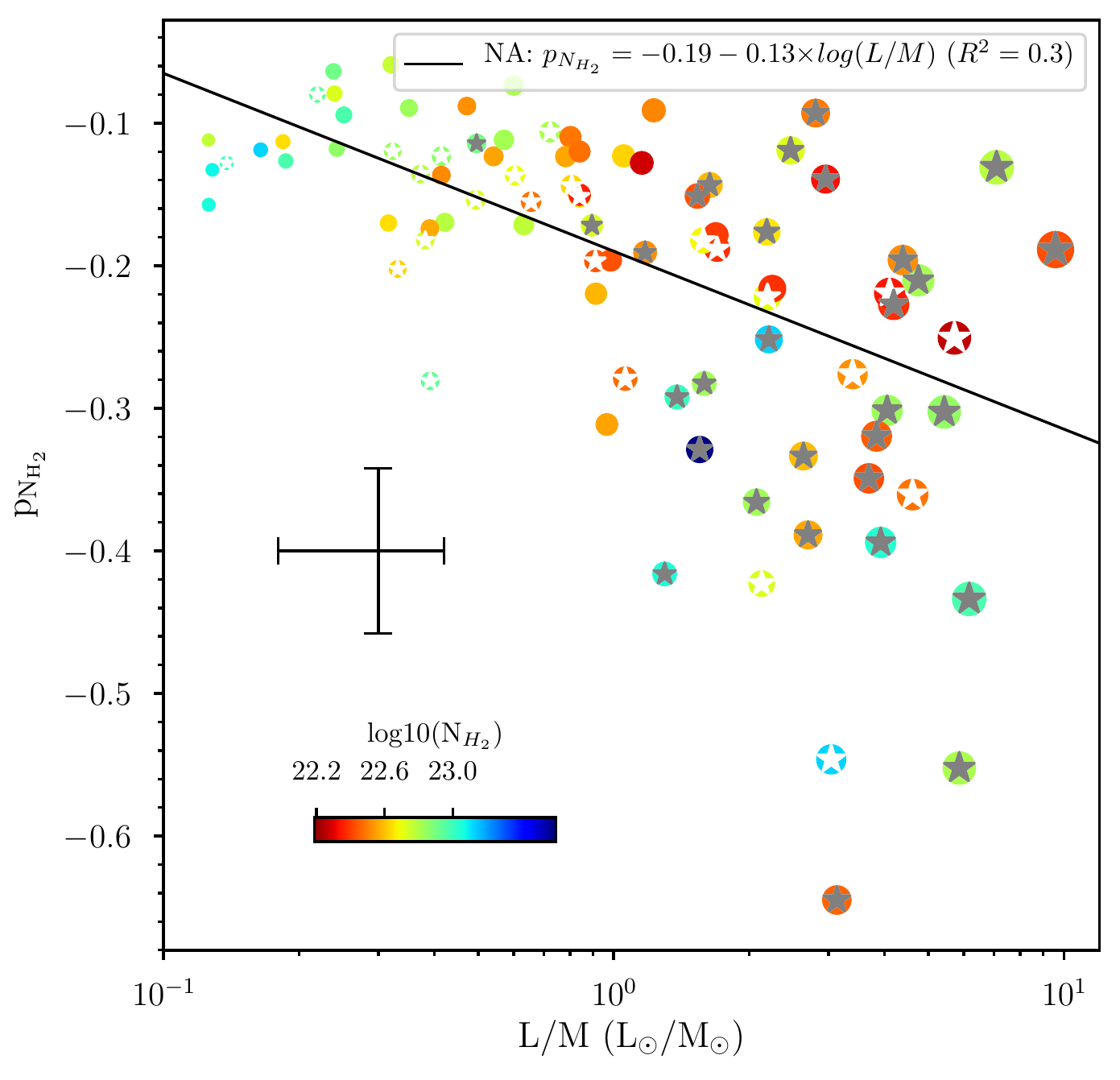}
       \caption{Plot of $L/M$ and \pnhtcd. The meanings of different dots are similar to Fig.~\ref{figure:4} (a). The colors of dots indicate \nhtcd~of HMSCs. The size of the dot is proportional to the \dustt. The cross represents the typical error. The black line indicates the linear regression to the data.}
        \label{lm-compactness}
    \end{figure}

A similar positive correlation between \pnhtcd~and $L/M$ of HMSCs is shown in Fig.~\ref{lm-compactness}. The reason might be the tight relation between \dustt~and impacts of \hii~region, as mentioned, rather than star formation activities inside HMSCs. The strengthened heating effect working on AS leads to a higher \dustt~and $L/M$ ratio. This result indicates that even for the \pnhtcd, the environment of the clump should be checked carefully in advance when these values are used as evolutionary probes for early massive clumps. \\

\section{Turbulence and virial status} \label{TVS}
In this section, we compare the turbulence level in AS and NA and seek to understand whether turbulence plays a dominant role in shaping massive star formation. Recent modeling results based on the global hierarchical collapse in molecular clouds argue that chaotic, multiscale infall due to gravity could dominate at all scales \citep{vaz19}. By analyzing a sample of 70 \micron~quiet clumps, \citet{tra19} propose a scenario that is a continuous interplay between turbulence and  gravity. The turbulence creates multiscale structure and then gravity becomes dominant when the critical density threshold is reached. Based on the hypothesis that mass distributions of stars result from turbulent fragmentation, \citet{pad19} propose a scenario in which large, converging, inertial flows in the environments of supersonic turbulence assemble massive stars. These studies shed light on the particular importance of understanding the sources that drive turbulence, their relative importance at the clump scale and the impacts of turbulence on regulating star formation \citep{lo15}. \\

   \begin{figure}
    \centering 
   \includegraphics[width=0.4\textwidth]{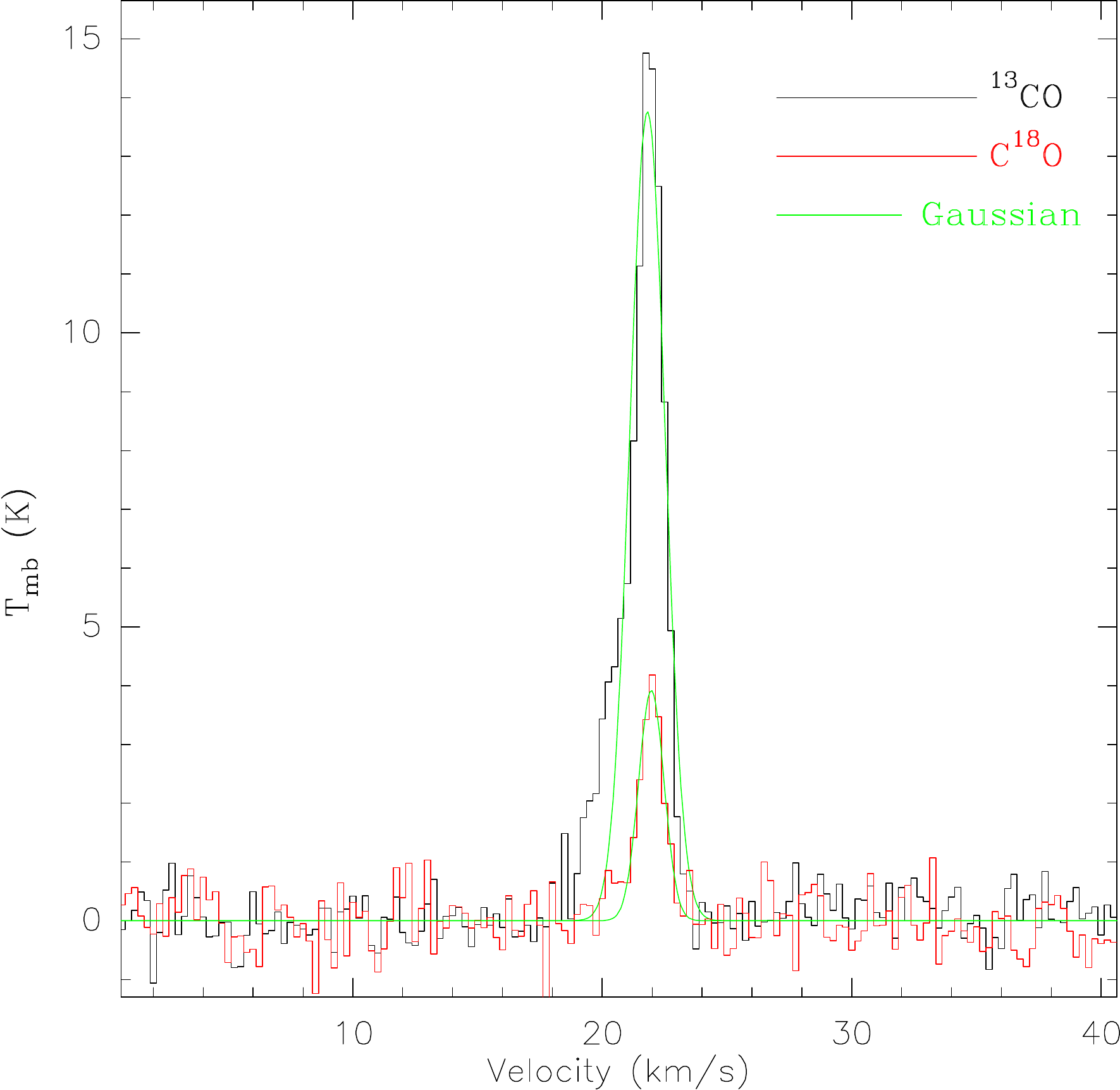}
       \caption{Example of Gaussian fitting to \tco~and \ceo~$J~=~2~-~1$ of the HMSC G015.2169$-$0.4267.}
        \label{fitting_example}
    \end{figure}

\subsection{Mach number estimation} \label{mach}
To show the turbulence level in HMSCs, we use SEDIGISM data to estimate the Mach number. SEDIGISM survey mapped the inner Galactic plane in the $J = 2 - 1$ rotational transition of \tco~and \ceo~with a spatial and a velocity resolution of about 30\arcsec~and 0.25~\kms, respectively. We extract the average spectra in one beam centered on the HMSC position from a spectral cube and then fit it with Gaussian using the CLASS package\footnote{\url{https://www.iram.fr/IRAMFR/GILDAS/doc/html/class-html/class.html}}. An example of Gaussian fitting to \tco~and \ceo~$J = 2 - 1$ spectra of the HMSC G015.2169$-$0.4267 is presented in Fig.~\ref{fitting_example}. We only use \ceotwo~rather than \tcotwo~to estimate the Mach number because \ceo~is likely to be optically thin and traces denser regions. \ceo~has a lower abundance than \tco~with a \tco/\ceo~abundance ratio of $\simeq$ 5 to 8 \citep{mil05, mat18, par18}.

The velocity dispersion $\sigma_{\rm v}$ is combined by a thermal component ${\sigma}_{\rm Therm} = (kT_{\rm ex}/m_{\rm C^{18}O})^{1/2}$ and nonthermal component ${\sigma}_{\rm NT}$ with equation ${\sigma_{\rm v}}^2 = {{\sigma}_{\rm Therm}}^2 + {{\sigma}_{\rm NT}}^2$. ${\sigma}_{\rm NT}/{\sigma}_{\rm Therm}$ for \ceo~lines of HMSCs in our cases are in a range of 10 to 40, suggesting the dominance of turbulence in HMSCs. We set ${\sigma}_{\rm NT}$ as the turbulence flow velocity, thus the Mach number $M = \sqrt{3}{\sigma}_{\rm NT}/c_{\rm s}$, where $\sqrt{3}$ is the correction from 1D to 3D and $c_{\rm s}= \sqrt{\gamma R_{\rm gas} T/m_{\rm mol}}$. The quantities $\gamma$, $R_{\rm gas}$, and $m_{\rm mol}$ are the adiabatic index, gas constant, and molar mass of the gas, respectively. \citet{ork17} use a similar method to estimate the Mach number with \tcoone~in the Orion B region. They assume that the opacity broadening, natural, and thermal widths of lines are negligible compared to turbulent broadening. We use ${\sigma}_{\rm NT}$ as the turbulent flow velocity rather than the velocity dispersion of spectra ${\sigma_{\rm v}}$ even if thermal widths make up only 2\% to 10\% width of the \ceotwo~lines in our HMSC sample. Opacity broadening is less important because \ceotwo~is more optically thin compared to \tcoone.

\begin{figure}
    \centering 
   \includegraphics[width=0.45\textwidth]{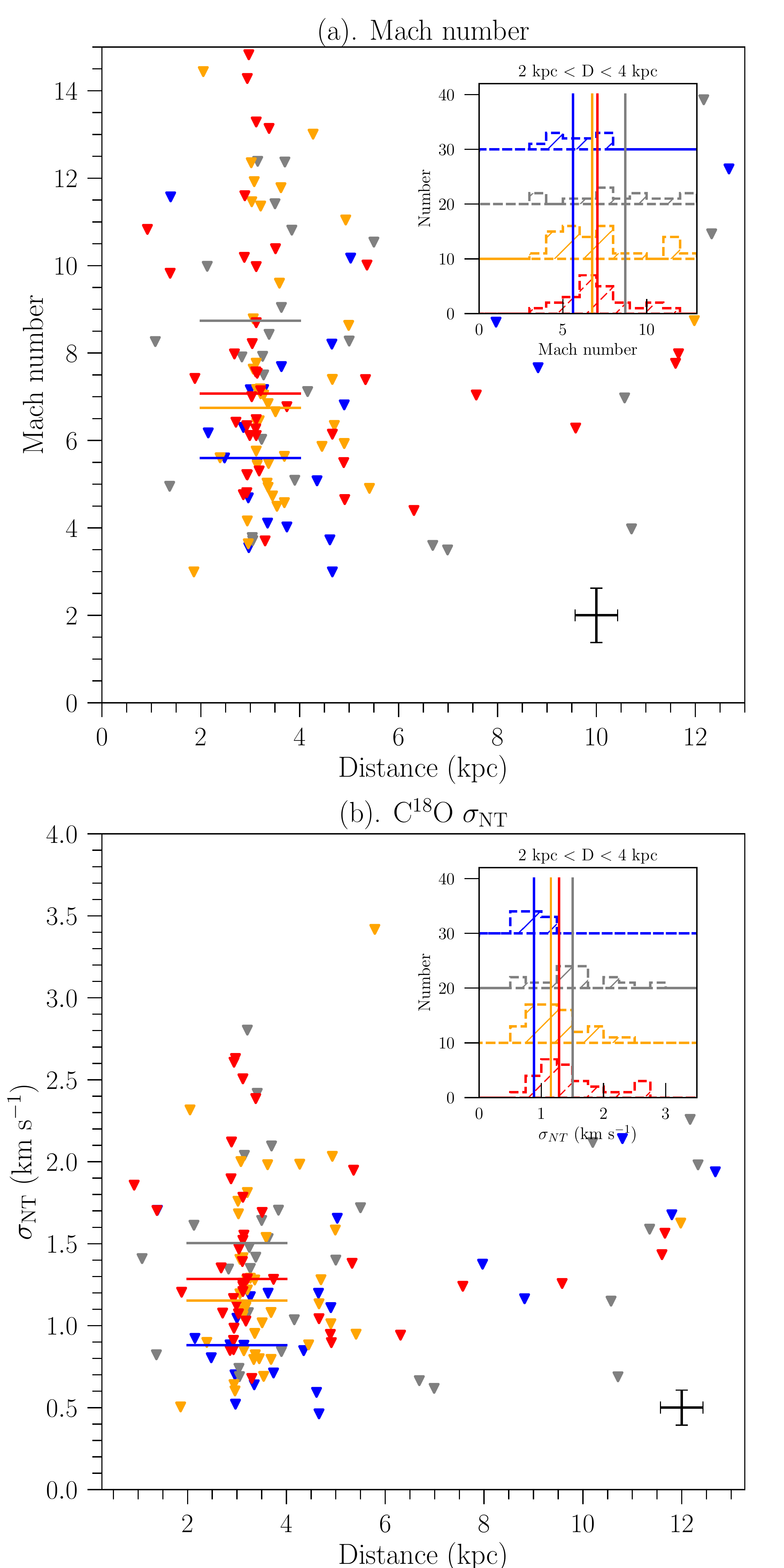}
       \caption{Turbulence. (a) and (b) show the Mach number and the non-thermal component of HMSCs, respectively. The red, orange, gray and blue triangles represent S-type AS, O-type AS, PA, and NA, respectively. The corresponding median values for different HMSCs in the 2 to 4~kpc range are shown in short lines with corresponding colors. The histograms in top right show the number distributions in the range of 2 to 4~kpc and their median values.}
        \label{turbulence_mach}
    \end{figure}

The Mach number, nonthermal component and their correlations with distance are shown in Fig.~\ref{turbulence_mach}. As shown in Sect.~\ref{DF}, HMSCs at a larger distance have a larger physical scale $r_{\rm pc}$. An increase in physical scale $r_{\rm pc}$ could increase $\sigma_{\rm v}$ according to Larson's law $\sigma_{\rm v} \propto {r_{\rm pc}}^{0.56 \pm 0.02}$ \kms~\citep{lar81}. An increasing trend of ${\sigma}_{\rm NT}$ from 2 kpc to 13~kpc shown in Fig.~\ref{turbulence_mach} (b) proves that the bias of distance exists when comparing $\sigma_{\rm v}$. Mach numbers or nonthermal components of the full sample studied do not show a clear difference between AS and NA. The ${\delta}_{S-AS,~NA}$ value indicates a slight difference (0.1$\sigma$) between S-type AS and NA in the full sample. To compare the Mach number more reliably, we extract HMSCs at a distance of 2 kpc to 4~kpc. More than half of \ceo~studied HMSCs (85 in all 135 HMSCs, including 11 NA, 16 PA, 30 O-type AS, and 28 S-type AS) are included. The HMSCs in 2 kpc to 4~kpc are better resolved and the numbers of different types of HMSCs are relatively statistically significant compared with other distance ranges with a width of 2~kpc. The median values of the Mach number in 2 kpc to 4~kpc are 5.6, 6.7, and 7.1 for NA, O-type AS, and S-type AS, respectively. The quantity $\delta \approx 1$  between S-type AS and NA, indicating that S-type AS have a Mach number statistically larger than that of NA. The turbulence level of HMSCs could be between supersonic regime (1.3 to 5) to hypersonic regime (5 to 10). The AS HMSCs are more likely to be hypersonic while NA types are more likely to be supersonic.  

\begin{figure}
    \centering 
   \includegraphics[width=0.45\textwidth]{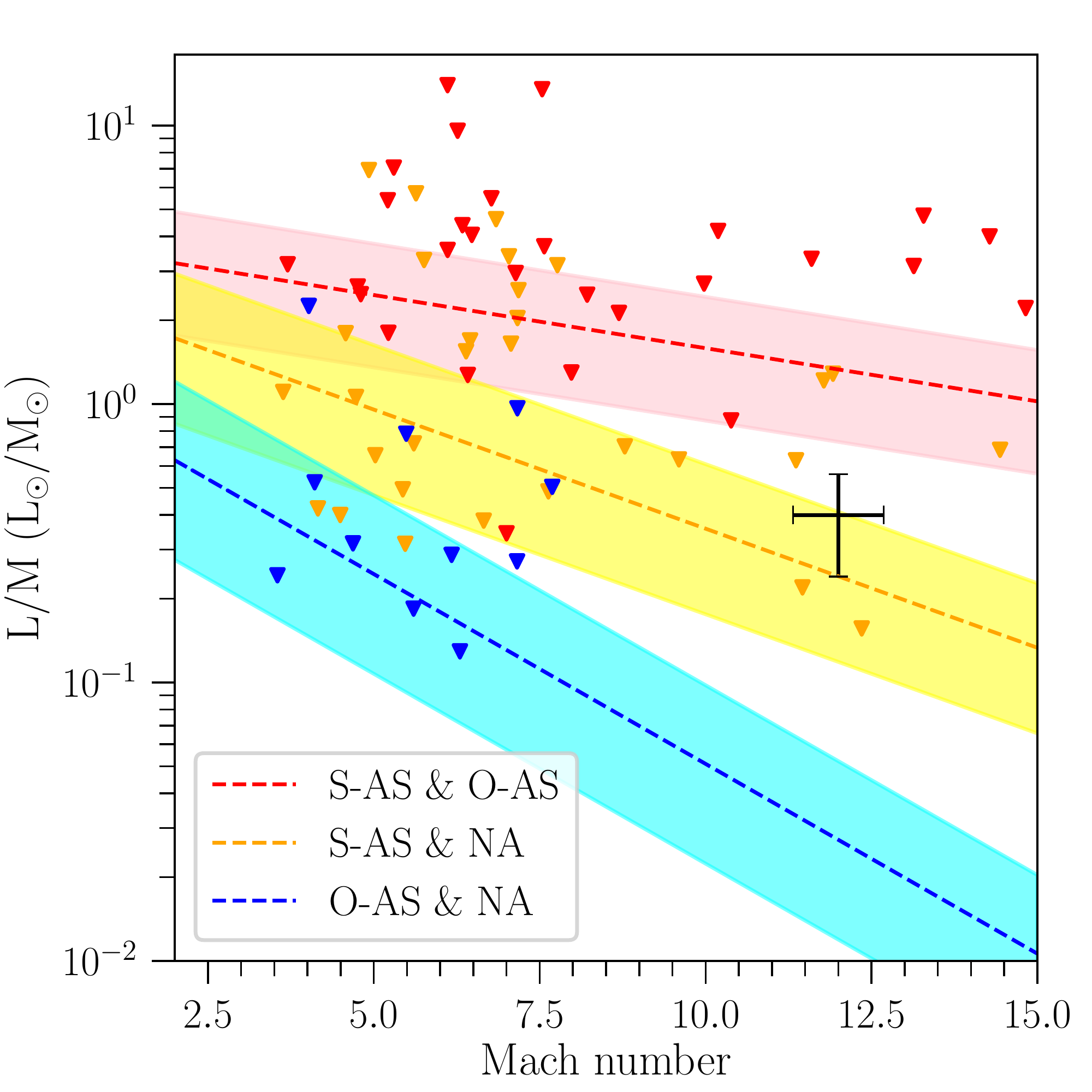}
  \caption{Turbulence and $L/M$ ratios of HMSCs in 2 kpc to 4~kpc. Typical errors of $L/M$ and Mach number are about 40\% of $L/M$ and 1 to 2, respectively. The uncertainties of \dustt~and line width are propagated to the calculations of Mach number. The triangles with different colors are similar to Fig.~\ref{turbulence_mach}. The blue, orange, and red dashed lines indicate the intervals of (1) O-type AS and NA, (2) S-type AS and NA, and (3) S-type AS and O-type AS. The cyan, yellow, and pink regions show the uncertainties of the boundaries. The red and blue lines roughly outline three regions: (i) low $L/M$ and Mach number, (ii) intermediate $L/M$ and Mach number, and (iii) high $L/M$ and Mach number.}
        \label{turbulence_relation}
    \end{figure}

A positive correlation between ${\sigma}_{\rm NT}$ and $L/M$ for presteller and protostellar dense clumps has been found with NH$_{3}$ lines of about 1000 clumps by \citet{mer19}. These authors propose that it could be due to the additional energy injection from the embedded protostars. As shown in Fig.~\ref{turbulence_relation}, NA, O-type AS and S-type AS HMSCs roughly occupy the (i) low Mach number - $L/M$ regime, (ii) intermediate Mach number - $L/M$ regime, and (iii) high Mach number - $L/M$ regime, respectively. The boundaries are estimated by the algorithm of support-vector clustering (SVC\footnote{\url{https://scikit-learn.org/stable/modules/svm.html}}). Our identified boundaries are just tentative visualizations for different regimes rather than a rigorous calculation of the margin due to the small size of the sample (less than 70 HMSCs). The roughly positive relation between $L/M$ and Mach number for our HMSCs between the range 2 kpc to 4~kpc has a mechanism different from that in star-forming clumps because of the absence of the embedded protostars. The external heating and turbulent energy injection by \hii~regions are the most likely reasons for the positive relations and these two impacts of \hii~regions on HMSCs are probably interconnected with each other.\\

\subsection{Virial status} \label{virial_cal}
We find that AS are more turbulent than NA. In this section, we analyze whether HMSCs are gravitationally bounded and search for the possible differences between AS and NA by estimating virial mass ratio, 
\begin{equation}
R_{\rm virial} = \frac{5}{{\alpha}{\beta}} \frac{r{\sigma}^2}{GM_{\rm clump}},
\end{equation}
where $\alpha$ and $\beta$ are factors related with the eccentricity and density profile, respectively \citep{li13}; $\alpha = (1+a/3)/(1+2a/5)$, where $a$ is for the power-law index of number density profile $n_{\rm H_{2}} \propto r^{a}$;  $\beta = arcsin e/e$ is the geometry factor determined by eccentricity $e$; and $a$ is approximately equal to \pnhtcd~$-~1$. As shown in Sect.~\ref{Com}, AS HMSCs have a lower value of \pnhtcd~than NA HMSCs, especially for the S-type AS HMSCs. The nature and limited resolution of PPMAP data could make us underestimate the real profile difference. To show the differences in the power-law index when calculating the virial mass ratio, we set a unique \pnhtcd~value of $-$0.6 for NA, which is the value for an isothermal clump in equilibrium \citep{li13}, whereas we set another value of $-$0.8 for AS HMSCs. The difference of 0.2 is estimated by the median values in Fig.~\ref{figure:4}.

\begin{figure*}
    \centering 
   \includegraphics[width=0.8\textwidth]{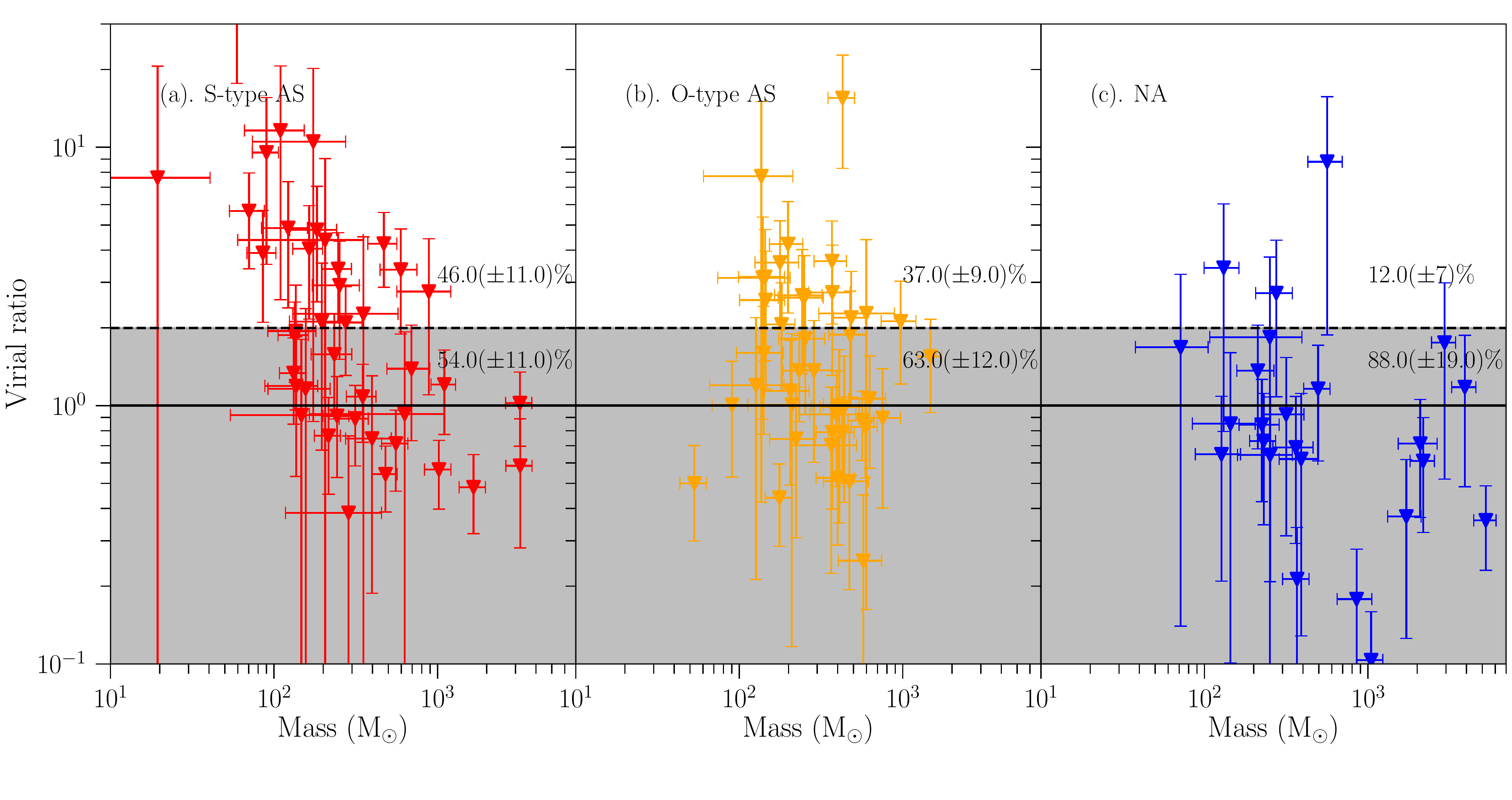}   
       \caption{Plots of virial ratios. The red, yellow, and blue triangles are similar to Fig.~\ref{turbulence_mach}. The black dashed lines and gray regions show the region of virial ratio <~2. The solid lines represent virial ratio = 1. The numbers report the fractions of clumps with a virial ratio $\geq$~2 and <~2. The uncertainties of clump mass and line width are propagated to the calculations of virial ratio.}
        \label{virial_ratios}
    \end{figure*}

\begin{figure}
    \centering 
  \includegraphics[width=0.4\textwidth]{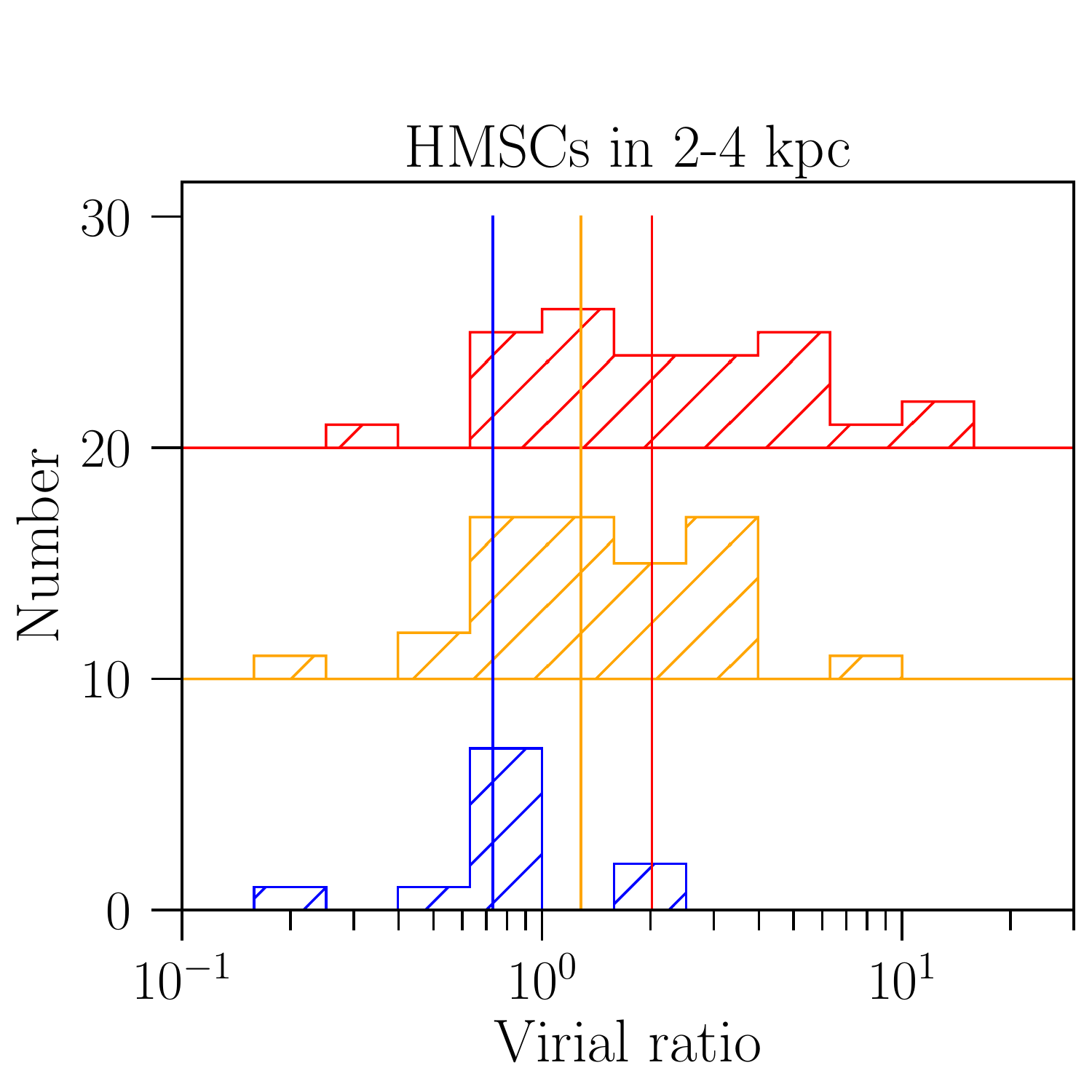}
       \caption{Number distribution of virial ratios in 2~kc to 4~kpc. The red, orange, and blue histograms represent S-type AS, O-type AS, and NA, respectively. The solid lines show the median values of corresponding distributions.}
        \label{virial_ratios24kpc}
    \end{figure}

The virial ratio and clump mass for all \ceo\  studied HMSCs are shown in Fig.~\ref{virial_ratios}. The trend that virial ratios decrease with clump mass is well observed by other similar  studies \citep{cse17b, wie18}. The fraction of HMSCs with a virial ratio of less than 2 decreases from about 90\% for NA to 50\% for S-type AS, which may imply a different virial status between AS and NA. A virial ratio of less than 2 suggests a gravitationally bound structure for the clump \citep{ber92}. We note that the surface pressure and magnetic field are not included in our virial calculation. For HMSCs with a virial ratio $\gg$ 1, these clumps could be unbound and transitory but still possible to be in bound status if we consider surface pressure or magnetic field. AS HMSCs are believed to be deeply impacted by nearby \hii~regions. The external pressure from \hii~regions could assist to confine the clumps. The $\delta$ values indicate that the difference between S-type AS and NA for all sample is at 0.9$\sigma$ level. An Anderson-Darling test shows that the null hypothesis that AS and NA have the same distributions can be rejected ($p$ values equal to 0.0018 for S-type AS and NA, and 0.015 for O-type AS and NA).  To avoid the bias of distance, we show the virial ratio distribution for HMSCs in the 2 to 4~kpc range in Fig. \ref{virial_ratios24kpc}, the median value for S-type AS, O-type AS, and NA are 2.0, 1.3 and 0.7, respectively. The $\delta$ value between S-type AS and NA is about 1.5, indicating a more significant difference of virial ratios compared to full sample.

The median virial ratios of AS and NA are in the transition regime between virialized to subvirial, even if the virial ratios for AS are statistically higher than NA. Recent simulations of \citet{ros19} on the collapse of subvirial ($R_{\rm virial} = 0.1$) and virialized ($R_{\rm virial} = 1$) turbulent massive cores (0.1 pc) suggest that the most massive stars likely form from subvirial massive cores by the rapid monolithic collapse without fragmentation, while other less massive stars likely form through slow collapse of highly turbulent cores with a larger number of fragments. We note that AS HMSCs are not isolated and are probably being compressed by \hii~regions. The pressure from \hii~regions is ignored in the calculations. Furthermore, the virial ratios indicate a similar virial status when move to the massive end (>~300~\msun) rather than totally different virial status like in the simulations of \citet{ros19}. Therefore, we propose that even if a different virial status probably exists between AS and NA, it is hard to find out whether it could dominate the star-forming process such as fragmentation (see Sect. \ref{dis_msf}) in HMSCs. \\

%
\section{Discussion} \label{discussion}
The HMSCs are cold and dense clumps and probably at the earliest stage of HMSF. Even in such short-living objects, we observed a systematic difference between different HMSCs, depending on their environment (associated or not with an \hii~region). This sample of HMSCs - \hii~regions combinations could be a useful database for future case studies about the impact of \hii~regions on HMSF, especially for very early stages. \\

\subsection{\dustt~and L/M ratio} \label{temperature and L/M}
We observed a higher \dustt~for AS than that of NA (about 4 to 6~K). The \dustt~difference could be larger when considering the details of environment such as whether the HMSCs are in shells created by the expansion of \hii~regions or whether there are any implications of compression of \hii~region on HMSCs. The decreasing trend of \dustt~from S-type AS to NA could be well explained with a weaker external heating of \hii~regions. Ultraviolet radiation could photodissociate the molecular gas at the surface of HMSCs, which has been shown by the higher detection rate of the typical photodissociation tracer \cth~for AS HMSCs (Zhang et al., in prep.). Furthermore, AS with \cth~detection have a higher \dustt~than AS without \cth~detection (Zhang et al., in prep.). These facts reveal the tight relation between photodissociation and external heating.
   
For HMSCs, the inner heating effect of potentially embedded low mass protostars could be negligible compared with the feedback from nearby \hii~regions. This leads to a more complex explanation for $L/M$ ratio of HMSCs. The AS HMSCs have a $L/M$ higher than NA about 3 to 4~\lsun/\msun. Furthermore, AS could have a $L/M$ ratio of > 10~\lsun/\msun~in extreme cases. 
  
We point out that $L/M$ ratio alone is not a reliable evolutionary probe for early massive clumps without knowledge of the environment of the clump. This is particularly true for the very early clumps (IR quiescent). 

\citet{mol16} used CH$_{3}$C$_{2}$H to estimate the $L/M$ intervals for different evolutionary stages of massive clumps (see Sect.~\ref{DPP}). We carried out an independent check to see whether the $L/M$ ratios of HMSCs could be significantly changed under the effect of \hii~regions. We took advantage of $L/M$ ratio and CH$_{3}$C$_{2}$H kinematic temperature data of dense massive clump (> 1000~\msun) in \citet{gia17}. Our check shows that the heating effect of \hii~region makes the $L/M$ ratios of IR quiescent massive clumps increase from $\lesssim$~0.5~\lsun/\msun~to $\gtrsim$~2~\lsun/\msun, even to $\gtrsim$~10~\lsun/\msun~in extreme cases (see details in Appendix \ref{appendix-check}). The improved $L/M$ ratios of IR quiescent massive clumps overlap the $L/M$ regimes of clumps with weak IR sources, such as 70~\micron~compact sources. CH$_{3}$C$_{2}$H is a good thermometer for dense gas ($\gtrsim$~10$^5$~cm$^{3}$), thus it traces mostly the inner region of the clump rather than the envelope. Under the effect of the \hii~region, the increase of the inner CH$_{3}$C$_{2}$H temperature is limited compared to the $L/M$ for IR quiescent clumps. In extreme cases, for example, the clump located at the edge of an \hii~region, the inner gas temperature of clump could be significantly increased to >~30~K owing to the impact of the \hii~region. 

We explore whether it is possible that the $L/M$ properties of star-forming clumps could be significantly changed by the presence of a nearby \hii~region. \citet{19pit} studied the $L/M$ ratios of Carina Nebula and RCW 64 by fitting SED pixel by pixel to Herschel and APEX far-IR to submillimeter data. They find that the $L/M$ ratio of the pixels in the clump show a complex feature. A number of clumps with ongoing star formation activities show some pixels with a higher $L/M$ or \dustt~at lower \nhtcd, which is inconsistent with the expected positive correlation between $L/M$ and \nhtcd~in star formation theory. We propose that this could be due to the feedback of the Carina and RCW 64 regions that heats the surface and surrounding of the star-forming clump, making some pixels at the surface to have a higher $L/M$ ratio or \dustt~with lower \nhtcd, for example, clump BYF 77 and BYF 109 in Fig. 21 of \citet{19pit}. The $L/M$ properties of starless clumps are more easily significantly changed due to the external heating and the absence of internal heating sources. 

In a word, the suggestions in \citet{mol16} that $L/M$ ratio could be used as an evolutionary probe of massive clumps at very early stages should be carefully treated, especially when there are impacts from \hii~regions. A detailed check of the environment of clumps is crucial before using $L/M$ ratios because the $L/M$ could be partly created by external heating of nearby \hii~regions rather than forming stars inside the clump. The single dish observation cannot resolve the compact cores embedded in HMSCs. Observations by ALMA toward a series of HMSCs in different environment are needed to address whether the embedded cores remain higher \dustt~and $L/M$ when they are associated with nearby \hii~regions.\\

\subsection{Nature of HMSCs in different environments}
We have shown that a large part of HMSCs are associated with \hii~regions (about 60\% to 80\%), which means that the \hii~regions are common environments for HMSCs. A simple look at a series of maps indicates that about half of the AS HMSCs probably show the morphology features of interaction with \hii~regions such as the compressed layer. A systematic difference has been found between the properties of HMSCs associated and nonassociated with \hii~regions. A short summary of the main differences we found for the properties of AS compared to NA is given in Table~\ref{summary_difference}. The quantity \dustt~is one of the key differences that has been investigated. A higher \dustt~causes a higher $L/M$ ratio. When \dustt~>~20~K, evaporation of CO isotopologs dramatically increases and changes the chemistry in HMSCs, for example, more CO promotes the destruction of \nthp~(Zhang et al., in prep.). Using chemistry as the evolutionary probe of very early massive clumps, especially using single dish observations with low resolution, creates bias. Recent observations toward Carina nebular reveal similar findings that the environmental changes due to the HMSF have modified the chemical composition of nearby dense clumps \citep{con19}. Besides, AS are expected to have different density structures and kinematics compared to NA. The compression effect of ionized gas makes HMSCs located close to \hii~regions become more centrally peaked. At the same time, the additional energy injection from \hii~regions makes HMSCs more turbulent. 

Even if some properties of HMSCs indicate a Galactocentric gradient with a large scatter, it is still possible that some properties of HMSCs are predominated by the local environment rather than the Galactic scale environments, for example, the \dustt~changes among different types of HMSCs are more significant than the changes among different Galactocentric distances. Some implications from other studies also indicate that the very early HMSF is more likely impacted by the local environment rather than the Galactic-scale environment. \citet{djo19} used the ratio of Lyman luminosity to clump mass $L_{\rm Ly}/M_{\rm clump}$ and ratio of bolometric luminosity to clump mass $L_{\rm bol}/M_{\rm clump}$ as the proxies for massive star formation efficiency (massive SFE) and the overall SFE for UC\hii~regions. They find a much significantly decreasing trend among the Galactocentric distance for $L_{\rm bol}/M_{\rm clump}$ than that of $L_{\rm Ly}/M_{\rm clump}$, suggesting that HMSF is relatively independent compared to low or overall star formation.

   \begin{table}[ht]
  \tiny
   \centering 
    \caption{Summary of the properties of AS compared to NA.} 
    \label{summary_difference} 
    \begin{tabular}{c c } 
    \hline\hline 
             Properties   & AS compared to NA     \\ 
    \hline 
      \dustt~and $L/M$   &   higher ($\delta \approx$ 1 to 2)    \\
      Density profile    &   steeper ($\delta \approx$ 1)  \\
      Turbulence (Mach number)  &   higher ($\delta \approx$ 1)                 \\
      Virial ratios             &   higher ($\delta \approx$ 1)            \\
     \hline                                         
      \end{tabular}
      \end{table}

The S-type AS HMSCs are excellent samples for studying the triggered HMSF at very early stages because they are in a compressed structure or being strongly photoionized/photodissociated. We want to understand how the \hii~region could impact the following HMSF in HMSCs according to the differences we have found.\\

\subsection{Massive star formation} \label{dis_msf}
Fragmentation is a key process that could determine the initial mass of cores (or fragments) embedded in HMSCs \citep{lin19, svo19, san19, lou19, li19}. According to the fragmentation theory of molecular clump in the simplest case \citep{jea02}, which is a self-gravitating medium without turbulent motions that is nonmagnetic, isothermal, infinite, and homogeneous, the clump fragments into smaller cores with mass on the order of the Jeans mass,  
\begin{eqnarray}
\label{jeans_simple}
M_{\rm J}~\propto~c_{\rm s}^{3}{n_{\rm H_{2}}}^{-1/2}~\propto~T^{3/2}{n_{\rm H_{2}}}^{-1/2},
\end{eqnarray}
thus a higher temperature or/and lower density generally lead to a massive fragment. The results of Mach number show that HMSCs are predominated by turbulence from supersonic to hypersonic. With the assumptions that turbulence could be treated as additional pressure and that the thermal component ${\sigma}_{\rm Therm}$ only occupies a limited proportion compared to the nonthermal component ${\sigma}_{\rm NT}$ \citep{mac04}, Eq.~\ref{jeans_simple} can be transformed to
\begin{eqnarray}
\label{jeans_turbulent}
M_{\rm J}~\propto {{\sigma}_{\rm NT}}^2.
\end{eqnarray}
The Jeans mass is higher for a more turbulent clump. Thus, the fragmented cores in AS HMSCs are probably more massive than those of NA HMSCs owing to more turbulent properties. The limited fragmentation at 0.01~pc scale compared to the thermal fragmentation (Eq.~\ref{jeans_simple}) has been found in the most massive clump surrounding the \hii~region RCW 120 \citep{fig18}. The fragmented cores are significantly more massive than the expected values with Eq.~\ref{jeans_simple}. It is well explained with the additional turbulence from \hii~region by \citet{fig18}. 

Our previous results have shown that AS HMSCs have a steeper density profile owing to the compression of \hii~region. The question is whether a steeper density profile promotes the formation of a massive central object \citep{mcg16}. The simulations of \citet{gir11} reveal that the cores/clumps with an initial density profile that is strongly centrally peaked are more likely to form one or a few massive stars in the center surrounded by a number of low mass stars, while cores/clumps with a flatter density profile are more likely to form a large number of low mass stars. Furthermore, these simulations suggest that the initial density profile of star-forming clumps/cores could be the most important factor for fragmentation and evolution of HMSF regions. \citet{pal14} observed a number of massive dense cores to study the relation between fragmentation and density structure of massive cores and they find that steeper density profiles tend to show lower fragmentation. These results probably indicate that AS more easily form massive stars by limiting fragmentation. 

\begin{figure}
    \centering 
   \includegraphics[width=0.45\textwidth]{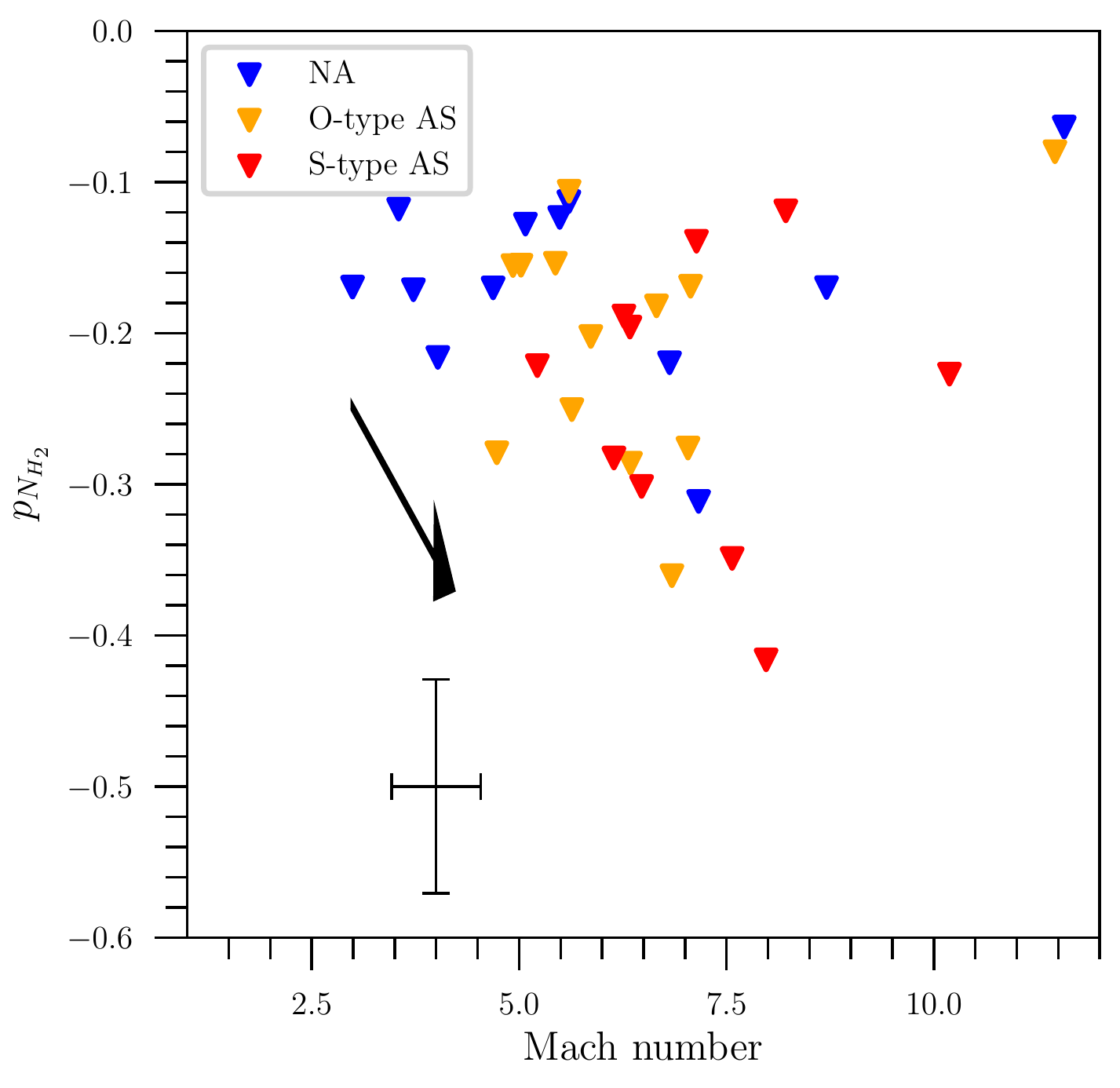}

       \caption{Power-law index of column density profile \pnhtcd~and Mach number. Only HMSCs with good density profile fitting are included; see details in Sect.~\ref{Com_cal}. The blue, orange, and red triangles are similar to Fig.~\ref{turbulence_mach}. The arrow indicates the trend.}
        \label{pvalue_mach}
    \end{figure}

The properties of more turbulent and steeper profiles of AS HMSCs are both implications of the impacts of \hii~regions and should interconnect with each other well. Such a positive correlation is shown in Fig.~\ref{pvalue_mach}. Therefore, the combined effects of more turbulent and more centrally peaked properties of AS probably increase the mass of the fragmented cores and promote HMSF.

As a pilot test to study the possible impacts of \hii~regions on fragmentation in HMSCs, we select two HMSCs with similar distance (3~kpc), single-dish derived clump mass ($\simeq$~300~\msun), and \nhtcd~($6\times10^{22}$~cm$^{-2}$), but one is S-type AS at the shell structure and another is NA far away from \hii~regions (Zhang et al., in prep.). The S-type AS HMSC has a higher $L/M$ ratio of 2.12~$\pm$~0.8~\lsun/\msun~and Mach number 8.7 $\pm$~0.5, while the NA HMSC has a lower $L/M$ ratio 0.27~$\pm$~0.08~\lsun/\msun~and Mach number 5.6~$\pm$~2.3. The ALMA observations (ALMA Cycle 4, Project code: 2016.1.01346.S, PI: Thushara Pillai) with a resolution of 1.5\arcsec~(0.02~pc) show that even though the numbers of fragmented cores are similar between AS and NA (between two to four), the fragmented cores are much denser in S-type AS compared to NA (Zhang et al., in prep.). This pair of AS and NA probably suggests that the impacts of the \hii~region could favor the formation of massive fragments in clumps at the earliest evolutionary stages.   

Recent interferometric studies of a sample of eight massive clumps (0.5~pc scale) close to the \hii~regions have revealed a possible link between the impacts of \hii~regions and asymmetric spatial features that high density regions (fragmented cores) drift to off-center position of clump, indicating that the thick envelope of the clump has been deeply compressed (see Fig. 8 in the paper of \citealt{zha19}). This probe of impacts of \hii~regions on massive clumps has a drawback that the projection effect probably minimizes the drifted features. For example, if the impacts of \hii~regions are along the line of sight, the impacted clump still shows a symmetric feature. The density profile could be more reliable than the ways suggested by \citet{zha19}. We try to search the relation between density profiles of these eight massive clumps as well as the mass or number of fragmented cores. No clear trend between the density profile power-law index \pnhtcd~and the mass or number of fragmented cores at 0.02~pc scale has been found, which could be due to the limited number of the sample and/or the broad difference in the properties of these massive clumps, for example, starless or star forming.

Magnetic field is another key factor in the fragmentation and the collapse of clumps \citep{dal19}. Studies suggest that a stronger magnetic field could inhibit the fragmentation \citep{fon16, beu18}. The relative importance of magnetic field compared to turbulence could be estimated by Alfv{\'e}n Mach number $M_{\rm A} = {\sigma}_{\rm NT}/{\rm v_{\rm A}}$, where ${\rm v_A} = B/\sqrt{4\pi \rho}$ is Alfv{\'e}n speed and $B$ is the total magnetic field. Observations toward massive starless cores/clumps indicate a $B$ value on the order of 0.1 to 1~mG \citep{tan13,tam19}. Even if we can expect that the clumps close to \hii~region have a higher $B$ value, whether it should impact the fragmentation requires dedicated observations.

Following our findings, we propose that the properties of AS HMSCs inherited from the impact of \hii~regions, probably favor the formation of high-mass stars. High-resolution interferometric observations are really needed to ascertain this. \\ 
 
\section{Conclusions} \label{conclusions}
We investigate the statistical differences between HMSCs close to \hii~regions (AS or AS HMSCs) and those far away from \hii~regions (NA or NA HMSCs), based on the following two selection criteria used to ascertain the association: the separations between HMSCs and \hii~regions in the plane of the sky are $< 2R_{\rm eff}$ of the \hii~region and the velocity differences are $< 7$~\kms. The results show that more than half of the HMSCs are AS HMSCs. Among these AS HMSCs, 30\% to 50\% of clumps are probably in a compressed structure or photodissociated/photoionized structure, indicating the powerful feedback of \hii~regions on HMSCs. This sample of HMSCs constitutes a useful database for future studies dedicated to the impact of \hii~regions on the earliest HMSF. 

We found the following systematic differences between AS and NA:
\begin{enumerate}
        \item The AS HMSCs are warmer and more luminous than NA types because of the external heating of \hii~regions. The increment in $L/M$ of AS HMSCs, which could be up to 10~\lsun/\msun, can cause a large uncertainty when $L/M$ ratio is used as the evolutionary probe of massive clumps at very early stages without  knowledge of the environment.
        \item A steeper column density profile, which means a more centrally peaked density structure, is found in AS compared to NA. The compression by \hii~region could be the possible reason for this difference.
        \item \ceotwo~lines reveal that AS have a higher Mach number than NA, indicating that AS are more turbulent.  We propose that this is due to the additional energy injection from \hii~regions. The AS HMSCs probably have a greater virial ratio compared to NA HMSCs if we do not consider the surface pressure and magnetic field.
\end{enumerate}

The implications of impacts of \hii~regions on HMSCs, such as positive correlations between the $L/M$ ratio,
the Mach number, and the power-law index of column density profile, are intercorrelated. We propose that the combined effects of more turbulent and steeper density profiles probably make AS HMSCs favor the formation of high-mass stars by limiting fragmentation. Future interferometric surveys toward a statistically significant number of HMSCs in different environments are needed to ascertain this.

\begin{acknowledgements}
      AZ thanks the support of the Institut Universitaire de France. We want to thank the anonymous referee for the constructive comments that helped to improve the quality of the paper. The SEDIGISM data is acquired with the Atacama Pathfinder EXperiment (APEX) under programmes 092.F-9315(A) and 193.C0584(A). APEX is a collaboration between the Max-Planck-Institut für Radioastronomie, the European Southern Observatory, and the Onsala Space Observatory. \textit{Herschel} Hi-GAL data processing, map production and source catalogue generation is the result of a multi-year effort thanks to Contracts I/038/080/0 and I/029/12/0 from ASI (Agenzia Spaziale Italiana). 
\end{acknowledgements}

\bibliographystyle{aa} 
\bibliography{hmsc}

\begin{thebibliography}{134}
\expandafter\ifx\csname natexlab\endcsname\relax\def\natexlab#1{#1}\fi

\bibitem[{{Aguirre} {et~al.}(2011){Aguirre}, {Ginsburg}, {Dunham}, {Drosback},
  {Bally}, {Battersby}, {Bradley}, {Cyganowski}, {Dowell}, {Evans}, {Glenn},
  {Harvey}, {Rosolowsky}, {Stringfellow}, {Walawender}, \& {Williams}}]{agu11}
{Aguirre}, J.~E., {Ginsburg}, A.~G., {Dunham}, M.~K., {et~al.} 2011, \apjs,
  192, 4

\bibitem[{{Anderson} {et~al.}(2015){Anderson}, {Armentrout}, {Johnstone},
  {Bania}, {Balser}, {Wenger}, \& {Cunningham}}]{and15}
{Anderson}, L.~D., {Armentrout}, W.~P., {Johnstone}, B.~M., {et~al.} 2015,
  \apjs, 221, 26

\bibitem[{{Anderson} {et~al.}(2018){Anderson}, {Armentrout}, {Luisi}, {Bania},
  {Balser}, \& {Wenger}}]{and18}
{Anderson}, L.~D., {Armentrout}, W.~P., {Luisi}, M., {et~al.} 2018, \apjs, 234,
  33

\bibitem[{{Anderson} {et~al.}(2014){Anderson}, {Bania}, {Balser}, {Cunningham},
  {Wenger}, {Johnstone}, \& {Armentrout}}]{and14}
{Anderson}, L.~D., {Bania}, T.~M., {Balser}, D.~S., {et~al.} 2014, \apjs, 212,
  1

\bibitem[{{Anderson} {et~al.}(2011){Anderson}, {Bania}, {Balser}, \&
  {Rood}}]{and11}
{Anderson}, L.~D., {Bania}, T.~M., {Balser}, D.~S., \& {Rood}, R.~T. 2011,
  \apjs, 194, 32

\bibitem[{{Andr{\'e}} {et~al.}(2014){Andr{\'e}}, {Di Francesco},
  {Ward-Thompson}, {Inutsuka}, {Pudritz}, \& {Pineda}}]{andr14}
{Andr{\'e}}, P., {Di Francesco}, J., {Ward-Thompson}, D., {et~al.} 2014,
  Protostars and Planets VI, 27

\bibitem[{{Baldeschi} {et~al.}(2017){Baldeschi}, {Elia}, {Molinari}, {Pezzuto},
  {Schisano}, {Gatti}, {Serra}, {Merello}, {Benedettini}, {Di Giorgio}, \&
  {Liu}}]{bal17}
{Baldeschi}, A., {Elia}, D., {Molinari}, S., {et~al.} 2017, \mnras, 466, 3682

\bibitem[{{Bania} {et~al.}(2010){Bania}, {Anderson}, {Balser}, \&
  {Rood}}]{ban10}
{Bania}, T.~M., {Anderson}, L.~D., {Balser}, D.~S., \& {Rood}, R.~T. 2010,
  \apjl, 718, L106

\bibitem[{{Barnes} {et~al.}(2015){Barnes}, {Muller}, {Indermuehle},
  {O'Dougherty}, {Lowe}, {Cunningham}, {Hernandez}, \& {Fuller}}]{bar15}
{Barnes}, P.~J., {Muller}, E., {Indermuehle}, B., {et~al.} 2015, \apj, 812, 6

\bibitem[{{Bertoldi}(1989)}]{ber89}
{Bertoldi}, F. 1989, \apj, 346, 735

\bibitem[{{Bertoldi} \& {McKee}(1992)}]{ber92}
{Bertoldi}, F. \& {McKee}, C.~F. 1992, \apj, 395, 140

\bibitem[{{Beuther} {et~al.}(2018){Beuther}, {Soler}, {Vlemmings}, {Linz},
  {Henning}, {Kuiper}, {Rao}, {Smith}, {Sakai}, {Johnston}, {Walsh}, \&
  {Feng}}]{beu18}
{Beuther}, H., {Soler}, J.~D., {Vlemmings}, W., {et~al.} 2018, \aap, 614, A64

\bibitem[{{Bock} {et~al.}(1999){Bock}, {Large}, \& {Sadler}}]{boc99}
{Bock}, D.~C.-J., {Large}, M.~I., \& {Sadler}, E.~M. 1999, \aj, 117, 1578

\bibitem[{{Bonnell} {et~al.}(2007){Bonnell}, {Larson}, \& {Zinnecker}}]{bon07}
{Bonnell}, I.~A., {Larson}, R.~B., \& {Zinnecker}, H. 2007, Protostars and
  Planets V, 149

\bibitem[{{Bronfman} {et~al.}(2000){Bronfman}, {Casassus}, {May}, \&
  {Nyman}}]{bro00}
{Bronfman}, L., {Casassus}, S., {May}, J., \& {Nyman}, L.~{\r{A}}. 2000, \aap,
  358, 521

\bibitem[{{Busquet} {et~al.}(2016){Busquet}, {Estalella}, {Palau}, {Liu},
  {Zhang}, {Girart}, {de Gregorio-Monsalvo}, {Pillai}, {Anglada}, \&
  {Ho}}]{bus16}
{Busquet}, G., {Estalella}, R., {Palau}, A., {et~al.} 2016, \apj, 819, 139

\bibitem[{{Butler} \& {Tan}(2012)}]{but12}
{Butler}, M.~J. \& {Tan}, J.~C. 2012, \apj, 754, 5

\bibitem[{{Carey} {et~al.}(2009){Carey}, {Noriega-Crespo}, {Mizuno}, {Shenoy},
  {Paladini}, {Kraemer}, {Price}, {Flagey}, {Ryan}, {Ingalls}, {Kuchar},
  {Pinheiro Gon{\c{c}}alves}, {Indebetouw}, {Billot}, {Marleau}, {Padgett},
  {Rebull}, {Bressert}, {Ali}, {Molinari}, {Martin}, {Berriman}, {Boulanger},
  {Latter}, {Miville-Deschenes}, {Shipman}, \& {Testi}}]{car09}
{Carey}, S.~J., {Noriega-Crespo}, A., {Mizuno}, D.~R., {et~al.} 2009, \pasp,
  121, 76

\bibitem[{{Churchwell} {et~al.}(2009){Churchwell}, {Babler}, {Meade},
  {Whitney}, {Benjamin}, {Indebetouw}, {Cyganowski}, {Robitaille}, {Povich},
  {Watson}, \& {Bracker}}]{chu09}
{Churchwell}, E., {Babler}, B.~L., {Meade}, M.~R., {et~al.} 2009, \pasp, 121,
  213

\bibitem[{{Churchwell} {et~al.}(2006){Churchwell}, {Povich}, {Allen}, {Taylor},
  {Meade}, {Babler}, {Indebetouw}, {Watson}, {Whitney}, {Wolfire}, {Bania},
  {Benjamin}, {Clemens}, {Cohen}, {Cyganowski}, {Jackson}, {Kobulnicky},
  {Mathis}, {Mercer}, {Stolovy}, {Uzpen}, {Watson}, \& {Wolff}}]{chu06}
{Churchwell}, E., {Povich}, M.~S., {Allen}, D., {et~al.} 2006, \apj, 649, 759

\bibitem[{{Clark} {et~al.}(2007){Clark}, {Klessen}, \& {Bonnell}}]{cla07}
{Clark}, P.~C., {Klessen}, R.~S., \& {Bonnell}, I.~A. 2007, \mnras, 379, 57

\bibitem[{{Contreras} {et~al.}(2019){Contreras}, {Rebolledo}, {Breen}, {Green},
  \& {Burton}}]{con19}
{Contreras}, Y., {Rebolledo}, D., {Breen}, S.~L., {Green}, A.~J., \& {Burton},
  M.~G. 2019, \mnras, 483, 1437

\bibitem[{{Csengeri} {et~al.}(2018){Csengeri}, {Bontemps}, {Wyrowski},
  {Belloche}, {Menten}, {Leurini}, {Beuther}, {Bronfman}, {Commer{\c c}on},
  {Chapillon}, {Longmore}, {Palau}, {Tan}, \& {Urquhart}}]{cse18}
{Csengeri}, T., {Bontemps}, S., {Wyrowski}, F., {et~al.} 2018, \aap, 617, A89

\bibitem[{{Csengeri} {et~al.}(2017{\natexlab{a}}){Csengeri}, {Bontemps},
  {Wyrowski}, {Megeath}, {Motte}, {Sanna}, {Wienen}, \& {Menten}}]{cse17b}
{Csengeri}, T., {Bontemps}, S., {Wyrowski}, F., {et~al.} 2017{\natexlab{a}},
  \aap, 601, A60

\bibitem[{{Csengeri} {et~al.}(2017{\natexlab{b}}){Csengeri}, {Bontemps},
  {Wyrowski}, {Motte}, {Menten}, {Beuther}, {Bronfman}, {Commer{\c c}on},
  {Chapillon}, {Duarte-Cabral}, {Fuller}, {Henning}, {Leurini}, {Longmore},
  {Palau}, {Peretto}, {Schuller}, {Tan}, {Testi}, {Traficante}, \&
  {Urquhart}}]{cse17}
{Csengeri}, T., {Bontemps}, S., {Wyrowski}, F., {et~al.} 2017{\natexlab{b}},
  \aap, 600, L10

\bibitem[{{Csengeri} {et~al.}(2014){Csengeri}, {Urquhart}, {Schuller}, {Motte},
  {Bontemps}, {Wyrowski}, {Menten}, {Bronfman}, {Beuther}, {Henning}, {Testi},
  {Zavagno}, \& {Walmsley}}]{cse14}
{Csengeri}, T., {Urquhart}, J.~S., {Schuller}, F., {et~al.} 2014, \aap, 565,
  A75

\bibitem[{{Cyganowski} {et~al.}(2017){Cyganowski}, {Brogan}, {Hunter}, {Smith},
  {Kruijssen}, {Bonnell}, \& {Zhang}}]{cyg17}
{Cyganowski}, C.~J., {Brogan}, C.~L., {Hunter}, T.~R., {et~al.} 2017, \mnras,
  468, 3694

\bibitem[{{Dall'Olio} {et~al.}(2019){Dall'Olio}, {Vlemmings}, {Persson},
  {Alves}, {Beuther}, {Girart}, {Surcis}, {Torrelles}, \& {Van
  Langevelde}}]{dal19}
{Dall'Olio}, D., {Vlemmings}, W.~H.~T., {Persson}, M.~V., {et~al.} 2019, \aap,
  626, A36

\bibitem[{{Deharveng} {et~al.}(2010){Deharveng}, {Schuller}, {Anderson},
  {Zavagno}, {Wyrowski}, {Menten}, {Bronfman}, {Testi}, {Walmsley}, \&
  {Wienen}}]{deh10}
{Deharveng}, L., {Schuller}, F., {Anderson}, L.~D., {et~al.} 2010, \aap, 523,
  A6

\bibitem[{{Djordjevic} {et~al.}(2019){Djordjevic}, {Thompson}, {Urquhart}, \&
  {Forbrich}}]{djo19}
{Djordjevic}, J.~O., {Thompson}, M.~A., {Urquhart}, J.~S., \& {Forbrich}, J.
  2019, \mnras, 487, 1057

\bibitem[{{Duronea} {et~al.}(2017){Duronea}, {Cappa}, {Bronfman}, {Borissova},
  {Gromadzki}, \& {Kuhn}}]{dur17}
{Duronea}, N.~U., {Cappa}, C.~E., {Bronfman}, L., {et~al.} 2017, \aap, 606, A8

\bibitem[{{Elia} {et~al.}(2017){Elia}, {Molinari}, {Schisano}, {Pestalozzi},
  {Pezzuto}, {Merello}, {Noriega-Crespo}, {Moore}, {Russeil}, {Mottram},
  {Paladini}, {Strafella}, {Benedettini}, {Bernard}, {Di Giorgio}, {Eden},
  {Fukui}, {Plume}, {Bally}, {Martin}, {Ragan}, {Jaffa}, {Motte}, {Olmi},
  {Schneider}, {Testi}, {Wyrowski}, {Zavagno}, {Calzoletti}, {Faustini},
  {Natoli}, {Palmeirim}, {Piacentini}, {Piazzo}, {Pilbratt}, {Polychroni},
  {Baldeschi}, {Beltr{\'a}n}, {Billot}, {Cambr{\'e}sy}, {Cesaroni},
  {Garc{\'\i}a-Lario}, {Hoare}, {Huang}, {Joncas}, {Liu}, {Maiolo}, {Marsh},
  {Maruccia}, {M{\`e}ge}, {Peretto}, {Rygl}, {Schilke}, {Thompson},
  {Traficante}, {Umana}, {Veneziani}, {Ward-Thompson}, {Whitworth}, {Arab},
  {Band ieramonte}, {Becciani}, {Brescia}, {Buemi}, {Bufano}, {Butora},
  {Cavuoti}, {Costa}, {Fiorellino}, {Hajnal}, {Hayakawa}, {Kacsuk}, {Leto}, {Li
  Causi}, {Marchili}, {Martinavarro-Armengol}, {Mercurio}, {Molinaro},
  {Riccio}, {Sano}, {Sciacca}, {Tachihara}, {Torii}, {Trigilio}, {Vitello}, \&
  {Yamamoto}}]{eli17}
{Elia}, D., {Molinari}, S., {Schisano}, E., {et~al.} 2017, \mnras, 471, 100

\bibitem[{{Ellsworth-Bowers} {et~al.}(2015){Ellsworth-Bowers}, {Rosolowsky},
  {Glenn}, {Ginsburg}, {Evans}~II, {Battersby}, {Shirley}, \&
  {Svoboda}}]{ell15}
{Ellsworth-Bowers}, T.~P., {Rosolowsky}, E., {Glenn}, J., {et~al.} 2015, \apj,
  799, 29

\bibitem[{{Elmegreen}(2011)}]{elm11}
{Elmegreen}, B.~G. 2011, in EAS Publications Series, Vol.~51, EAS Publications
  Series, ed. C.~{Charbonnel} \& T.~{Montmerle}, 45--58

\bibitem[{{Figueira} {et~al.}(2018){Figueira}, {Bronfman}, {Zavagno}, {Louvet},
  {Lo}, {Finger}, \& {Rod{\'o}n}}]{fig18}
{Figueira}, M., {Bronfman}, L., {Zavagno}, A., {et~al.} 2018, \aap, 616, L10

\bibitem[{{Fleming} {et~al.}(2010){Fleming}, {France}, {Lupu}, \&
  {McCandliss}}]{fle10}
{Fleming}, B., {France}, K., {Lupu}, R.~E., \& {McCandliss}, S.~R. 2010, \apj,
  725, 159

\bibitem[{{Fontani} {et~al.}(2018){Fontani}, {Commer{\c c}on}, {Giannetti},
  {Beltr{\'a}n}, {S{\'a}nchez-Monge}, {Testi}, {Brand}, \& {Tan}}]{fon18}
{Fontani}, F., {Commer{\c c}on}, B., {Giannetti}, A., {et~al.} 2018, \aap, 615,
  A94

\bibitem[{{Fontani} {et~al.}(2016){Fontani}, {Commer{\c{c}}on}, {Giannetti},
  {Beltr{\'a}n}, {S{\'a}nchez-Monge}, {Testi}, {Brand}, {Caselli}, {Cesaroni},
  {Dodson}, {Longmore}, {Rioja}, {Tan}, \& {Walmsley}}]{fon16}
{Fontani}, F., {Commer{\c{c}}on}, B., {Giannetti}, A., {et~al.} 2016, \aap,
  593, L14

\bibitem[{{Giannetti} {et~al.}(2013){Giannetti}, {Brand}, {S{\'a}nchez-Monge},
  {Fontani}, {Cesaroni}, {Beltr{\'a}n}, {Molinari}, {Dodson}, \&
  {Rioja}}]{gia13}
{Giannetti}, A., {Brand}, J., {S{\'a}nchez-Monge}, {\'A}., {et~al.} 2013, \aap,
  556, A16

\bibitem[{{Giannetti} {et~al.}(2017){Giannetti}, {Leurini}, {Wyrowski},
  {Urquhart}, {Csengeri}, {Menten}, {K{\"o}nig}, \& {G{\"u}sten}}]{gia17}
{Giannetti}, A., {Leurini}, S., {Wyrowski}, F., {et~al.} 2017, \aap, 603, A33

\bibitem[{{Ginsburg} {et~al.}(2016){Ginsburg}, {Henkel}, {Ao}, {Riquelme},
  {Kauffmann}, {Pillai}, {Mills}, {Requena-Torres}, {Immer}, {Testi}, {Ott},
  {Bally}, {Battersby}, {Darling}, {Aalto}, {Stanke}, {Kendrew}, {Kruijssen},
  {Longmore}, {Dale}, {Guesten}, \& {Menten}}]{gin16}
{Ginsburg}, A., {Henkel}, C., {Ao}, Y., {et~al.} 2016, \aap, 586, A50

\bibitem[{{Girichidis} {et~al.}(2011){Girichidis}, {Federrath}, {Banerjee}, \&
  {Klessen}}]{gir11}
{Girichidis}, P., {Federrath}, C., {Banerjee}, R., \& {Klessen}, R.~S. 2011,
  \mnras, 413, 2741

\bibitem[{{Helfand} {et~al.}(2006){Helfand}, {Becker}, {White}, {Fallon}, \&
  {Tuttle}}]{hel06}
{Helfand}, D.~J., {Becker}, R.~H., {White}, R.~L., {Fallon}, A., \& {Tuttle},
  S. 2006, \aj, 131, 2525

\bibitem[{{Heyer} \& {Brunt}(2004)}]{hey04}
{Heyer}, M.~H. \& {Brunt}, C.~M. 2004, \apjl, 615, L45

\bibitem[{{Hou} \& {Gao}(2014)}]{hou14}
{Hou}, L.~G. \& {Gao}, X.~Y. 2014, \mnras, 438, 426

\bibitem[{{Humphreys} \& {Larsen}(1995)}]{hum95}
{Humphreys}, R.~M. \& {Larsen}, J.~A. 1995, \aj, 110, 2183

\bibitem[{{Imai} {et~al.}(2017){Imai}, {Sugitani}, {Miao}, {Fukuda},
  {Watanabe}, {Kusune}, \& {Pickles}}]{ima17}
{Imai}, R., {Sugitani}, K., {Miao}, J., {et~al.} 2017, \apj, 845, 99

\bibitem[{{Ingallinera} {et~al.}(2014){Ingallinera}, {Trigilio}, {Umana},
  {Leto}, {Noriega-Crespo}, {Flagey}, {Paladini}, {Agliozzo}, \&
  {Buemi}}]{ing14}
{Ingallinera}, A., {Trigilio}, C., {Umana}, G., {et~al.} 2014, \mnras, 437,
  3626

\bibitem[{{Jackson} {et~al.}(2004){Jackson}, {Simon}, {Shah}, {Rathborne},
  {Heyer}, {Clemens}, \& {Bania}}]{jac04}
{Jackson}, J.~M., {Simon}, R., {Shah}, R., {et~al.} 2004, in Astronomical
  Society of the Pacific Conference Series, Vol. 317, Milky Way Surveys: The
  Structure and Evolution of our Galaxy, ed. D.~{Clemens}, R.~{Shah}, \&
  T.~{Brainerd}, 49

\bibitem[{{Jeans}(1902)}]{jea02}
{Jeans}, J.~H. 1902, Philosophical Transactions of the Royal Society of London
  Series A, 199, 1

\bibitem[{{Juvela} {et~al.}(2018){Juvela}, {Malinen}, {Montillaud}, {Pelkonen},
  {Ristorcelli}, \& {T{\'o}th}}]{juv18}
{Juvela}, M., {Malinen}, J., {Montillaud}, J., {et~al.} 2018, \aap, 614, A83

\bibitem[{{Kauffmann} \& {Pillai}(2010)}]{kau10}
{Kauffmann}, J. \& {Pillai}, T. 2010, \apjl, 723, L7

\bibitem[{{Kendrew} {et~al.}(2016){Kendrew}, {Beuther}, {Simpson}, {Csengeri},
  {Wienen}, {Lintott}, {Povich}, {Beaumont}, \& {Schuller}}]{ken16}
{Kendrew}, S., {Beuther}, H., {Simpson}, R., {et~al.} 2016, \apj, 825, 142

\bibitem[{{Kendrew} {et~al.}(2012){Kendrew}, {Simpson}, {Bressert}, {Povich},
  {Sherman}, {Lintott}, {Robitaille}, {Schawinski}, \& {Wolf-Chase}}]{ken12}
{Kendrew}, S., {Simpson}, R., {Bressert}, E., {et~al.} 2012, \apj, 755, 71

\bibitem[{{Klessen}(2001)}]{kle01}
{Klessen}, R.~S. 2001, \apj, 556, 837

\bibitem[{{Kruijssen} {et~al.}(2015){Kruijssen}, {Dale}, \& {Longmore}}]{kru15}
{Kruijssen}, J.~M.~D., {Dale}, J.~E., \& {Longmore}, S.~N. 2015, \mnras, 447,
  1059

\bibitem[{{Kruijssen} {et~al.}(2014){Kruijssen}, {Longmore}, {Elmegreen},
  {Murray}, {Bally}, {Testi}, \& {Kennicutt}}]{kru14}
{Kruijssen}, J.~M.~D., {Longmore}, S.~N., {Elmegreen}, B.~G., {et~al.} 2014,
  \mnras, 440, 3370

\bibitem[{{Krumholz} {et~al.}(2009){Krumholz}, {Klein}, {McKee}, {Offner}, \&
  {Cunningham}}]{kru09}
{Krumholz}, M.~R., {Klein}, R.~I., {McKee}, C.~F., {Offner}, S.~S.~R., \&
  {Cunningham}, A.~J. 2009, Science, 323, 754

\bibitem[{{Krumholz} \& {McKee}(2008)}]{kru08}
{Krumholz}, M.~R. \& {McKee}, C.~F. 2008, \nat, 451, 1082

\bibitem[{{Larson}(1981)}]{lar81}
{Larson}, R.~B. 1981, \mnras, 194, 809

\bibitem[{{Li} {et~al.}(2013){Li}, {Kauffmann}, {Zhang}, \& {Chen}}]{li13}
{Li}, D., {Kauffmann}, J., {Zhang}, Q., \& {Chen}, W. 2013, \apjl, 768, L5

\bibitem[{{Li}(2018)}]{ligx18}
{Li}, G.-X. 2018, \mnras, 477, 4951

\bibitem[{{Li} {et~al.}(2018){Li}, {Li}, {Yuan}, {Huang}, \& {Ren}}]{li18}
{Li}, H., {Li}, J.-Z., {Yuan}, J.-H., {Huang}, Y.-F., \& {Ren}, Z.-Y. 2018,
  Research in Astronomy and Astrophysics, 18, 122

\bibitem[{{Li} {et~al.}(2019){Li}, {Zhang}, {Pillai}, {Stephens}, {Wang}, \&
  {Li}}]{li19}
{Li}, S., {Zhang}, Q., {Pillai}, T., {et~al.} 2019, \apj, 886, 130

\bibitem[{{Lin} {et~al.}(2019){Lin}, {Csengeri}, {Wyrowski}, {Urquhart},
  {Schuller}, {Weiss}, \& {Menten}}]{lin19}
{Lin}, Y., {Csengeri}, T., {Wyrowski}, F., {et~al.} 2019, \aap, 631, A72

\bibitem[{{Liu} {et~al.}(2018){Liu}, {Tan}, {Cheng}, \& {Kong}}]{liu18}
{Liu}, M., {Tan}, J.~C., {Cheng}, Y., \& {Kong}, S. 2018, \apj, 862, 105

\bibitem[{{Lo} {et~al.}(2015){Lo}, {Wiles}, {Redman}, {Cunningham}, {Bains},
  {Jones}, {Burton}, \& {Bronfman}}]{lo15}
{Lo}, N., {Wiles}, B., {Redman}, M.~P., {et~al.} 2015, \mnras, 453, 3245

\bibitem[{{Louvet} {et~al.}(2019){Louvet}, {Neupane}, {Garay}, {Russeil},
  {Zavagno}, {Guzman}, {Gomez}, {Bronfman}, \& {Nony}}]{lou19}
{Louvet}, F., {Neupane}, S., {Garay}, G., {et~al.} 2019, \aap, 622, A99

\bibitem[{{Lu} {et~al.}(2019){Lu}, {Zhang}, {Kauffmann}, {Pillai}, {Ginsburg},
  {Mills}, {Kruijssen}, {Longmore}, {Battersby}, {Liu}, \& {Gu}}]{lu19}
{Lu}, X., {Zhang}, Q., {Kauffmann}, J., {et~al.} 2019, \apj, 872, 171

\bibitem[{{Lumsden} {et~al.}(2013){Lumsden}, {Hoare}, {Urquhart}, {Oudmaijer},
  {Davies}, {Mottram}, {Cooper}, \& {Moore}}]{lum13}
{Lumsden}, S.~L., {Hoare}, M.~G., {Urquhart}, J.~S., {et~al.} 2013, \apjs, 208,
  11

\bibitem[{{Mac Low} \& {Klessen}(2004)}]{mac04}
{Mac Low}, M.-M. \& {Klessen}, R.~S. 2004, Reviews of Modern Physics, 76, 125

\bibitem[{{Makai} {et~al.}(2017){Makai}, {Anderson}, {Mascoop}, \&
  {Johnstone}}]{mak17}
{Makai}, Z., {Anderson}, L.~D., {Mascoop}, J.~L., \& {Johnstone}, B. 2017,
  \apj, 846, 64

\bibitem[{{Marsh} \& {Whitworth}(2019)}]{marsh19}
{Marsh}, K.~A. \& {Whitworth}, A.~P. 2019, \mnras, 483, 352

\bibitem[{{Marsh} {et~al.}(2015){Marsh}, {Whitworth}, \& {Lomax}}]{mar15}
{Marsh}, K.~A., {Whitworth}, A.~P., \& {Lomax}, O. 2015, \mnras, 454, 4282

\bibitem[{{Marsh} {et~al.}(2017){Marsh}, {Whitworth}, {Lomax}, {Ragan},
  {Becciani}, {Cambr{\'e}sy}, {Di Giorgio}, {Eden}, {Elia}, {Kacsuk},
  {Molinari}, {Palmeirim}, {Pezzuto}, {Schneider}, {Sciacca}, \&
  {Vitello}}]{mar17}
{Marsh}, K.~A., {Whitworth}, A.~P., {Lomax}, O., {et~al.} 2017, \mnras, 471,
  2730

\bibitem[{{Marshall} \& {Kerton}(2019)}]{marshall19}
{Marshall}, B. \& {Kerton}, C.~R. 2019, \mnras, 489, 4809

\bibitem[{{Mattern} {et~al.}(2018){Mattern}, {Kauffmann}, {Csengeri},
  {Urquhart}, {Leurini}, {Wyrowski}, {Giannetti}, {Barnes}, {Beuther},
  {Bronfman}, {Duarte-Cabral}, {Henning}, {Kainulainen}, {Menten}, {Schisano},
  \& {Schuller}}]{mat18}
{Mattern}, M., {Kauffmann}, J., {Csengeri}, T., {et~al.} 2018, \aap, 619, A166

\bibitem[{{McGuire} {et~al.}(2016){McGuire}, {Fuller}, {Peretto}, {Zhang},
  {Traficante}, {Avison}, \& {Jimenez-Serra}}]{mcg16}
{McGuire}, C., {Fuller}, G.~A., {Peretto}, N., {et~al.} 2016, \aap, 594, A118

\bibitem[{{McKee} \& {Tan}(2003)}]{mck03}
{McKee}, C.~F. \& {Tan}, J.~C. 2003, \apj, 585, 850

\bibitem[{{Merello} {et~al.}(2019){Merello}, {Molinari}, {Rygl}, {Evans},
  {Elia}, {Schisano}, {Traficante}, {Shirley}, {Svoboda}, \&
  {Goldsmith}}]{mer19}
{Merello}, M., {Molinari}, S., {Rygl}, K.~L.~J., {et~al.} 2019, \mnras, 483,
  5355

\bibitem[{{Milam} {et~al.}(2005){Milam}, {Savage}, {Brewster}, {Ziurys}, \&
  {Wyckoff}}]{mil05}
{Milam}, S.~N., {Savage}, C., {Brewster}, M.~A., {Ziurys}, L.~M., \& {Wyckoff},
  S. 2005, \apj, 634, 1126

\bibitem[{{Misiriotis} {et~al.}(2006){Misiriotis}, {Xilouris},
  {Papamastorakis}, {Boumis}, \& {Goudis}}]{mis06}
{Misiriotis}, A., {Xilouris}, E.~M., {Papamastorakis}, J., {Boumis}, P., \&
  {Goudis}, C.~D. 2006, \aap, 459, 113

\bibitem[{{Molinari} {et~al.}(2019){Molinari}, {Baldeschi}, {Robitaille},
  {Morales}, {Schisano}, {Traficante}, {Merello}, {Molinaro}, {Vitello},
  {Sciacca}, \& {Liu}}]{mol19}
{Molinari}, S., {Baldeschi}, A., {Robitaille}, T.~P., {et~al.} 2019, \mnras,
  486, 4508

\bibitem[{{Molinari} {et~al.}(2016){Molinari}, {Merello}, {Elia}, {Cesaroni},
  {Testi}, \& {Robitaille}}]{mol16}
{Molinari}, S., {Merello}, M., {Elia}, D., {et~al.} 2016, \apjl, 826, L8

\bibitem[{{Molinari} {et~al.}(2010){Molinari}, {Swinyard}, {Bally}, {Barlow},
  {Bernard}, {Martin}, {Moore}, {Noriega-Crespo}, {Plume}, {Testi}, {Zavagno},
  {Abergel}, {Ali}, {Andr{\'e}}, {Baluteau}, {Benedettini}, {Bern{\'e}},
  {Billot}, {Blommaert}, {Bontemps}, {Boulanger}, {Brand}, {Brunt}, {Burton},
  {Campeggio}, {Carey}, {Caselli}, {Cesaroni}, {Cernicharo}, {Chakrabarti},
  {Chrysostomou}, {Codella}, {Cohen}, {Compiegne}, {Davis}, {de Bernardis}, {de
  Gasperis}, {Di Francesco}, {di Giorgio}, {Elia}, {Faustini}, {Fischera},
  {Fukui}, {Fuller}, {Ganga}, {Garcia-Lario}, {Giard}, {Giardino}, {Glenn},
  {Goldsmith}, {Griffin}, {Hoare}, {Huang}, {Jiang}, {Joblin}, {Joncas},
  {Juvela}, {Kirk}, {Lagache}, {Li}, {Lim}, {Lord}, {Lucas}, {Maiolo},
  {Marengo}, {Marshall}, {Masi}, {Massi}, {Matsuura}, {Meny}, {Minier},
  {Miville-Desch{\^e}nes}, {Montier}, {Motte}, {M{\"u}ller}, {Natoli}, {Neves},
  {Olmi}, {Paladini}, {Paradis}, {Pestalozzi}, {Pezzuto}, {Piacentini},
  {Pomar{\`e}s}, {Popescu}, {Reach}, {Richer}, {Ristorcelli}, {Roy}, {Royer},
  {Russeil}, {Saraceno}, {Sauvage}, {Schilke}, {Schneider-Bontemps},
  {Schuller}, {Schultz}, {Shepherd}, {Sibthorpe}, {Smith}, {Smith},
  {Spinoglio}, {Stamatellos}, {Strafella}, {Stringfellow}, {Sturm}, {Taylor},
  {Thompson}, {Tuffs}, {Umana}, {Valenziano}, {Vavrek}, {Viti}, {Waelkens},
  {Ward-Thompson}, {White}, {Wyrowski}, {Yorke}, \& {Zhang}}]{mol10}
{Molinari}, S., {Swinyard}, B., {Bally}, J., {et~al.} 2010, \pasp, 122, 314

\bibitem[{{Motogi} {et~al.}(2019){Motogi}, {Hirota}, {Machida}, {Yonekura},
  {Honma}, {Takakuwa}, \& {Matsushita}}]{mot19}
{Motogi}, K., {Hirota}, T., {Machida}, M.~N., {et~al.} 2019, \apjl, 877, L25

\bibitem[{{Motte} {et~al.}(2018){Motte}, {Bontemps}, \& {Louvet}}]{mot18}
{Motte}, F., {Bontemps}, S., \& {Louvet}, F. 2018, \araa, 56, 41

\bibitem[{{Myers}(2009)}]{mye09}
{Myers}, P.~C. 2009, \apj, 700, 1609

\bibitem[{{Olmi} {et~al.}(2018){Olmi}, {Elia}, {Schisano}, \&
  {Molinari}}]{olm18}
{Olmi}, L., {Elia}, D., {Schisano}, E., \& {Molinari}, S. 2018, \mnras, 480,
  1831

\bibitem[{{Orkisz} {et~al.}(2017){Orkisz}, {Pety}, {Gerin}, {Bron},
  {Guzm{\'a}n}, {Bardeau}, {Goicoechea}, {Gratier}, {Le Petit}, {Levrier},
  {Liszt}, {{\"O}berg}, {Peretto}, {Roueff}, {Sievers}, \& {Tremblin}}]{ork17}
{Orkisz}, J.~H., {Pety}, J., {Gerin}, M., {et~al.} 2017, \aap, 599, A99

\bibitem[{{Padoan} {et~al.}(2019){Padoan}, {Pan}, {Juvela}, {Haugb{\o}lle}, \&
  {Nordlund}}]{pad19}
{Padoan}, P., {Pan}, L., {Juvela}, M., {Haugb{\o}lle}, T., \& {Nordlund},
  {\r{A}}. 2019, arXiv e-prints, arXiv:1911.04465

\bibitem[{{Palau} {et~al.}(2014){Palau}, {Estalella}, {Girart}, {Fuente},
  {Fontani}, {Commer{\c{c}}on}, {Busquet}, {Bontemps}, {S{\'a}nchez-Monge},
  {Zapata}, {Zhang}, {Hennebelle}, \& {di Francesco}}]{pal14}
{Palau}, A., {Estalella}, R., {Girart}, J.~M., {et~al.} 2014, \apj, 785, 42

\bibitem[{{Palmeirim} {et~al.}(2017){Palmeirim}, {Zavagno}, {Elia}, {Moore},
  {Whitworth}, {Tremblin}, {Traficante}, {Merello}, {Russeil}, {Pezzuto},
  {Cambr{\'e}sy}, {Baldeschi}, {Bandieramonte}, {Becciani}, {Benedettini},
  {Buemi}, {Bufano}, {Bulpitt}, {Butora}, {Carey}, {Costa}, {Deharveng}, {Di
  Giorgio}, {Eden}, {Hajnal}, {Hoare}, {Kacsuk}, {Leto}, {Marsh}, {M{\`e}ge},
  {Molinari}, {Molinaro}, {Noriega-Crespo}, {Schisano}, {Sciacca}, {Trigilio},
  {Umana}, \& {Vitello}}]{pal17}
{Palmeirim}, P., {Zavagno}, A., {Elia}, D., {et~al.} 2017, \aap, 605, A35

\bibitem[{{Paradis} {et~al.}(2012){Paradis}, {Paladini}, {Noriega-Crespo},
  {M{\'e}ny}, {Pia centini}, {Thompson}, {Marshall}, {Veneziani}, {Bernard}, \&
  {Molinari}}]{par12}
{Paradis}, D., {Paladini}, R., {Noriega-Crespo}, A., {et~al.} 2012, \aap, 537,
  A113

\bibitem[{{Paron} {et~al.}(2018){Paron}, {Areal}, \& {Ortega}}]{par18}
{Paron}, S., {Areal}, M.~B., \& {Ortega}, M.~E. 2018, \aap, 617, A14

\bibitem[{{Pillai} {et~al.}(2019){Pillai}, {Kauffmann}, {Zhang}, {Sanhueza},
  {Leurini}, {Wang}, {Sridharan}, \& {K{\"o}nig}}]{pil19}
{Pillai}, T., {Kauffmann}, J., {Zhang}, Q., {et~al.} 2019, \aap, 622, A54

\bibitem[{{Pitts} {et~al.}(2019){Pitts}, {Barnes}, \& {Varosi}}]{19pit}
{Pitts}, R.~L., {Barnes}, P.~J., \& {Varosi}, F. 2019, \mnras, 484, 305

\bibitem[{{Rathborne} {et~al.}(2006){Rathborne}, {Jackson}, \& {Simon}}]{rat06}
{Rathborne}, J.~M., {Jackson}, J.~M., \& {Simon}, R. 2006, \apj, 641, 389

\bibitem[{{Reid} {et~al.}(2016){Reid}, {Dame}, {Menten}, \&
  {Brunthaler}}]{rei16}
{Reid}, M.~J., {Dame}, T.~M., {Menten}, K.~M., \& {Brunthaler}, A. 2016, \apj,
  823, 77

\bibitem[{{Rigby} {et~al.}(2016){Rigby}, {Moore}, {Plume}, {Eden}, {Urquhart},
  {Thompson}, {Mottram}, {Brunt}, {Butner}, {Dempsey}, {Gibson}, {Hatchell},
  {Jenness}, {Kuno}, {Longmore}, {Morgan}, {Polychroni}, {Thomas}, {White}, \&
  {Zhu}}]{rig16}
{Rigby}, A.~J., {Moore}, T.~J.~T., {Plume}, R., {et~al.} 2016, \mnras, 456,
  2885

\bibitem[{{Rosen} {et~al.}(2019){Rosen}, {Li}, {Zhang}, \& {Burkhart}}]{ros19}
{Rosen}, A.~L., {Li}, P.~S., {Zhang}, Q., \& {Burkhart}, B. 2019, \apj, 887,
  108

\bibitem[{{Samal} {et~al.}(2014){Samal}, {Zavagno}, {Deharveng}, {Molinari},
  {Ojha}, {Paradis}, {Tig{\'e}}, {Pandey}, \& {Russeil}}]{sam14}
{Samal}, M.~R., {Zavagno}, A., {Deharveng}, L., {et~al.} 2014, \aap, 566, A122

\bibitem[{{Sanhueza} {et~al.}(2019){Sanhueza}, {Contreras}, {Wu}, {Jackson},
  {Guzm{\'a}n}, {Zhang}, {Li}, {Lu}, {Silva}, {Izumi}, {Liu}, {Miura},
  {Tatematsu}, {Sakai}, {Beuther}, {Garay}, {Ohashi}, {Saito}, {Nakamura},
  {Saigo}, {Veena}, {Nguyen-Luong}, \& {Tafoya}}]{san19}
{Sanhueza}, P., {Contreras}, Y., {Wu}, B., {et~al.} 2019, \apj, 886, 102

\bibitem[{{Sanna} {et~al.}(2019){Sanna}, {K{\"o}lligan}, {Moscadelli},
  {Kuiper}, {Cesaroni}, {Pillai}, {Menten}, {Zhang}, {Caratti o Garatti},
  {Goddi}, {Leurini}, \& {Carrasco-Gonz{\'a}lez}}]{san18}
{Sanna}, A., {K{\"o}lligan}, A., {Moscadelli}, L., {et~al.} 2019, \aap, 623,
  A77

\bibitem[{{Schilke}(2015)}]{sch15}
{Schilke}, P. 2015, in EAS Publications Series, Vol. 75-76, EAS Publications
  Series, 227--235

\bibitem[{{Schuller} {et~al.}(2017){Schuller}, {Csengeri}, {Urquhart},
  {Duarte-Cabral}, {Barnes}, {Giannetti}, {Hernandez}, {Leurini}, {Mattern},
  {Medina}, {Agurto}, {Azagra}, {Anderson}, {Beltr{\'a}n}, {Beuther},
  {Bontemps}, {Bronfman}, {Dobbs}, {Dumke}, {Finger}, {Ginsburg}, {Gonzalez},
  {Henning}, {Kauffmann}, {Mac-Auliffe}, {Menten}, {Montenegro-Montes},
  {Moore}, {Muller}, {Parra}, {Perez-Beaupuits}, {Pettitt}, {Russeil},
  {S{\'a}nchez-Monge}, {Schilke}, {Schisano}, {Suri}, {Testi}, {Torstensson},
  {Venegas}, {Wang}, {Wienen}, {Wyrowski}, \& {Zavagno}}]{sch17}
{Schuller}, F., {Csengeri}, T., {Urquhart}, J.~S., {et~al.} 2017, \aap, 601,
  A124

\bibitem[{{Schuller} {et~al.}(2009){Schuller}, {Menten}, {Contreras},
  {Wyrowski}, {Schilke}, {Bronfman}, {Henning}, {Walmsley}, {Beuther},
  {Bontemps}, {Cesaroni}, {Deharveng}, {Garay}, {Herpin}, {Lefloch}, {Linz},
  {Mardones}, {Minier}, {Molinari}, {Motte}, {Nyman}, {Reveret}, {Risacher},
  {Russeil}, {Schneider}, {Testi}, {Troost}, {Vasyunina}, {Wienen}, {Zavagno},
  {Kovacs}, {Kreysa}, {Siringo}, \& {Wei{\ss}}}]{sch09}
{Schuller}, F., {Menten}, K.~M., {Contreras}, Y., {et~al.} 2009, \aap, 504, 415

\bibitem[{{Simpson} {et~al.}(2012){Simpson}, {Povich}, {Kendrew}, {Lintott},
  {Bressert}, {Arvidsson}, {Cyganowski}, {Maddison}, {Schawinski}, {Sherman},
  {Smith}, \& {Wolf-Chase}}]{sim12}
{Simpson}, R.~J., {Povich}, M.~S., {Kendrew}, S., {et~al.} 2012, \mnras, 424,
  2442

\bibitem[{{Stil} {et~al.}(2006){Stil}, {Taylor}, {Dickey}, {Kavars}, {Martin},
  {Rothwell}, {Boothroyd}, {Lockman}, \& {McClure-Griffiths}}]{sti06}
{Stil}, J.~M., {Taylor}, A.~R., {Dickey}, J.~M., {et~al.} 2006, \aj, 132, 1158

\bibitem[{{Svoboda} {et~al.}(2016){Svoboda}, {Shirley}, {Battersby},
  {Rosolowsky}, {Ginsburg}, {Ellsworth-Bowers}, {Pestalozzi}, {Dunham},
  {Evans}, {Bally}, \& {Glenn}}]{svo16}
{Svoboda}, B.~E., {Shirley}, Y.~L., {Battersby}, C., {et~al.} 2016, \apj, 822,
  59

\bibitem[{{Svoboda} {et~al.}(2019){Svoboda}, {Shirley}, {Traficante},
  {Battersby}, {Fuller}, {Zhang}, {Beuther}, {Peretto}, {Brogan}, \&
  {Hunter}}]{svo19}
{Svoboda}, B.~E., {Shirley}, Y.~L., {Traficante}, A., {et~al.} 2019, \apj, 886,
  36

\bibitem[{{Tamaoki} {et~al.}(2019){Tamaoki}, {Sugitani}, {Nguyen-Luong},
  {Nakamura}, {Kusune}, {Nagayama}, {Watanabe}, {Nishiyama}, \&
  {Tamura}}]{tam19}
{Tamaoki}, S., {Sugitani}, K., {Nguyen-Luong}, Q., {et~al.} 2019, \apjl, 875,
  L16

\bibitem[{{Tan} {et~al.}(2013){Tan}, {Kong}, {Butler}, {Caselli}, \&
  {Fontani}}]{tan13}
{Tan}, J.~C., {Kong}, S., {Butler}, M.~J., {Caselli}, P., \& {Fontani}, F.
  2013, \apj, 779, 96

\bibitem[{{Tang} {et~al.}(2018){Tang}, {Liu}, {Qin}, {Kim}, {Wu}, {Tatematsu},
  {Yuan}, {Wang}, {Parsons}, \& {Koch}}]{tan18}
{Tang}, M., {Liu}, T., {Qin}, S.-L., {et~al.} 2018, \apj, 856, 141

\bibitem[{{Taylor} \& {Cordes}(1993)}]{tay93}
{Taylor}, J.~H. \& {Cordes}, J.~M. 1993, \apj, 411, 674

\bibitem[{{Thompson} {et~al.}(2012){Thompson}, {Urquhart}, {Moore}, \&
  {Morgan}}]{tho12}
{Thompson}, M.~A., {Urquhart}, J.~S., {Moore}, T.~J.~T., \& {Morgan}, L.~K.
  2012, \mnras, 421, 408

\bibitem[{{Traficante} {et~al.}(2017){Traficante}, {Fuller}, {Billot},
  {Duarte-Cabral}, {Merello}, {Molinari}, {Peretto}, \& {Schisano}}]{tra17}
{Traficante}, A., {Fuller}, G.~A., {Billot}, N., {et~al.} 2017, \mnras, 470,
  3882

\bibitem[{{Traficante} {et~al.}(2020){Traficante}, {Fuller}, {Duarte-Cabral},
  {Elia}, {Heyer}, {Molinari}, {Peretto}, \& {Schisano}}]{tra19}
{Traficante}, A., {Fuller}, G.~A., {Duarte-Cabral}, A., {et~al.} 2020, \mnras,
  491, 4310

\bibitem[{{Tremblin} {et~al.}(2014{\natexlab{a}}){Tremblin}, {Anderson},
  {Didelon}, {Raga}, {Minier}, {Ntormousi}, {Pettitt}, {Pinto}, {Samal},
  {Schneider}, \& {Zavagno}}]{tre14}
{Tremblin}, P., {Anderson}, L.~D., {Didelon}, P., {et~al.} 2014{\natexlab{a}},
  \aap, 568, A4

\bibitem[{{Tremblin} {et~al.}(2015){Tremblin}, {Audit}, {Minier}, {Schmidt}, \&
  {Schneider}}]{tre15}
{Tremblin}, P., {Audit}, E., {Minier}, V., {Schmidt}, W., \& {Schneider}, N.
  2015, Highlights of Astronomy, 16, 590

\bibitem[{{Tremblin} {et~al.}(2014{\natexlab{b}}){Tremblin}, {Schneider},
  {Minier}, {Didelon}, {Hill}, {Anderson}, {Motte}, {Zavagno}, {Andr{\'e}}, \&
  {Arzoumanian}}]{tre14c}
{Tremblin}, P., {Schneider}, N., {Minier}, V., {et~al.} 2014{\natexlab{b}},
  \aap, 564, A106

\bibitem[{{Urquhart} {et~al.}(2018){Urquhart}, {K{\"o}nig}, {Giannetti},
  {Leurini}, {Moore}, {Eden}, {Pillai}, {Thompson}, {Braiding}, {Burton},
  {Csengeri}, {Dempsey}, {Figura}, {Froebrich}, {Menten}, {Schuller}, {Smith},
  \& {Wyrowski}}]{urq18}
{Urquhart}, J.~S., {K{\"o}nig}, C., {Giannetti}, A., {et~al.} 2018, \mnras,
  473, 1059

\bibitem[{{V{\'a}zquez-Semadeni} {et~al.}(2019){V{\'a}zquez-Semadeni}, {Palau},
  {Ballesteros-Paredes}, {G{\'o}mez}, \& {Zamora-Avil{\'e}s}}]{vaz19}
{V{\'a}zquez-Semadeni}, E., {Palau}, A., {Ballesteros-Paredes}, J.,
  {G{\'o}mez}, G.~C., \& {Zamora-Avil{\'e}s}, M. 2019, \mnras, 490, 3061

\bibitem[{{Veltchev} {et~al.}(2013){Veltchev}, {Donkov}, \& {Klessen}}]{vel13}
{Veltchev}, T.~V., {Donkov}, S., \& {Klessen}, R.~S. 2013, \mnras, 432, 3495

\bibitem[{{Wenger} {et~al.}(2019){Wenger}, {Dickey}, {Jordan}, {Balser},
  {Armentrout}, {Anderson}, {Bania}, {Dawson}, {McClure-Griffiths}, \&
  {Shea}}]{wen19}
{Wenger}, T.~V., {Dickey}, J.~M., {Jordan}, C.~H., {et~al.} 2019, \apjs, 240,
  24

\bibitem[{{Wienen} {et~al.}(2015){Wienen}, {Wyrowski}, {Menten}, {Urquhart},
  {Csengeri}, {Walmsley}, {Bontemps}, {Russeil}, {Bronfman}, {Koribalski}, \&
  {Schuller}}]{wie15}
{Wienen}, M., {Wyrowski}, F., {Menten}, K.~M., {et~al.} 2015, \aap, 579, A91

\bibitem[{{Wienen} {et~al.}(2018){Wienen}, {Wyrowski}, {Menten}, {Urquhart},
  {Walmsley}, {Csengeri}, {Koribalski}, \& {Schuller}}]{wie18}
{Wienen}, M., {Wyrowski}, F., {Menten}, K.~M., {et~al.} 2018, \aap, 609, A125

\bibitem[{{Wright} {et~al.}(2010){Wright}, {Eisenhardt}, {Mainzer}, {Ressler},
  {Cutri}, {Jarrett}, {Kirkpatrick}, {Padgett}, {McMillan}, {Skrutskie},
  {Stanford}, {Cohen}, {Walker}, {Mather}, {Leisawitz}, {Gautier}, {McLean},
  {Benford}, {Lonsdale}, {Blain}, {Mendez}, {Irace}, {Duval}, {Liu}, {Royer},
  {Heinrichsen}, {Howard}, {Shannon}, {Kendall}, {Walsh}, {Larsen}, {Cardon},
  {Schick}, {Schwalm}, {Abid}, {Fabinsky}, {Naes}, \& {Tsai}}]{wri10}
{Wright}, E.~L., {Eisenhardt}, P. R.~M., {Mainzer}, A.~K., {et~al.} 2010, \aj,
  140, 1868

\bibitem[{{Yan} {et~al.}(2016){Yan}, {Xu}, {Zhang}, {Lu}, {Chen}, \&
  {Tang}}]{yan16}
{Yan}, Q.-z., {Xu}, Y., {Zhang}, B., {et~al.} 2016, \aj, 152, 117

\bibitem[{{Yuan} {et~al.}(2018){Yuan}, {Li}, {Wu}, {Ellingsen}, {Henkel},
  {Wang}, {Liu}, {Liu}, {Zavagno}, {Ren}, \& {Huang}}]{yua18}
{Yuan}, J., {Li}, J.-Z., {Wu}, Y., {et~al.} 2018, \apj, 852, 12

\bibitem[{{Yuan} {et~al.}(2017){Yuan}, {Wu}, {Ellingsen}, {Evans}, {Henkel},
  {Wang}, {Liu}, {Liu}, {Li}, \& {Zavagno}}]{yua17}
{Yuan}, J., {Wu}, Y., {Ellingsen}, S.~P., {et~al.} 2017, \apjs, 231, 11

\bibitem[{{Zavagno} {et~al.}(2010){Zavagno}, {Russeil}, {Motte}, {Anderson},
  {Deharveng}, {Rod{\'o}n}, {Bontemps}, {Abergel}, {Baluteau}, {Sauvage},
  {Andr{\'e}}, {Hill}, \& {White}}]{zav10}
{Zavagno}, A., {Russeil}, D., {Motte}, F., {et~al.} 2010, \aap, 518, L81

\bibitem[{{Zhang} {et~al.}(2019){Zhang}, {Csengeri}, {Wyrowski}, {Li},
  {Pillai}, {Menten}, {Hatchell}, {Thompson}, \& {Pestalozzi}}]{zha19}
{Zhang}, C.-P., {Csengeri}, T., {Wyrowski}, F., {et~al.} 2019, \aap, 627, A85

\bibitem[{{Zhang} {et~al.}(2016){Zhang}, {Wu}, {Li}, {Yuan}, {Liu}, {Dong}, \&
  {Huang}}]{zha16}
{Zhang}, S.-J., {Wu}, Y., {Li}, J.~Z., {et~al.} 2016, \mnras, 458, 4222

\end{thebibliography}

\begin{appendix}
\section{Details about \vlsr~determination and KDA solutions} \label{appendix-KDA}
In this section, we describe the details of \vlsr~determination and KDA solutions mentioned in Sect. \ref{VD}.
In the Hi-GAL distance project, a series of inner Galactic surveys are used. These include SEDIGISM, JCMT \tco/\ceo~$J = 3 - 2$ HARP Inner Milky-Way Plane Survey (CHIMPS, $|b|$~<~0.5\degree~and 28\degree~<~$l$~<~46\degree, with an angular resolution of 15\arcsec; \citealt{rig16}), \tcoone~ FCRAO Galactic Ring Survey (FCRAO-GRS, 18\degree~<~$l$~<~55\degree~with a resolution of 46\arcsec;~\citealt{jac04}), The Three-mm Ultimate Mopra Milky Way Survey (ThrUMMS \co/\tco/\ceo~ $J=1-0$ with a resolution of 72\arcsec; \citealt{bar15}). Hi-GAL distance project identified the \vlsr~components of Hi-GAL compact sources with the observed lines. Generally, we take the \vlsr~components from the spectral lines tracing denser gas. For example, we take the \ceotwo~rather than \tcotwo~in SEDIGISM. For HMSCs that are not included in the Hi-GAL distance project (82 in 463), we keep the \vlsr~identified by Y17 or update it with the results of \citet{urq18}.

For KDA solutions, we follow the work flows in \citet{urq18}, which test several criteria one by one to ascertain distance. For our cases, we firstly check whether the associated z-distance to the Galactic midplane for far distance is larger than 120 pc. If this is the case, the near distance is chosen to be the distance of HMSCs. Secondly, we check if there is any \hi~SA feature. A clear \hi~SA feature indicates that the clump is at near distance. Finally, if there are no \hi~data or the \hi~SA feature is inconclusive, we check our GLIMPSE 24/8/4.5 \micron~RGB images and find out whether these images have a clear extinction feature. The clump has a higher probability to be in near distance if it shows extinction. The extinction criterion is carefully used when HMSC is associated with an \hii~region. The photodissociation effect of \hii~region could cover the extinction features of the clumps. Thus we also check the size of associated \hii~regions; if the size of the \hii~region is too huge  (several dozens of pc) when put the HMSC at far distance, we put HMSC in near distance.\\

\section{Morphology of surroundings of HMSCs} \label{appendix-morphology}
\begin{figure*}
       \centering
  \includegraphics[width=0.97\textwidth]{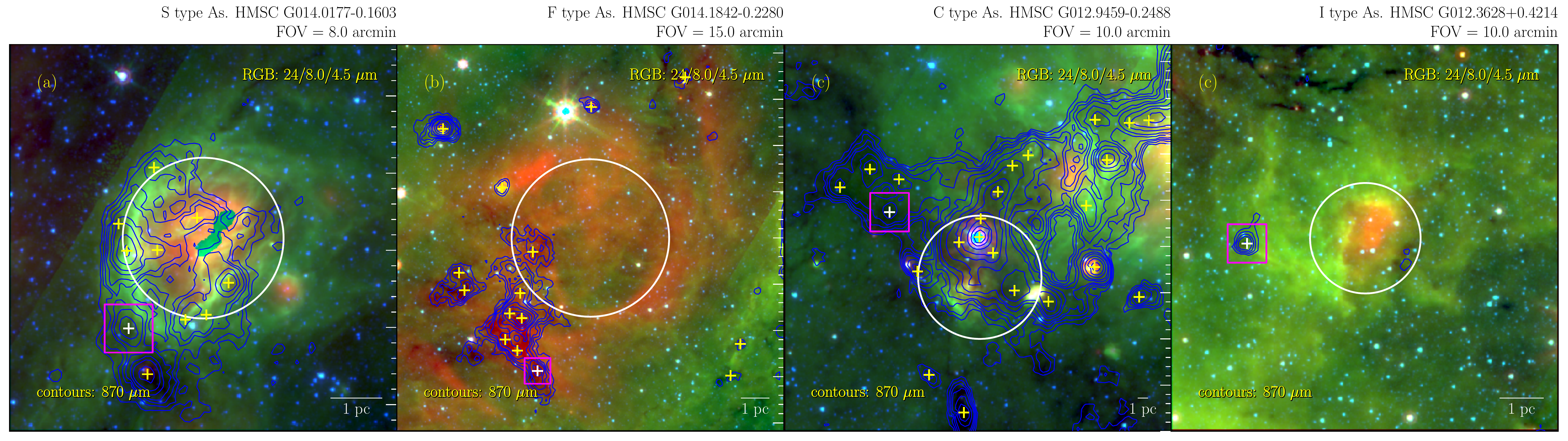}
       \caption{Example of a more detailed classification of the surroundings of HMSC. Four types of morphology are shown for AS. The RGB images are constructed in the same methods as in Fig.~\ref{velocity_determination}. The solid circles indicate the \hii~regions. Yellow crosses and blue contours represent the ATLASGAL clumps identified by \citet{cse14} and the ATLASGAL 870~\micron~emission, respectively. The white crosses with purple squares indicate the HMSCs.}
        \label{type-example}
    \end{figure*}

We introduce the subtypes for the HMSCs without significant interaction with \hii~regions in our series of maps. They belong to O-type AS, O-type PA, or NA HMSCs, which are partly described in Sect. \ref{DMDM}. Basically, there are three subtypes for these HMSCs, which are \textbf{F (filament) type}, \textbf{C (clumpy) type}, and \textbf{I (isolated) type} HMSCs. Observational studies have shown that the massive protostars or protoclusters located at the center (``spoke'') of converging filaments (``hub'') could accrete mass along the filaments \citep{mye09, andr14, yua18}. We classify the HMSC as \textbf{F-type} if the HMSC is in an elongated molecular cloud structure (several pc) and the morphology of the structure is not in accordance with the ionized front of \hii~region. When there are several ATLASGAL clumps close to the HMSC and their combined morphology does not indicate an elongated or filamentary structure, we classify it as \textbf{C-type} HMSC. Finally, if the HMSC is relatively isolated from the surrounding molecular material and there is no obvious connection with other ATLASGAL clumps, we classify it as \textbf{I-type}. An example of the classification for morphology of the surrounding of AS HMSCs is shown in Fig.~\ref{type-example}. A short summary is shown in Table~\ref{environ_table}. The most common morphologies for the surroundings of AS, PA, and NA are S-type (54\%), S-type (36\%), and F-type (57\%), respectively. I-type clumps are less common for AS ($\simeq$~10\%) compared to NA ($\simeq$~30\%), suggesting that HMSCs associated with \hii~regions are usually hosted in a larger molecular cloud complex. Further analyses of this classification and its meaning are ongoing (Zhang et al., in prep.)\\

   \begin{table}[ht]
  \tiny
   \centering 
    \caption{Environmental morphology statistics.} 
    \label{environ_table} 
    \begin{tabular}{ccccc} 
    \hline\hline 
 Types & S-type & F-type & C-type & I-type \\
   \hline 

AS HMSCs (193) & 104 (54\%) & 50 (26\%) & 23 (12\%) & 16 (8\%) \\
PA HMSCs (64)   & 23 (36\%) & 18 (28\%) & 10 (16\%) & 13 (20\%) \\
NA HMSCs (81)   & -         & 46 (57\%) &  7 (9\%)  & 28 (34\%) \\
     \hline                                         
      \end{tabular}
      \end{table}

\section{Detailed descriptions of other properties.}   \label{appendix-properties}
In this section, we describe in more detail the similar properties of the AS and NA HMSCs; these properties consist of \nhtcd, $n_{\rm H_{2}}$, $M_{\rm clump}$, size, and eccentricity.

\begin{figure}
       \centering
  \includegraphics[width=0.45\textwidth]{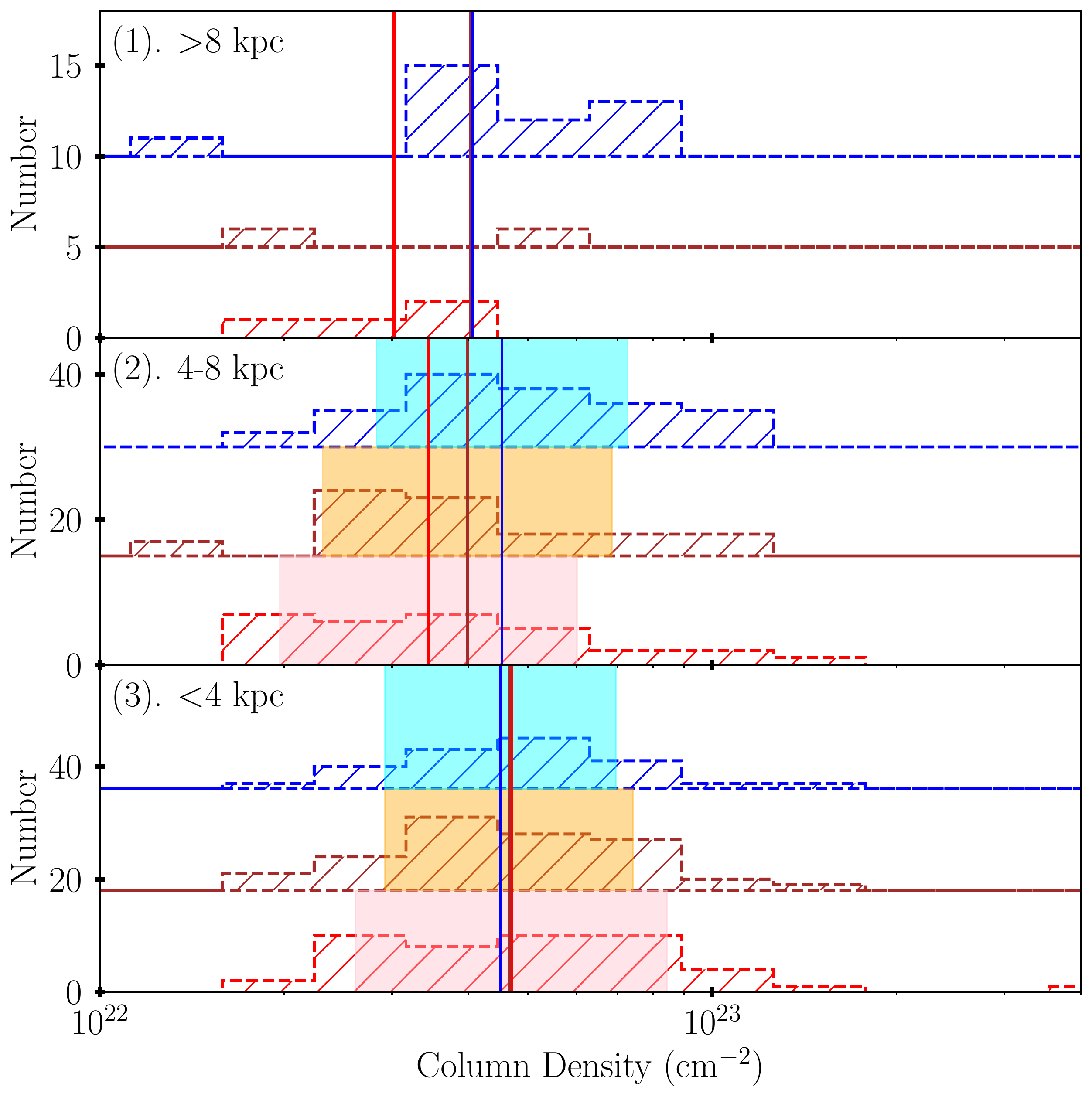}
       \caption{Number distributions of \nhtcd~of HMSCs. The meanings of histograms, lines, and color-filled regions are similar to Fig.~\ref{properties:temperature} but for \nhtcd. }
    \label{properties:Nh2}
  \end{figure}

\paragraph{\textbf{(a). Column density \nhtcd.}} \nhtcd~likely have a flat relation with distance in Fig.~\ref{distance-scatter} and Fig.~\ref{properties:Nh2}. The K-S tests indicate that S-type AS have a distribution similar to NA and O-type AS. The difference between S-type AS and NA at nearby distance are smaller than 0.1$\sigma$ level (see Fig.~\ref{figure:3}). Normally, the uncertainties of \nhtcd~is around 15\%, thus the uncertainties are around $1 \times 10^{22}$~cm$^{-2}$ if we set \nhtcd~as an acceptable value of $5 \times 10^{22}$~cm$^{-2}$. The uncertainties are larger than or around the \nhtcd~difference between different types.
	
The corresponding surface density of the median \nhtcd~derived by Y17 is a magnitude lower than the minimum surface density of 1~g~cm$^{-2}$ for massive star formation derived by \citet{kru08}. The reason is the beam dilution effect \citep{yua17}. Hi-GAL and ATLASGAL combined images with a resolution of about 36\arcsec~cannot resolve the inner cores.

\begin{figure}
       \centering
  \includegraphics[width=0.45\textwidth]{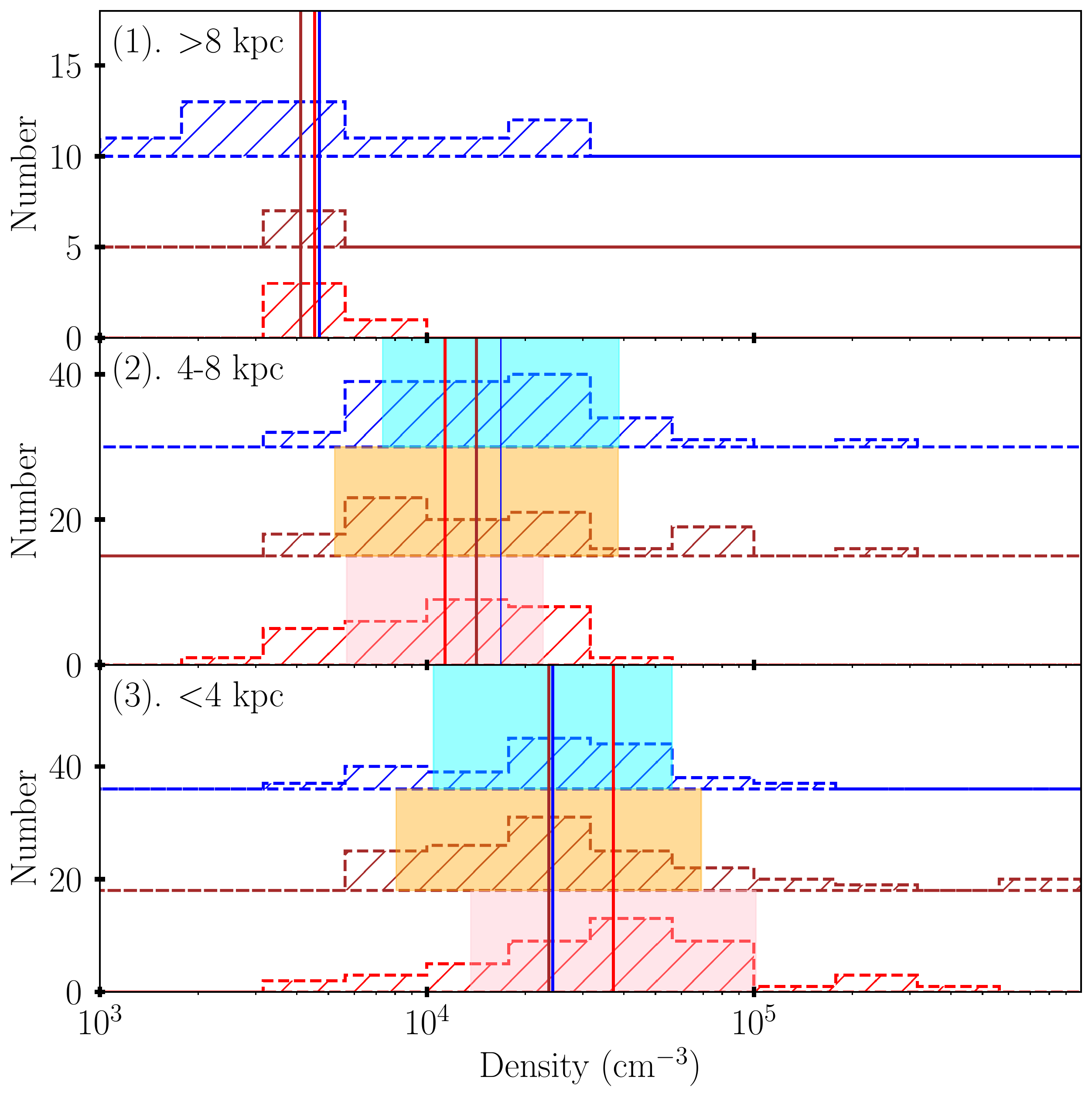}
       \caption{Number distributions of $n_{\rm H_{2}}$ of HMSCs. The meanings of histograms, lines, and color-filled regions are similar to Fig.~\ref{properties:temperature} but for $n_{\rm H_{2}}$. }
    \label{properties:n_H2}
  \end{figure}

\paragraph{\textbf{(b). Density $n_{\rm H_{2}}$.}} There is a decreasing trend of $n_{\rm H_{2}}$ with distance. The distance distribution of NA is more concentrated at 4 kpc to 8 kpc (intermediate distance) compared to that of AS, thus a larger number of NA tend to be in the range of intermediate $n_{\rm H_{2}}$ to create a distribution of $n_{\rm H_{2}}$ that is different from that of AS. Therefore, the K-S test to full sample is unreliable due to the distance bias. K-S tests to AS and NA in each distance bin are not performed because of the small size of sample. The $\delta$ values indicate that there is a weak trend that HMSCs become denser (at 0.2$\sigma$ to 0.4$\sigma$ level) from NA to S-type AS at near distance range, but this trend is inverse at larger distance as shown in Fig.~\ref{figure:3} and Fig.~\ref{properties:n_H2}. The uncertainty of $n_{\rm H_{2}}$ is larger than or similar to the difference between different HMSCs.

\begin{figure}
       \centering
   \includegraphics[width=0.45\textwidth]{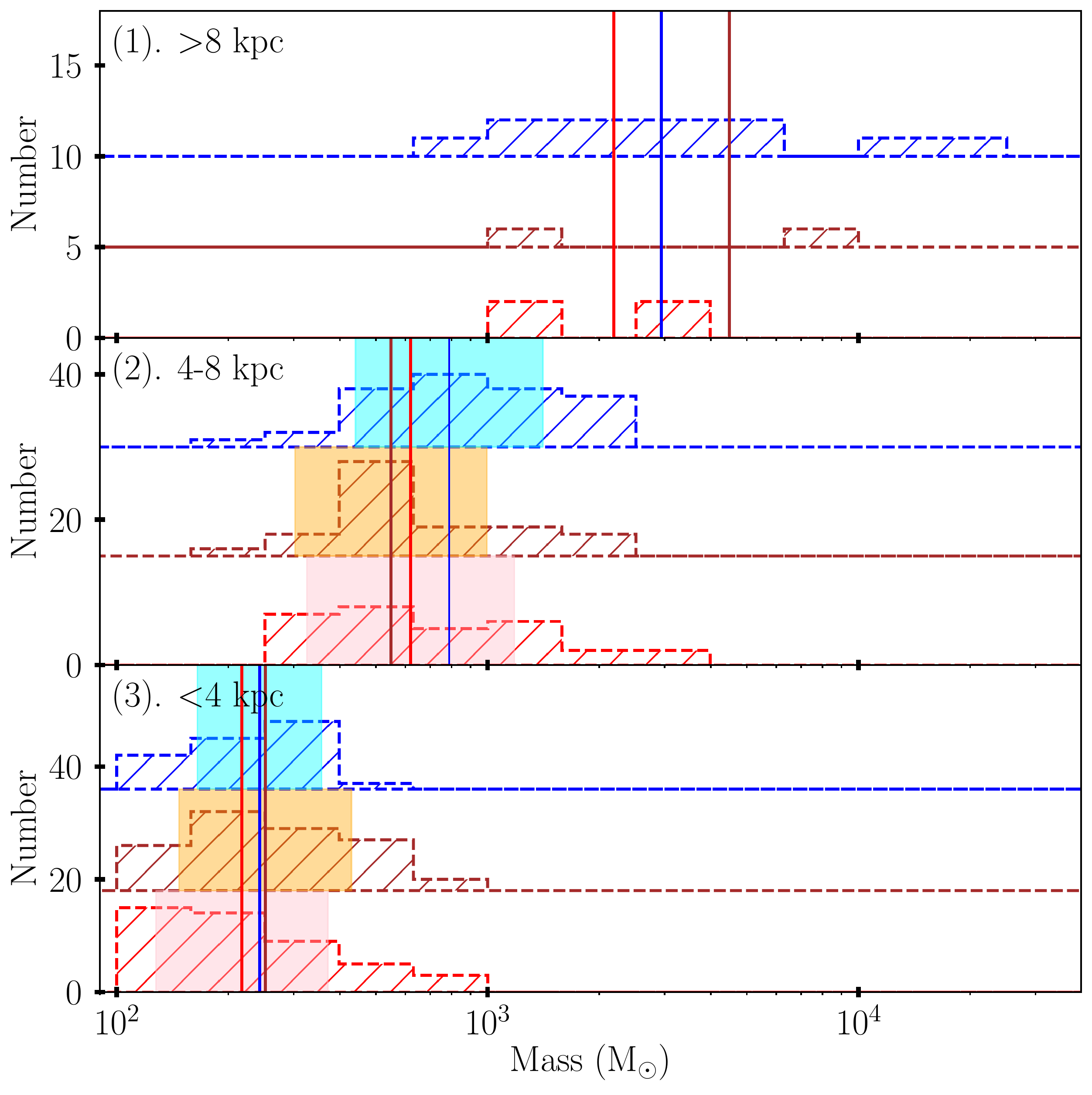}
       \caption{Number distributions of $M_{\rm clump}$ of HMSCs. The meanings of histograms, lines, and color-filled regions are similar to Fig.~\ref{properties:temperature} but for $M_{\rm clump}$. }
    \label{properties:Mclump}
  \end{figure}

\paragraph{\textbf{(c). Clump mass $M_{\rm clump}$.}} An increasing trend of $M_{\rm clump}$ with distance is clearly shown in Fig.~\ref{distance-scatter} and Fig.~\ref{properties:Mclump}. The reason is that our sample of HMSCs has a physical size increasing with distance while \nhtcd~is similar. Mass distributions of clumps (clump mass function; CMF) are largely discussed by a number of research \citep{kle01, cla07, vel13, wie15}, and their results show that the high-mass end of mass distribution could be described by a power law $dN/dM_{\rm clump} \propto~{M_{\rm clump}}^{\alpha}$, where $dN$ is the number of clumps within mass bin $dM_{\rm clump}$. We could not directly perform such a fitting to our mass distribution owing to the bias that mass increases with distance and because the number of sources in different distance bins is limited. \citet{rat06} obtained a power-law index of $\alpha \simeq-2.1\pm 0.4$ for the cores (>~100~\msun) without 8~\micron~emission in IRDC at 2 kpc to 8~kpc. A tentative fitting to all HMSCs and AS HMSCs ($M_{\rm clump} > 500$~\msun~and the mass completeness at 4~kpc is about 450~\msun) results in $\alpha \simeq-2.38 \pm 0.49$ and $-2.23 \pm 0.32$ (see Fig.~\ref{ClumpMassFunction}), which is similar to the results of \citet{rat06}. This tentative fitting sheds light on the facts that the commonly existing distance bias could produce an unreal CMF. Including more distant clumps, which are generally more massive, could make mass distribution flatter at high-mass end. We propose that distance bias should be seriously considered when studying CMF.

The question arises whether environmental factors impact the mass function of massive clumps \citep{liu18}. \citet{wie15} suggested that power-law index ($-$2.2 $\pm$ 0.1) is independent from the Galactic scale environment and evolutionary stages by studying the massive end (>~1000~\msun) of ATLASGAL clumps at different Galactocentric distances for various kinds of sources. The question regarding whether different types of HMSCs could have a similar or different CMF need to be confirmed in a larger sample with a similar distance \citep{olm18}.

The $\delta$ values show that NA are slight massive than S-type AS at 0.3$\sigma$ to 0.5$\sigma$ level in the same distance bins according to Fig.~\ref{figure:3}. The uncertainty of clump mass, which is about 20\% to 40\%,  mainly comes from the uncertainty in distance and column density. The mass difference between different HMSCs is less than its uncertainty.

\begin{figure}
       \centering
    \includegraphics[width=0.45\textwidth]{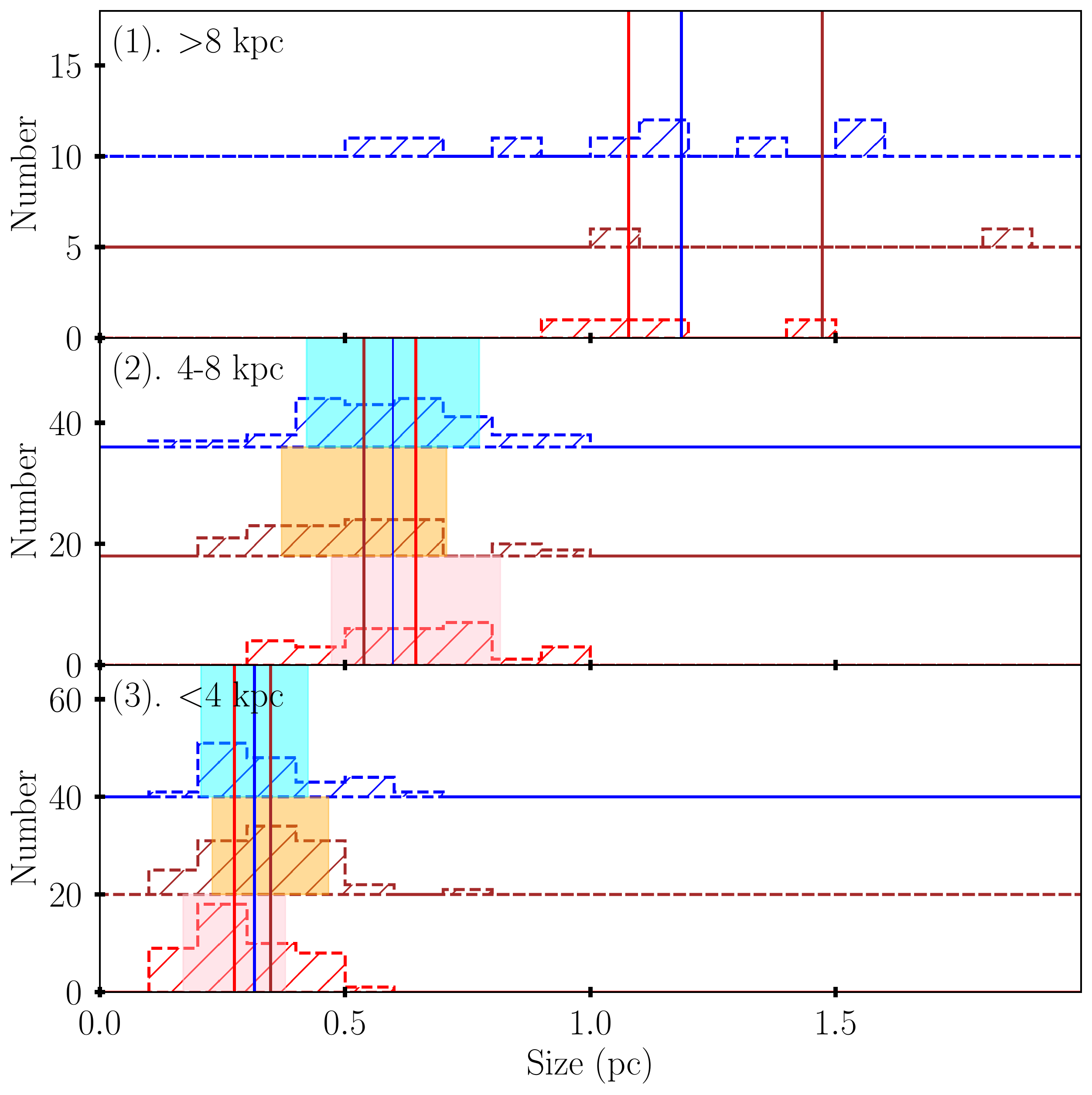}  
       \caption{Number distributions of size of HMSCs. The meanings of histograms, lines, and color-filled regions are similar to Fig.~\ref{properties:temperature} but for size. }
    \label{properties:Size}
  \end{figure}

\paragraph{\textbf{(d) Size.}} The sizes of our HMSCs are in core-to-clump regime (0.1 to 1~pc) and increase with the distance as shown in Fig.~\ref{distance-scatter} and Fig.~\ref{properties:Size}. The impact of \hii~regions on the size of HMSCs is complex. There is no consistent differences between AS and NA in different distance bins. S-type AS are smaller than NA at 1$\sigma$ level in <~4~kpc bin while they are larger at 4 to 8~kpc. Furthermore, there is no consistent changes from S-type AS to O-type AS and then NA.

\paragraph{\textbf{(e) Eccentricity.}} The eccentricity of HMSCs is calculated with $e_{\rm clump} = \sqrt{a^2-b^2}/a$, where $a$ and $b$ are the semimajor axis and semiminor axis of the ellipse defining the shape of clump \citep{zha16}. K-S tests show eccentricity distributions are similar between S-type AS, O-type AS, and NA. Furthermore, $\delta$ values in different distance bins show an inconsistent trend. \\

         \begin{figure}
       \centering
   \includegraphics[width=0.45\textwidth]{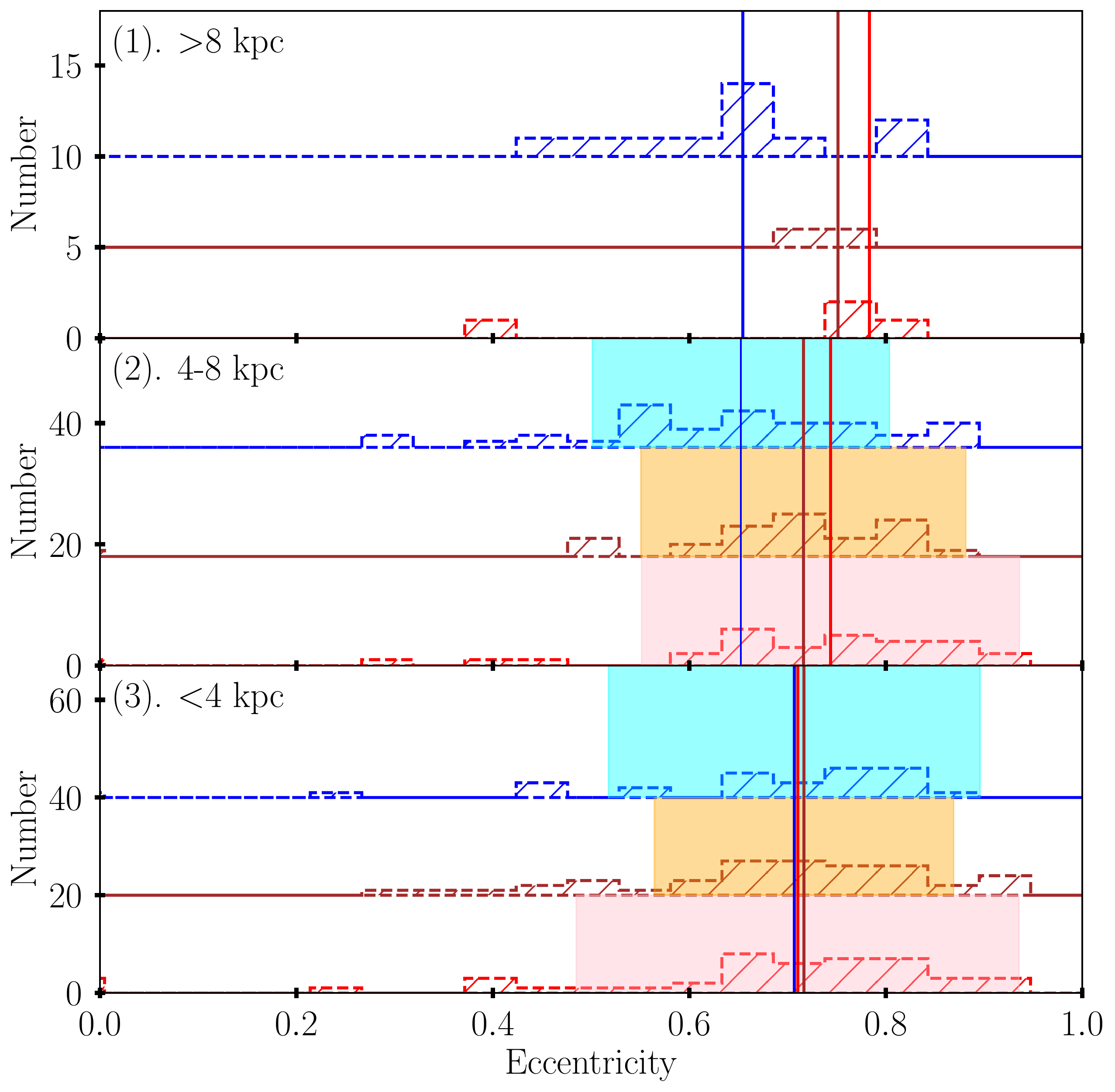}
       \caption{Number distributions of eccentricity of HMSCs. The meanings of histograms, lines, and color-filled regions are similar to Fig.~\ref{properties:temperature} but for eccentricity. }
    \label{properties:Ecc_ratio}
  \end{figure}

   \begin{figure}
   \includegraphics[width=0.45\textwidth]{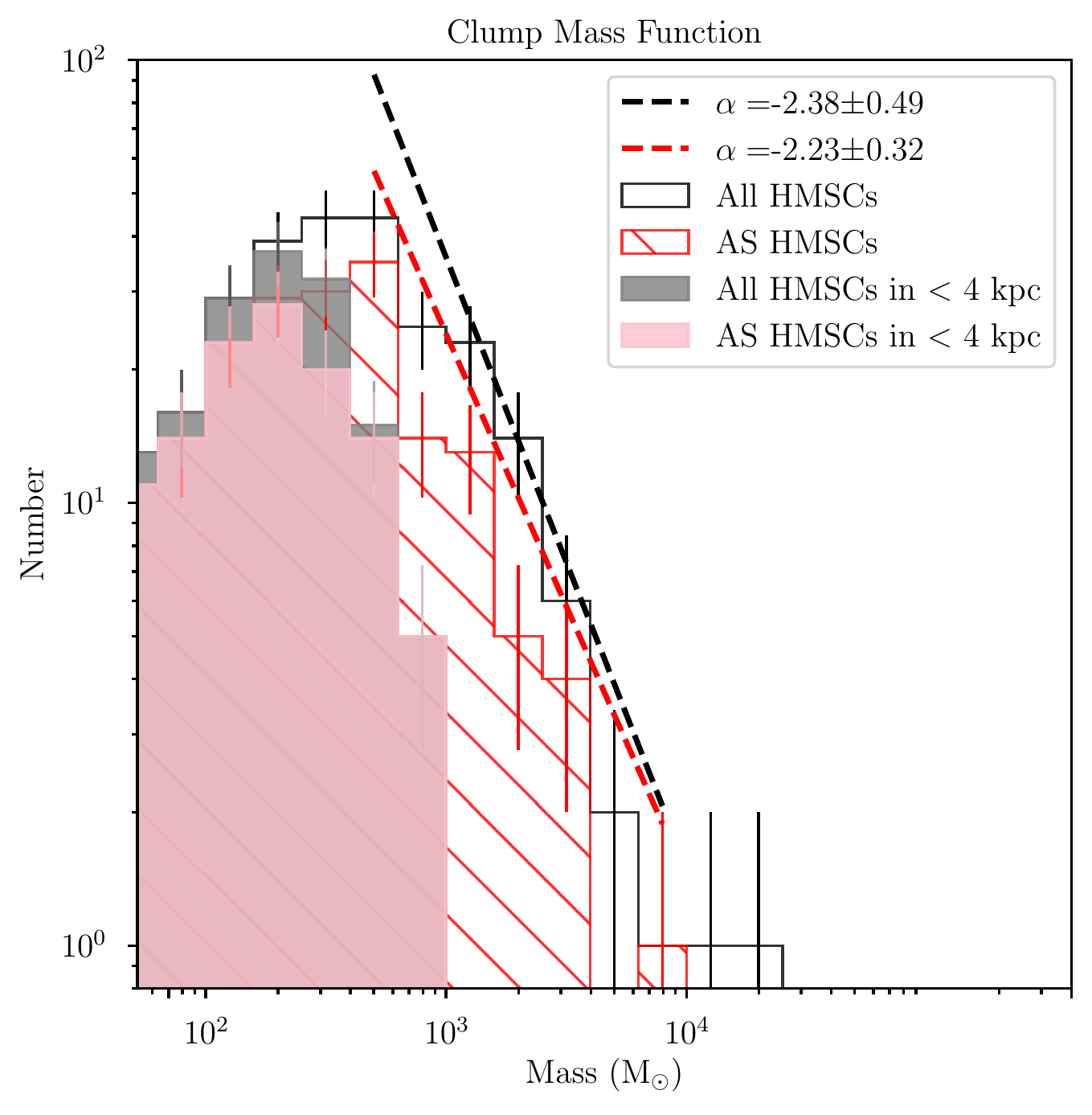}
       \caption{Clump mass function. The red hatched and black histograms show the number distributions of all selected AS HMSCs and all selected HMSCs, respectively. The black and red dashed lines indicate the fittings with power-law index of $-$2.38 and $-$2.23, respectively. The gray and pink histograms show the number distributions of all HMSCs and AS HMSCs in < 4 kpc, respectively.}
        \label{ClumpMassFunction}
    \end{figure}

\section{Independent check on CH$_{3}$C$_{2}$H kinematic temperature and $L/M$ ratio relations}  \label{appendix-check}
\begin{figure*}
    \centering 
   \includegraphics[width=0.8\textwidth]{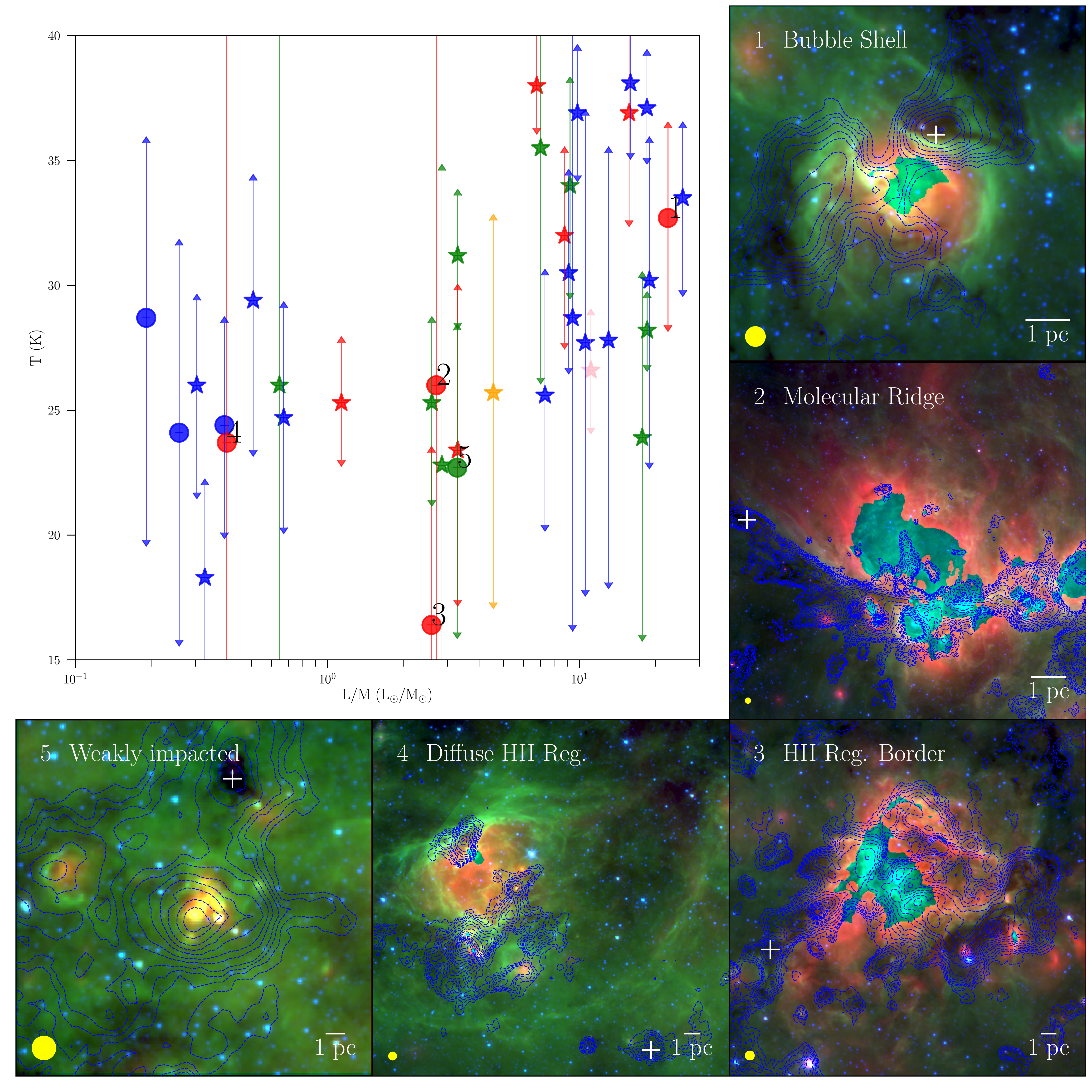}
       \caption{CH$_{3}$C$_{2}$H temperature, $L/M$ ratio, and environments. The top left plot shows $L/M$ and CH$_{3}$C$_{2}$H gas temperature of massive clumps. The dots and stars represent the 70~\micron~weak clumps and IR weak clumps, respectively. The red, green, pink, orange, and blue labels represent S-type associated, O-type associated, S-type possibly associated, O-type possibly associated, and nonassociated clumps, respectively. $L/M$ ratio, CH$_{3}$C$_{2}$H temperature and its upper and lower limits are taken from \citet{gia17}. Other plots show the environments of all 70~\micron~weak clumps associated (or possibly associated) with \hii~region by \textit{Spitzer} RGB images (constructed in the same methods as Fig. \ref{velocity_determination}). The contours and yellow dots in the bottom left of the figures show the ATLASGAL 870~\micron~continuum and its resolution.}
        \label{lm-gas-temperature}
    \end{figure*}

About 100 ATLASGAL massive clumps in the inner Galaxy are included in the survey of \citet{gia17}. The selected massive clumps are separated into four categories according to star formation indicators observed in IR and radio images: (1) 70~\micron~weak clumps: These clumps are weak at 70~\micron~thus they are considered as IR quiescent. They represent the clumps at the earliest evolutionary stages similar to our HMSCs. (2) IR weak clumps. These clumps are associated with 70~\micron~compact sources or weak 24~\micron~sources. (3) IR bright clumps. These clumps are associated with strong IR sources. (4) Compact \hii~regions. We use 70~\micron~weak massive clumps and IR weak massive clumps to carry out the check. We cross match the clumps with \hii~region in a method similar to that in Sect.~\ref{HRCM}. The clumps are classified as associated, possibly associated, and nonassociated clumps. The results are shown in Fig.~\ref{lm-gas-temperature}. Five IR quiescent massive clumps associated with \hii~regions are shown in Fig. \ref{lm-gas-temperature}. Source 1 is located at the edge of a bubble; ATLASGAL 870~\micron~continuum emission contours indicate that the clump is being compressed. Source 2 is located in the famous molecular ridge of the NGC 6334 \hii~region; 870~\micron~continuum shows that the ridge is being compressed. Source 3 is located at the edge of the \hii~region with complex morphology and intense star formation. Source 4 is located at the edge of a very diffuse bubble. The surrounding diffuse 24~\micron~emission indicates that this clump is weakly impacted by the \hii~region. Its $L/M$ ratio is quite similar to the source that is nonassociated with \hii~regions. Source 5 is possibly associated with \hii~region owing to the lack of \hii~region velocity.

It is significant that the $L/M$ ratio and CH$_{3}$C$_{2}$H kinematic temperature of Source 1 are shifted to the level of IR weak clumps. Furthermore, $L/M$ ratios of other sources associated with \hii~regions are also shifted to the $L/M$ regime of IR weak clumps. 

\end{appendix}

\end{document}